\documentclass[breaklinks=true]{aa}

\usepackage{silence}
\usepackage{graphicx}
\usepackage{lscape}
\usepackage{txfonts}
\usepackage{stackengine}
\usepackage{subcaption}
\usepackage{geometry}
\usepackage{float}
\usepackage{tabularx}
\usepackage{longtable}
\usepackage{booktabs}
\usepackage{gensymb}
\usepackage{textgreek}
\usepackage{textcomp}
\usepackage{multirow}
\usepackage{orcidlink}
\usepackage{hyperref}

\usepackage{natbib,twoopt}

\bibpunct{(}{)}{;}{a}{}{,}   

\hypersetup{
    colorlinks=true,
    linkcolor=red,
    citecolor=blue,
    urlcolor=blue
}

\makeatletter
\let\linenumbers\relax
\let\endlinenumbers\relax
\let\internallinenumbers\relax
\let\resetlinenumber\relax
\makeatother

\begin{document}

   \title{Discriminating blazar emission models with high-energy polarimetry:
Multi-band predictions and detectability} 
   
   \titlerunning{Discriminating blazar emission models with high-energy polarimetry}
    \authorrunning{S. Capecchiacci et al.}

\author{
  Sara Capecchiacci \inst{\ref*{physics_crete}, \ref*{forth}}\orcidlink{0009-0007-1918-577X} \and
  Ioannis Liodakis \inst{\ref*{forth}}\orcidlink{0000-0001-9200-4006} \and
  Haocheng Zhang \inst{\ref*{baltimore}, \ref*{nasa_goddard}}\orcidlink{0000-0001-9826-1759} \and
  Michela Negro \inst{\ref*{lousiana_stateuni}}\orcidlink{0000-0002-6548-5622} \and
  Jorge Otero-Santos \inst{\ref*{infn_padova}}\orcidlink{0000-0002-4241-5875}
}
\institute{
    Department of Physics, University of Crete, 70013, Heraklion, Greece \label{physics_crete} \and
  Institute of Astrophysics, Foundation for Research and Technology-Hellas, 70013, Heraklion, Greece \label{forth} \and
  University of Maryland Baltimore County Baltimore, MD 21250, USA \label{baltimore} \and
  NASA Goddard Space Flight Center Greenbelt, MD 20771, USA \label{nasa_goddard} \and
  Department of Physics and Astronomy, Louisiana State University, LA, USA \label{lousiana_stateuni} \and
  Istituto Nazionale di Fisica Nucleare, Sezione di Padova, 35131 Padova, Italy \label{infn_padova}
}

  \abstract{
    Polarimetric properties of blazars provide key constraints on the acceleration mechanisms powering their relativistic jets, the high-energy emission processes involved, and the composition of the jet itself.
    We present a multi-band polarimetric study spanning from soft X-rays to $\gamma$-ray energies, considering several bands for current and future missions (0.275 keV, 2-8 keV, 0.5-10 keV, 6-35 keV, 0.2-5 MeV, and 1-100 GeV).
    Our sample is drawn from the RoboPol monitoring programme, a statistically complete $\gamma$-ray sample including low-, intermediate-, and high-synchrotron-peaked blazars.

    Using spectral energy distribution fitting performed with the \texttt{Bjet\_MCMC} tool, we give predictions on the flux expected for each source in the selected energy bands.
    We model the polarimetric signatures under three competing emission scenarios: leptonic, hadronic, and hybrid.
    The detectability of each source was assessed by accounting for the instruments’ minimum detectable polarisation (MDP) and by computing the probability of the detection in a blind survey.
    We show that simultaneous multi-wavelength observations can effectively discriminate between competing emission models due to the difference in the expected polarisation degree.
    Finally, we derive sensitivity requirements for future $\gamma$-ray polarimetric missions aimed at increasing the number of detectable sources.

}

   \maketitle

\section{Introduction}
\label{sect_introduction}

About 10\% of active galactic nuclei (AGN) are characterised by strong, luminous jets extended along their polar axis \citep{angel1980, urry1995}.
Their non-thermal radiation, extended at all wavelengths, dominates their whole spectrum.
This phenomenon is thought to originate from particles being accelerated to relativistic energies, and then losing energy as they move away from the acceleration region \citep{blandford2019, hovatta2019, raiteri2025}.
When the jet is directed towards our line of sight, we classify these objects as blazars.

The spectral energy distribution (SED) of jetted AGN shows two humps: the synchrotron and the high-energy hump.
The peak frequency of the synchrotron bump, which arises from synchrotron radiation emitted by electrons and positrons, allows us to separate these astrophysical objects in three classes: low-synchrotron-peaked sources (LSP) present $\nu_{peak} < 10^{14}$ Hz, intermediate-synchrotron-peaked sources (ISP) have $10^{14}$ Hz $< \nu_{peak} < 10^{15}$ Hz, and high-synchrotron-peaked sources (HSP) present $\nu_{peak} > 10^{15}$ Hz.

The origin of the high-energy component of the SED is still under debate.
The main possible scenarios consider an emission dominated by leptonic and/or hadronic processes.

Under leptonic models, the most relevant emission process is the inverse Compton scattering of low-energy photons by the same non-thermal electrons that contribute to the low energy bump. 
These photons can be the synchrotron photons themselves, in which case the process is called synchrotron self Compton (hereafter SSC), or they can come from other sources, in which case the process is referred to as external Compton (EC), for example thermal photons being reprocessed within the broad line region (BLR), within molecular clouds or within the dusty torus that surrounds the central supermassive black hole.
The leptonic scenario generally predicts lower polarisation degrees (PD) at high energies with respect to other models. 
This is due to the fact that inverse-Compton scattering reduces the polarisation of the seed photons.
Therefore, SSC would typically be less polarised than the synchrotron emission, and EC would likely be unpolarised as a consequence of the unpolarised nature of the thermal photons \citep{bonometto1973,krawczynski2012}.

Under the scenario of hadronic emission, non-thermal protons are considered to contribute significantly to high-energy emission (in the X-ray and $\gamma$-ray bands). 
In particular, the emission is thought to consist of the primary proton synchrotron and/or synchrotron from the secondary charged particles produced through hadronic interactions \citep{mannheim1992,mucke2001}. 
In this scenario, the two emission peaks do not necessarily correlate, as they originate from different particle populations, and the high-energy component can be as highly polarised as the low-energy component, or even higher \cite[e.g.,][]{zhangboettch2013}.
Another consequence of this emission scenario is the production of high-energy neutrinos: if protons are accelerated at high energies, these may produce charged pions through the interaction with the strong local radiation field, which in turn would decay and emit neutrinos \citep{zhang2019}.

The correlation between the synchrotron component and the high-energy component that has been observed in many flaring events generally supports the leptonic scenario \citep{rani2013, liodakis2018, liodakis2019, jaeger2023}; however, the association of a very high-energy neutrino event with the flaring activity of blazar TXS~0506+056 \citep{icecube2018} suggests a hadronic contribution to the emission.

Due to the fundamental differences in the PD predicted by the different models, polarisation measurements at high energies can provide much stronger constraints on the jet composition than total intensity signatures, which are often indistinguishable \cite[e.g.,][]{boettcher2012modeling}. The current and future polarimetry missions will therefore play a crucial role in our deeper understanding of the nature of these jets.

High-energy polarimetric observations of numerous blazars with the Imaging X-ray Polarimetry Explorer (IXPE, \citealt{weisskopf2022imaging,soffitta2023}), often combined with observations in other bands, proved useful in putting constraints on their emission mechanisms. Observations of HSP blazars, which probe possible acceleration processes taking place within the jet, have often favoured energy-stratified scenarios \citep{liodakis2022polarized, di2022x, kim2024magnetic, maksym2024two, chen2024x, kouch2024ixpe, cape2025}. Other studies have supported scenarios of significant turbulence in the jet flow \citep{errando2024detection}, and some have focused on the structure of the magnetic field, finding a persistent component primarily orthogonal to the jet \citep{pacciani2025}.

Regarding the high-energy component, many studies favour a leptonic emission mechanism -- such as 
\citet{middei2022x} and \citet{agudo2025} for BL Lacertae, but also \citet{ehlert2022}, \citet{peirson2023bllac}, \citet{marshall2024observations}, and \citet{liodakis2025lsp} for other LSP and ISP sources -- although hadronic models have not been ruled out. Considering more complicated magnetic field configurations and more extended proton distributions in the jet further complicates the situation \citep{zhang2024,Tavecchio2025}, stressing the need for higher-energy polarisation observations.

A new generation of high-energy polarimeters has been proposed in recent years. These include proposed missions such as EXPO, StokeSAT, e-ASTROGAM \citep{eastrogam2018}, and AMEGO-X \citep{amegox2022}, as well as selected or approved missions currently under development, such as   COSI \citep{tomsick2014}, eXTP \citep{extp2019}, and GOSoX \citep{gosox2022}.
With these new instruments, combined with current missions such as IXPE, it will be crucial to target a large sample of blazars, in order to improve our understanding of their high-energy emission and possibly confirm or rule out one or more of the emission models proposed.

In this work, we aim to give predictions about the expected flux and PD of a large sample of blazars consisting of LSP, ISP, and HSP sources.
We assess their detectability by accounting for the instruments' minimum detectable polarisation function (MDP).
Our predictions are based on three different emission scenarios -- leptonic, hadronic, and hybrid -- and we show that multi-wavelength polarimetric observations provide a powerful means to discriminate between them.
We further derive sensitivity requirements for future $\gamma$-ray polarimetry missions to detect polarisation from blazars under different scenarios.

In Section \ref{sect_sources} we describe our sample and data.
In Section \ref{sect_data_analysis} we describe the procedures followed for our data analysis, and the polarisation models taken into account.
In Section \ref{sect_results} we discuss our results, and we summarise them and draw our conclusions in Section \ref{sect_conclusions}.

\section{Sample}
\label{sect_sources}

Our sample of blazars was selected from the Robopol monitoring programme \citep{blinov2019}.
The programme was carried out between 2013 and 2017 at the 1.3m telescope at the Skinakas Observatory in Crete, Greece, using the Robopol polarimeter \citep{robopol}.
By monitoring a large number of AGN, it obtained measurements of both the PD and the polarisation angle (PA) of these sources, with the primary science goal of exploring a possible link between rotations in the PA and $\gamma$-ray flaring activity \citep{blinov2016, blinov2018}.
It is an unbiased statistically complete $\gamma$-ray ($> 2\cdot  10^{-8} \ \rm ph/cm^2/s$) and optical (<17.5 R-band magnitude) flux-limited sample of blazars from the second Fermi-LAT source catalogue \citep{nolan2012}.
It includes all sources that are visible from the Skinakas observatory with an elevation of $>40^\circ$ for at least 120 consecutive days between June and November. A more detailed description of the sample can be found in \citealt{pavlidou2014}.
Our sample consists of a total of 62 sources, 36 of which are LSP, 13 are ISP, and 13 are HSP.
A detailed description of the sample and observation dates can be found in Appendix \ref{appendix_sample}.

\section{Data analysis}
\label{sect_data_analysis}

The process followed in this work will be described in detail in the following sub-sections.
The multiwavelength flux data were obtained from the Markarian Multiwavelength Data Center (MMDC) and the Space Science Data Center (SSDC) SED builder tool.
We started by fitting the frequency binned data (Sect. \ref{sect_sed_fitting}) with the \texttt{Bjet\_MCMC} tool \citep{bjetmcmc}, and we applied two shifts to the obtained SED in order to match current observations, as described in Sect. \ref{sect_match_obs}.
After shifting, we extracted the flux at the central frequencies of the bands considered, described in Sect. \ref{sect_pd_vs_flux}; we then proceeded to integrate the flux over the whole bands.

We took into account three different polarisation models (leptonic, hadronic, and hybrid) for each of our SED classes (LSP, ISP, and HSP), as explained in detail in Sect. \ref{sect_pol_models}.
We shifted our models in order to match current polarisation measurements (Sect. \ref{sect_match_obs}), and extracted the predicted PD at the frequencies of interest.

We took into account the stability of the PA to estimate the exposure time needed to obtain a polarisation detection without any depolarisation effect due to PA rotations (Sect. \ref{sect_exposure_time}).
We related the predicted flux and PD of our sources (Sect. \ref{sect_pd_vs_flux}) to assess the detectability of the sources through the MDP of each instrument.
We also derived estimates of the detection duty cycles of the instruments on our sources (Sect. \ref{sect_duty_cycle}), which represent the probability of detecting a given source in a blind survey; this is based on the source's flux and PD, and the instrument's MDP.

\subsection{SED fitting}
\label{sect_sed_fitting}

The flux data used to fit the SED of our sources were obtained from the MMDC and the SSDC SED builder tool. We selected $3 \sigma$ detections and removed upper limits from our dataset.
Since the data were not time-filtered, we took into account the average flux over frequency bins, and then fitted the data through the tool \texttt{Bjet\_MCMC}, considering a one-zone SSC model.
The choice of this model was made for simplicity and to reduce computational time.
Since we are only interested in the rough shape of the SED for estimating the flux in different bands, and not in the spectral parameters of the different models, any model would have produced similar results.

The SED obtained after fitting was then shifted to match current observations.
This is described in detail in Sect. \ref{sect_match_obs}.

\subsection{Polarisation models}
\label{sect_pol_models}

For each blazar class, we considered three different models: a purely leptonic model, a purely hadronic model, and a hybrid one.
The lower-energy spectral component consists of electron synchrotron emission in all three models, and does not take into account synchrotron self absorption.
All the models are built on the one-zone spectral fitting code by \citet{boettch2013}, which includes SSC and EC, synchrotron self absorption, synchrotron from electrons and protons, pair synchrotron from hadronic cascades, and radiative cooling processes.
Post-processing was obtained through the polarisation code by \citet{zhangboettch2013}.
Since previous multiwavelength studies show that optical and X-ray polarisation properties do not always correlate (the PD can be very different, \citealt{liodakis2022polarized}, \citealt{di2022x}, \citealt{cape2025}, and polarisation angle rotations can happen at different times, \citealt{digesu2023}, \citealt{middei2023}), the models include the semi-analytical multi-zone depolarisation effects obtained by \citet{zhang2024}.
These effects are due to the fact that inhomogeneous magnetic fields in the emission regions can affect the degree of polarisation of the SSC component.
Since the synchrotron peak does not include synchrotron self absorption, and since the polarisation codes do not consider Klein-Nishina effects, our results can be considered valid within the $10^{12}-10^{25}$ Hz frequency range.

In the leptonic model, the spectrum is dominated by electron and possibly positron emission.
This model takes into account SSC (polarised) and EC (unpolarised) as contributors to the high-energy emission, with seed photons coming from direct disk emission, accretion disk emission reprocessed by the BLR, and/or an isotropic external radiation field.
In the hadronic model, both primary electrons and protons are thought to be accelerated to high energies and to contribute significantly to the total emission.
For the high-energy peak of the spectrum, this model takes into account contributions from synchrotron emission of primary protons, and from secondary electron-positron pairs.
The hybrid model takes into account the same contributions as the hadronic model, but also includes a SSC component from primary electrons.

For LSP sources, the reference models are those obtained from the source BL Lac ($RA= 22h \ 02m \ 43.2s$, $Dec = +42^{\circ}\,16^{\prime}\,39.9^{\prime\prime}$, $z=0.069$) in \citet{liodakis2025}.
For HSP, we use Mrk~501 as a reference ($RA= 16h \ 53m \ 52.2s$, $Dec = +39^{\circ}\,45^{\prime}\,36.6^{\prime\prime}$, $z=0.03412338$).
For ISP sources, we use BZBJ0211+1051 ($RA= 02h \ 11m \ 13.18s$, $Dec = +10^{\circ}\,51^{\prime}\,34.81^{\prime\prime}$, $z=0.2$).
The leptonic, hadronic, and hybrid models obtained for LSP sources are reported in Fig. \ref{fig:bllac_model}, while the ones for ISP and HSP can be found in Appendix \ref{appendix_polarisation_models} (Fig. \ref{fig:bzb_models} and \ref{fig:mrk501_models}, respectively).

\begin{figure}[h!]
    \centering
    \includegraphics[width=\linewidth]{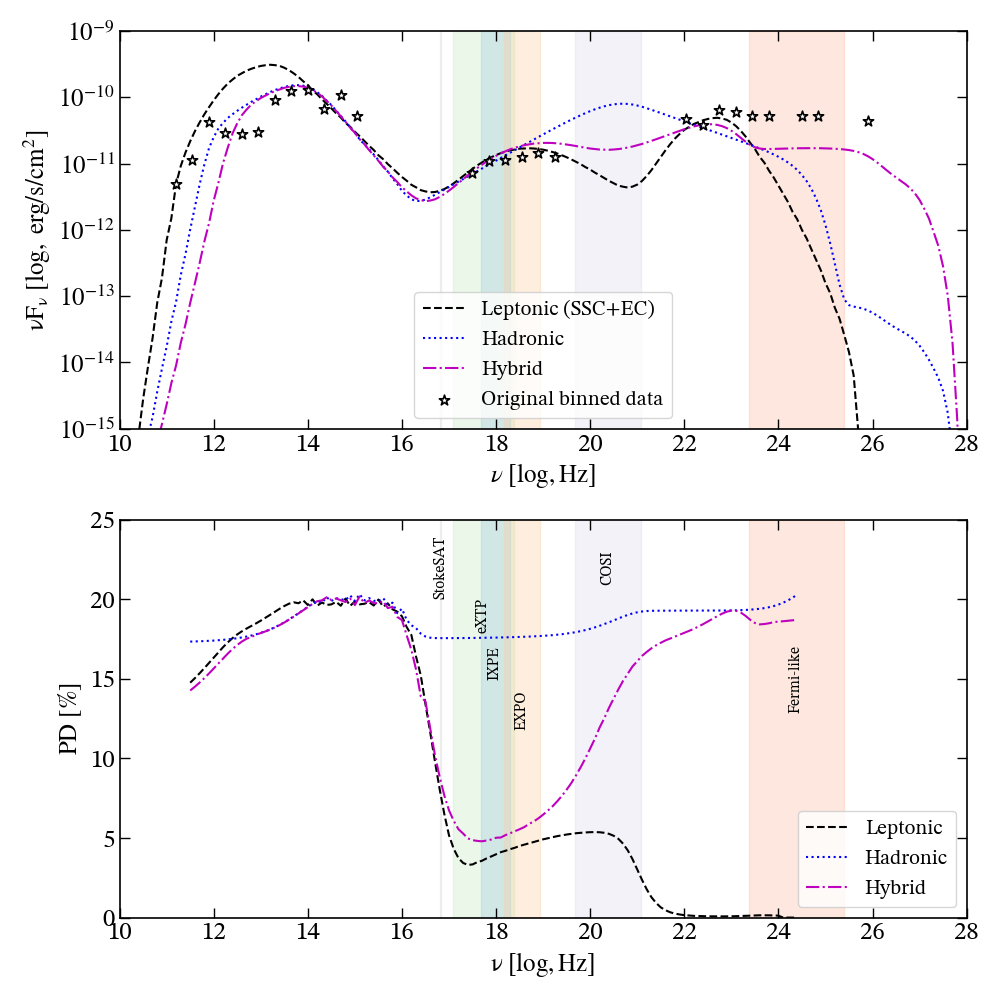}
    \caption{BL Lac model used as a reference for all our LSP sources.
    The upper panel shows the fitted SED obtained by applying the leptonic (dashed black line), hadronic (dotted blue line), and hybrid (dash-dotted magenta line) models, with the black stars representing the frequency-binned data.
    The lower panel shows the PD predicted by the three models.
    The shaded areas represent the observation bands of the considered instruments: from left to right, StokeSAT, eXTP, IXPE, EXPO, COSI, and a Fermi-like instrument.
    }
    \label{fig:bllac_model}
\end{figure}

\subsection{Matching with observations}
\label{sect_match_obs}

After fitting our data with \texttt{Bjet\_MCMC}, we applied two multiplicative scaling factors in frequency and flux to obtain a SED compatible with the flux observations at high energies.
The horizontal shift was made in order for the synchrotron peak to match the value reported in Fermi 4LAC-DR3 \citep{4lac2022}.
The vertical shift was applied to match the flux in the 0.3-10 keV band from the CAZ (CRTS, ATLAS, and ZTF) catalogue \citep{kouch2025}.
To obtain flux predictions, we extracted the monochromatic flux from our shifted SED at the central frequencies of the bands, and then integrated over the whole band using a power-law distribution with different photon indexes based on the class of the source (1.8 for LSP, 2.0 for ISP, and 2.3 for HSP).
For the Fermi band, we used the integrated flux values reported in the 4FGL DR4 catalogue \citep{ajello2020fourth, ballet2023}, given the difficulty of constraining the high-energy tail of the SED through direct fitting.
The fitting tool uses 100 frequency points in the range $\approx
 10^8 - 6 \cdot 10^{27}$ Hz, with logarithmic frequency steps of $\approx 0.2 \rm \ dex$.
 Therefore the uncertainty on the minimum frequency estimate can be approximated to $\approx 0.1 \rm \ dex$.

\begin{figure}[h!]
    \centering
    \includegraphics[width=\linewidth]{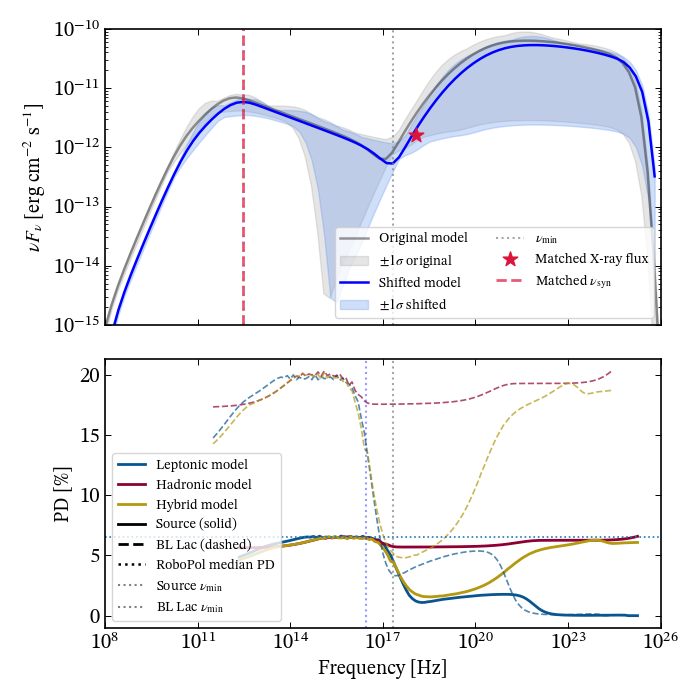}
    \caption{
    Example of the shift applied to our SED fit (upper panel) and polarisation models (lower panel). 
    In the example, an LSP (J1635+3808).
    In the upper panel, the grey line represents the original fitted SED, and the blue line the shifted SED, matched with the synchrotron peak frequency of the source reported in the 4LAC catalogue (dashed crimson line) and the X-ray flux reported in the CAZ catalogue (crimson star).
    The shaded areas represent the $1 \sigma$ error on the fit.
    The vertical dotted grey line represents the minimum between the two peaks used to shift the polarisation model (see lower panel, where the vertical dotted blue line represents the one of BL Lac).
    In the lower panel, the solid lines represent the shifted polarisation model, while the dashed lines the template model.
    The horizontal dotted black line represents the median PD measured by Robopol for the source.
    The blue lines represent the leptonic model, the red ones the hadronic, and the yellow the hybrid.
    }
    \label{fig:shifted_model}
\end{figure}

To obtain polarisation models compatible with current observations, we applied two multiplicative scaling factors in frequency and PD to the template models.
An example of this can be found in Fig. \ref{fig:shifted_model}.
We applied a frequency shift (i.e. a shift on the $x$ axis) based on the frequency of the minimum between the two bumps of the SED of our sources, represented by the vertical dash-dotted black lines in the plot.
For the template sources, the frequencies of the minimum between the two bumps were computed as the average of the three minimum frequencies of the models (for the LSP template, see Fig. \ref{fig:bllac_model}, upper panel); for the rest of the sources, they were computed from the fitted SED.
The models were horizontally shifted of a factor of $\nu_{1}/\nu_{2}$, where $\nu_{1}$ is the frequency of the minimum between the two bumps of the reference model source, and $\nu_{2}$ is the one of the source considered.
We applied a PD shift (i.e. a shift on the $y$ axis) so that the PD in the R band matches the median PD measured by Robopol during the monitoring programme; in the plot, the median PD is represented by the horizontal dotted black line.

Note that this approach assumes that the overall shape of the PD spectrum is preserved across sources, which is a reasonable approximation for synchrotron-dominated hadronic scenarios.
However, in leptonic models, depolarisation effects (e.g. due to SSC emission) and the presence of unpolarised EC components are not expected to scale linearly.
In hybrid scenarios, where the total emission arises from a combination of leptonic and hadronic processes, the relative contribution of each component can vary with energy, and the PD spectrum may not be preserved under simple scaling.
Therefore, the predicted PD values for leptonic and hybrid models should be considered approximate.
However, since we are interested in the properties of the population and the typical range of PD in different bands (shown with the error bars) is large, this approximation should not have any significant effect on our results.

After applying the shifts to our fitted SED and to our polarisation models, our predictions can be considered to be compatible with the observations carried out on our sources, in terms of both flux emission and PD.

\subsection{Estimation of optimal exposure time}
\label{sect_exposure_time}

To estimate the optimal exposure time required by each instrument to obtain a reliable polarisation measurement, we accounted for the fact that strong rotations of the PA can lead to depolarisation when averaged over long observations, and can therefore lead to a reduced observed PD or even to non-detections.
To minimise this effect, observations should be performed over time intervals during which the PA remains approximately stable.

We defined a PA rotation event as a monotonic variation of the PA with a total amplitude larger than $90^\circ$, and we excluded such events from the RoboPol polarisation light curves.
We then measured the median duration of continuous observation intervals, which we interpret as representative timescales of PA stability.
Using this approach, we find typical PA-stable time windows of approximately one month for ISP sources, about two weeks for HSP sources, and sometimes even shorter durations for LSP sources.

Based on these results, we chose to take into account two-week and four-week exposure times.
These timescales both ensure the stability of the PA and the minimum exposure time required to reach a detectable polarisation signal with the considered instruments, although blazars generally present high variability in both flux and PA, and therefore these estimates need to be treated with caution.
By restricting the effective exposure to these intervals, we ensure that the predicted polarisation detectability is not artificially modified by averaging over periods in which the PA undergoes large, rapid variations.

\subsection{PD prediction}
\label{sect_pd_vs_flux}

From the models obtained as explained in the previous subsections, we were able to extract PD values corresponding to our frequencies of interest. We took into account different instruments: StokeSAT (0.270-0.280 keV, in the soft X-ray range),
IXPE (in the 2-8 keV X-ray range, \citealt{weisskopf2022imaging}),
eXTP (in the 0.5-10 keV X-ray range, \citealt{extp2019}),
EXPO (in the 6-35 keV X-ray range),
and COSI (0.2-5 MeV, in the soft $\gamma$-ray range, \citealt{tomsick2014}).
We also considered, for a prediction on future mission requirements, a Fermi-like instrument in the 1-100 GeV range.

StokeSAT is a mission concept (NASA’s Pioneer-scale) in the soft-X energy range; due to the narrow energy range covered, we treat this band as monochromatic at 0.275 keV.
IXPE, a joint NASA-ASI mission launched on December 9, 2021, is currently operational, and provides measurements of X-ray polarisation within the 2-8 keV band.
eXTP (enhanced X-ray Timing and Polarimetry mission) is intended to be launched in 2030 and will observe the sky in the $0.5-10$ keV energy range; one of its many science goals is to study the accretion processes around the supermassive black holes of AGN.
EXPO (the Enhanced X-ray Polarimetry Observatory) is an ESA-proposed M-class mission, set to operate in the 6-35 keV energy range, with a similar sensitivity to eXTP in both hard and soft X-rays; the main goal of this satellite is to enable time-resolved polarisation measurements across the full X-ray band, which is crucial to discriminate between competing acceleration models and understand how energy is transferred from magnetic fields to particles in relativistic jets.
COSI (Compton Spectrometer and Imager) is a selected NASA gamma-ray mission expected to be launched in mid-2027, capable of imaging, spectroscopy, and polarimetry of astrophysical sources \citep{tomsick2021}; it will perform pioneering polarimetric studies of different kinds of astrophysical sources, including transients and several bright persistent sources, through its continuous all-sky monitoring operating mode.

After extracting the predicted PD in each of the bands of these instruments, we related it to the predicted flux of the source extracted from our fitted SED.
With these PD-flux pair predictions, we can assess the detectability of the sources through the minimum detectable polarisation function of each instrument.

The MDP generally depends on the exposure time of the observation and the flux of the source.
We compared our predictions on PD and flux with the MDP of the instruments in order to understand which sources are potentially detectable.
For COSI, we used the MDP estimated in \citealt{Latiolais2026}, which accounts for the dominant expected instrumental and astrophysical background.
We report our results on LSP sources for eXTP (MDP obtained through private communications) in Figure \ref{fig:mdp_extp_lsp}.
The rest of the plots can be found in Appendix \ref{appendix_mdp_plots}.

\begin{figure}[h!]
    \centering
    \includegraphics[width=\linewidth]{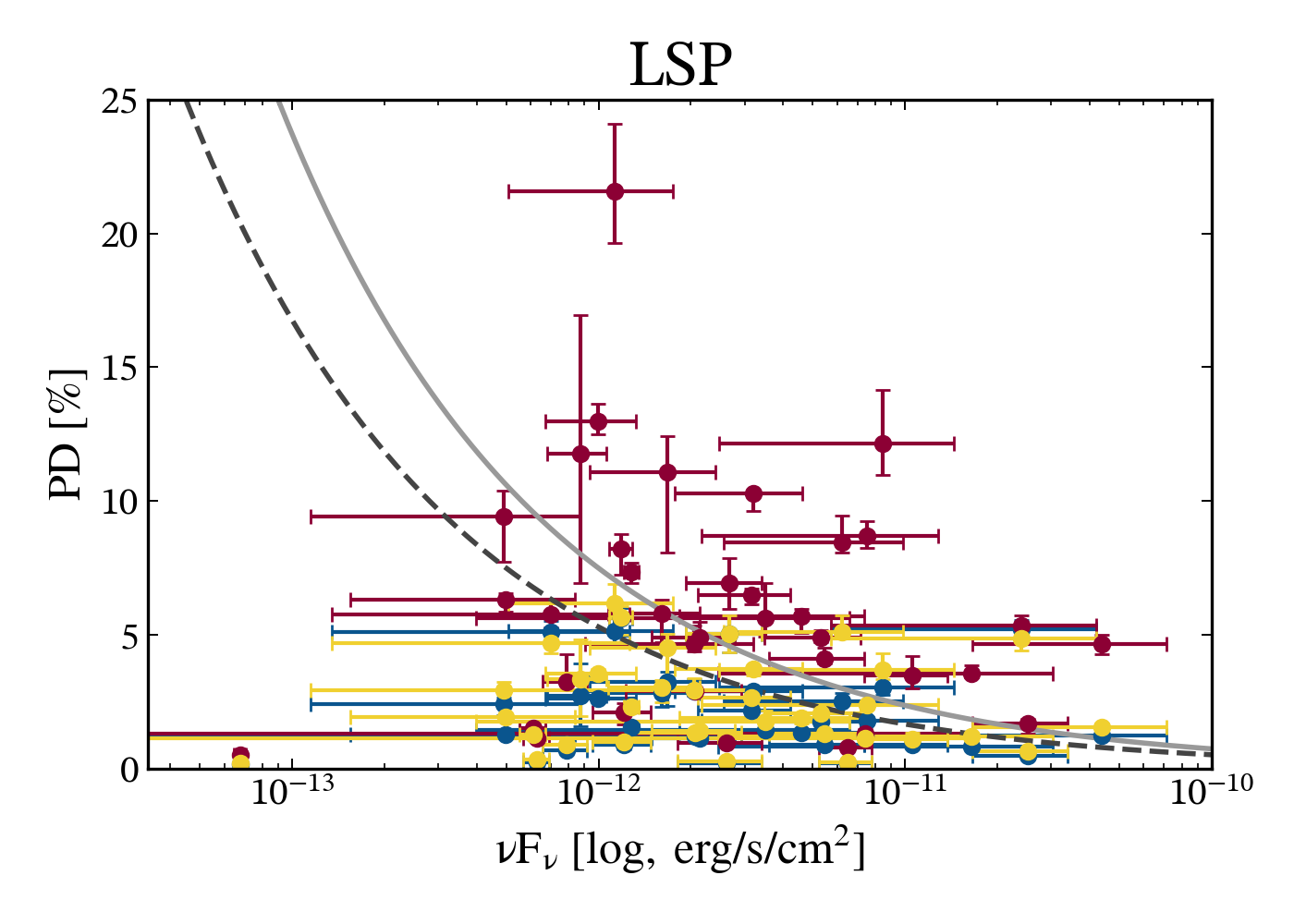}
    \caption{Predicted flux and PD for the LSP sources in the 0.5-10 keV band.
    The curves represent the MDP(99\%) of the eXTP instrument for two week exposures (solid light grey) and four week exposures (dashed dark grey).
    Blue points represent predictions from the leptonic model, yellow points from the hybrid model, and red points from the hadronic model.
    The sources that lie above the lines are detectable by eXTP.
    }
    \label{fig:mdp_extp_lsp}
\end{figure}

The error bars of each point were computed as follows.
We took the relative error (ratio between flux and standard deviation) of the CAZ values in the 0.3-10 keV band, and used the proportion to compute the error on the flux in the other bands.
From the Robopol light curves, we computed the median PD value and the $68 \%$ confidence levels; we then computed the ratio between the median PD and the confidence level amplitude, and used this proportion to build the error bars on the $x$ axis.

\subsection{Duty cycle estimation}
\label{sect_duty_cycle}

With the upcoming new generation of polarimeters, it is not only useful to predict the results of future observations, but also to identify which sources are expected to be readily detectable and which are likely to remain below the sensitivity limits.
This distinction is fundamental for future observation proposals, as it allows for an optimised selection of targets and exposure times, minimising the risk of non-detections.
For this reason, we studied the duty cycle of the instruments over each of our sources, which represents the probability of the source being detected in a blind survey.

To obtain an estimate of the duty cycles, we picked 1000 random values of X-ray flux from a Gaussian distribution centred on the CAZ reported value with the corresponding standard deviation, and 1000 random PD values using the CDF obtained from the Robopol light curves.
This process corresponds to picking 1000 possible observations of flux in the X-rays, and PD in the optical range.

Following the same procedure as explained in Section \ref{sect_match_obs}, we then shifted our fitted SED and our polarisation models using the extracted pairs as reference values of PD and flux, and therefore obtained 1000 shifted SED and polarisation models for each source.
We extracted the PD and flux values at our frequencies of interest and integrated the flux over the whole bands.
At this point, we have 1000 possible observations (PD-flux pairs) for each source, for every band.

Each instrument has a MDP function that strongly depends on the source's flux and the exposure time (the MDP curves of eXTP for exposure times of two weeks and four weeks are represented in Fig. \ref{fig:mdp_extp_lsp}, where the LSP sources lying above the MDP curves are detectable; the rest of the plots can be found in Appendix \ref{appendix_mdp_plots}).
We used this function to determine the detectability of the 1000 states (represented by the PD-flux pairs) of each source.
We then computed the percentage of observations that would be successful, which represents the duty cycle of our instrument over the single source.

\section{Results}
\label{sect_results}

In this section we present and discuss the results of this work.
In particular, we focus on the detectability of our sources in different bands (Sect. \ref{sect_duty_cycle_results}), on the difference between the results expected from the three emission models taken into account (Sect. \ref{sect_different_models}), and on the requirements for future missions that ensure increased duty cycles, which represent the probability of detecting a given source in a blind survey (Sect. \ref{sect_future_missions}).

\subsection{Detectability and duty cycle}
\label{sect_duty_cycle_results}

We assessed the polarisation detectability of the sources in the sample as explained in Sect. \ref{sect_pd_vs_flux}.
We report in Fig. \ref{fig:mdp_extp_lsp} the results for eXTP on the LSP sources; the rest of the plots can be found in Appendix \ref{appendix_mdp_plots}.

According to our results, for exposures of both 14 and 28 days StokeSAT will be able to detect most of the LSP, ISP, and HSP sources, regardless of the emission model considered.
For exposures of 14 and 28 days, IXPE is able to detect about 50\% of HSP sources under hadronic and hybrid models, about $1/3$ of them under leptonic models, and only a few ISP regardless of the emission model, while the detectability of the LSP sources strongly depends on the polarisation model (a larger number of detectable sources for the hadronic model, while leptonic and hybrid models lead to only BL~Lac being detectable).
Our predictions for HSP sources, such as Mrk~501, 1ES~1959+650, and PG~1553+113 in the 2-8 keV band, are consistent  with actual IXPE observations (\citealt{liodakis2022polarized, middei2023, chen2024x, errando2024detection, pacciani2025, cape2025}).
For both exposure times, eXTP will be able to detect most of our LSP sources under hadronic emission, but only a few under the leptonic and hybrid model, and most of the HSP and ISP sources regardless of the emission mechanism.
EXPO will be able to detect most of the sources of HSP and LSP classes, with the number of detectable LSP sources increasing drastically under the hadronic model, and about 50\% of the ISP regardless of the emission scenario.
According to our results, COSI will only be able to detect polarisation from 1ES~1959+650 (HSP).

\begin{figure}[h!]
  \centering
    \includegraphics[width=\linewidth]{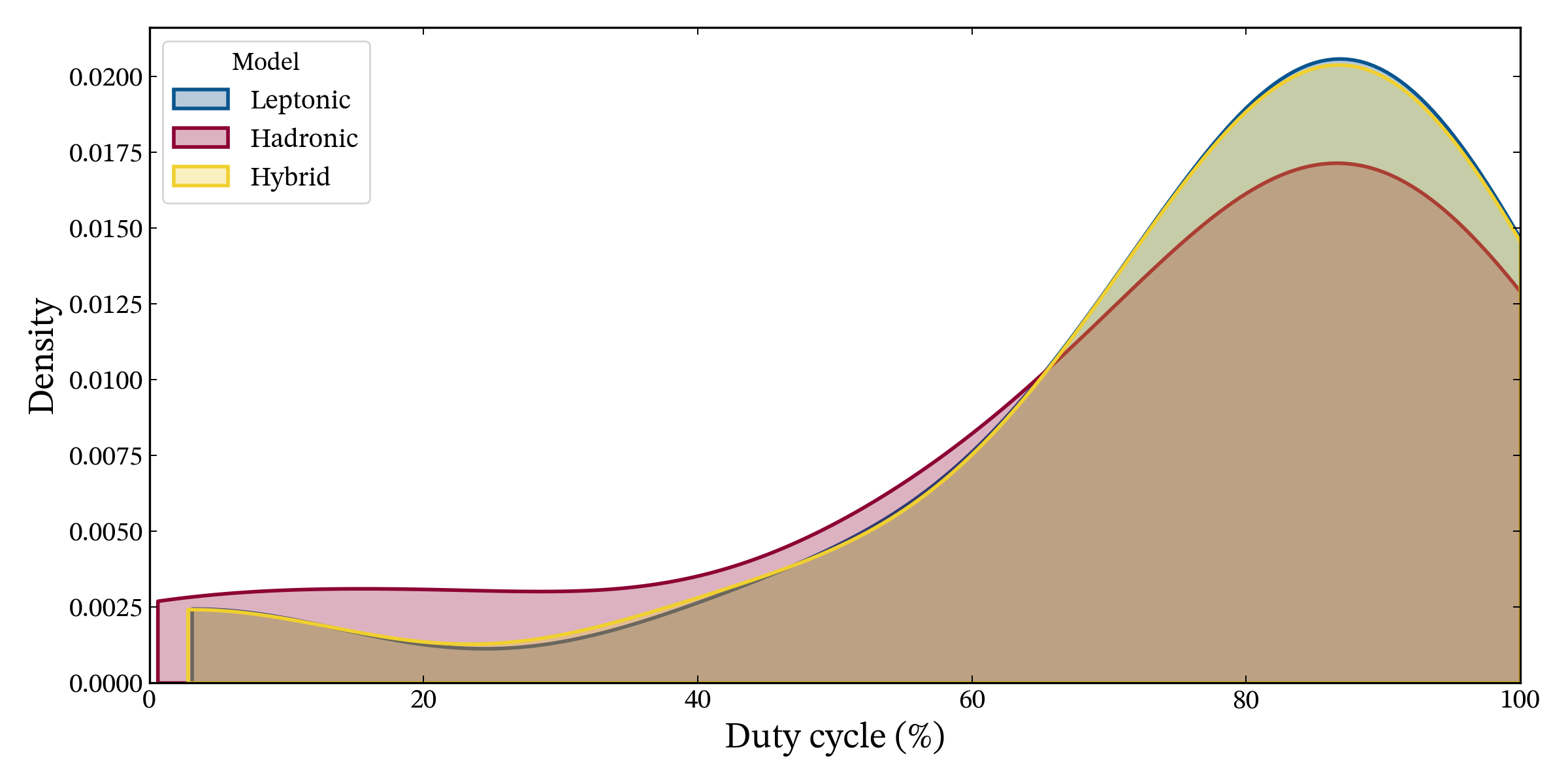}
    \caption{Kernel density distribution of the duty cycles for HSP sources in the 0.5-10 keV band (eXTP) for 28 day exposures.
    The blue curve represents the leptonic model, the red one the hadronic model, and the yellow one the hybrid model.}
    \label{fig:dutycycle_extp_hsp}
\end{figure}

We plot the duty cycles of our instrument using kernel density estimation, which represents the distribution through a continuous probability density curve.
Figure \ref{fig:dutycycle_extp_hsp} represents the case of HSP in the eXTP band for 28 day exposures.
The duty cycle plots for all other instruments can be found on the \href{https://doi.org/10.5281/zenodo.20327371}{Zenodo} platform.
Some of our instruments' duty cycles are close to $0.0\%$ for all sources of a given class; therefore, we only report relevant plots.

\subsection{Discerning between different models}
\label{sect_different_models}

\begin{figure}[h!]
  \centering

  \begin{subfigure}[t]{\columnwidth}
    \centering
    \includegraphics[width=0.85\linewidth]{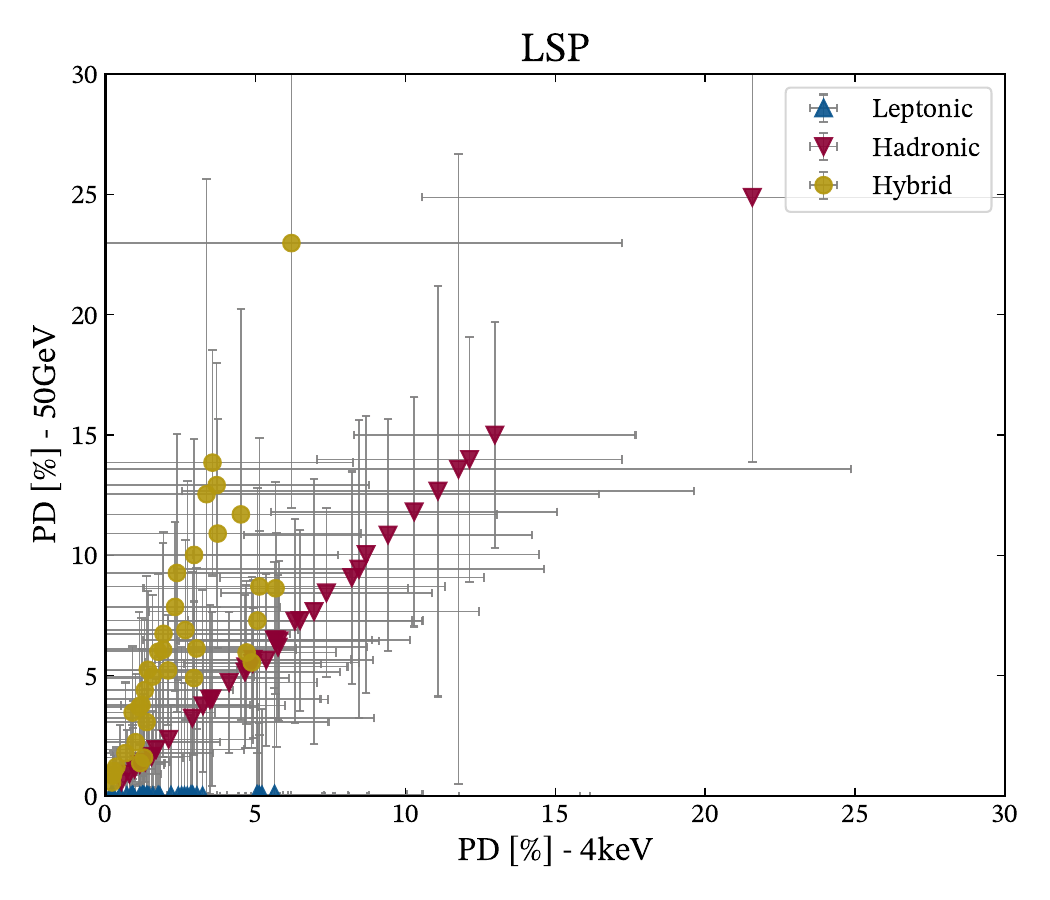}
  \end{subfigure}

  \begin{subfigure}[t]{\columnwidth}
    \centering
    \includegraphics[width=0.85\linewidth]{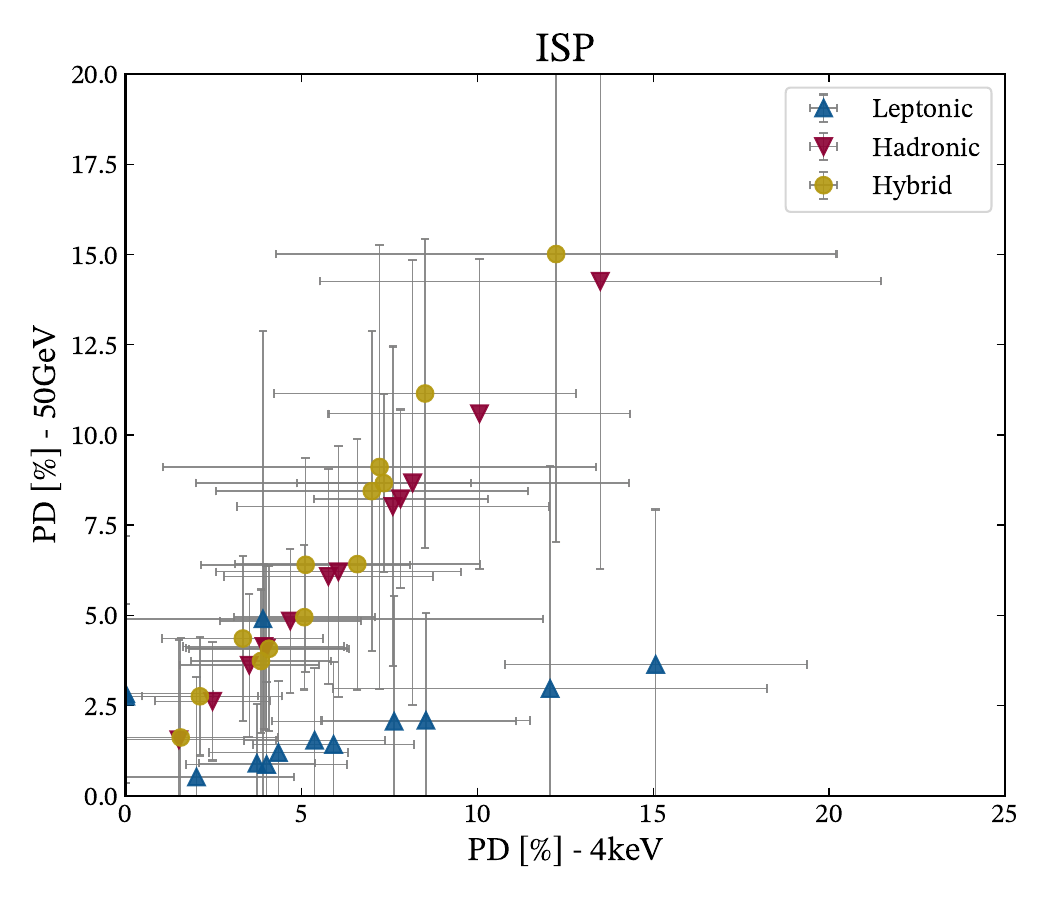}
  \end{subfigure}

  \begin{subfigure}[t]{\columnwidth}
    \centering
    \includegraphics[width=0.85\linewidth]{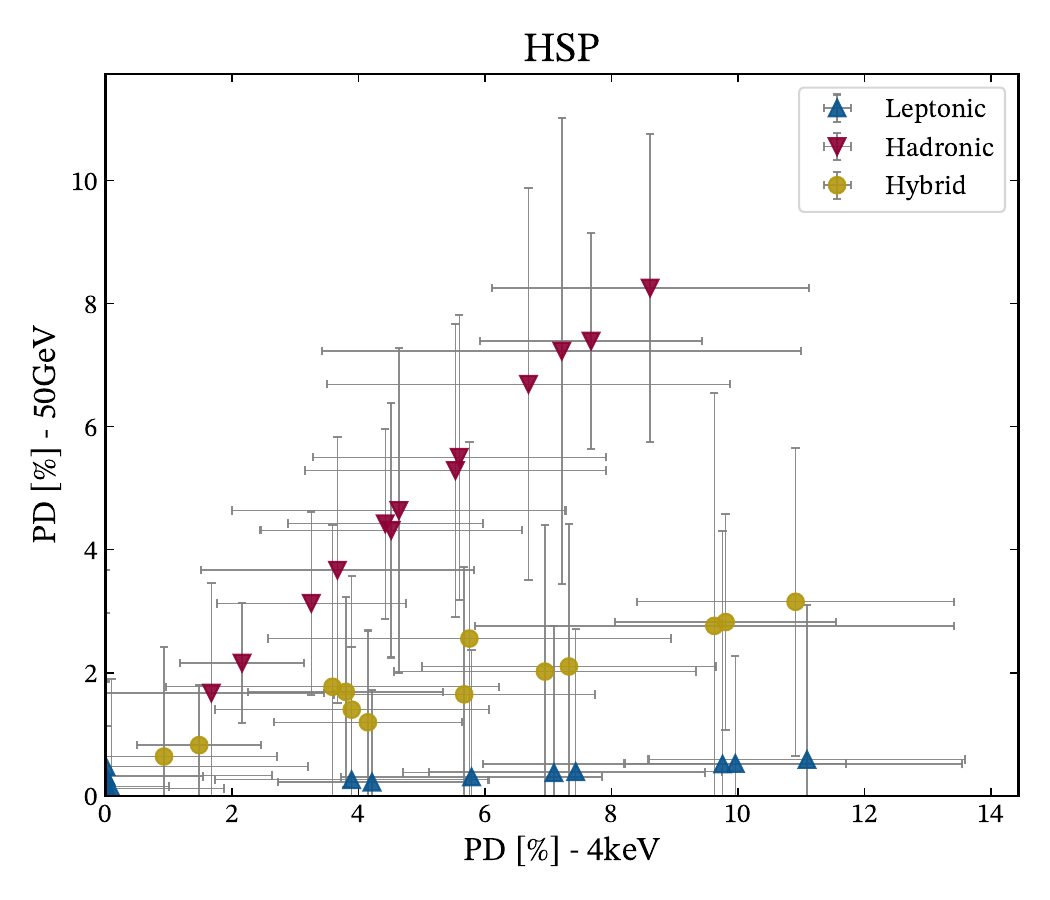}
  \end{subfigure}

  \caption{
  PD predictions in the 2-8 keV band and the 0.1-100 GeV band.
  Top to bottom: LSP, ISP, and HSP.
  Upward blue triangles represent the leptonic model, downward red triangles the hadronic one, and yellow circles the hybrid one.
  The three models give very distinct polarimetric predictions, especially for the HSP sources.
  Simultaneous observations of sources in these two bands allow us to discriminate between the different proposed emission scenarios.
  }
  \label{fig:ixpe_vs_fermi}
\end{figure}

Different polarisation models predict distinct PD across energy bands.
Observing a source simultaneously with multiple instruments will therefore allow us to distinguish between competing models.

In Figure \ref{fig:ixpe_vs_fermi} we represent the comparison between polarimetric predictions of the 2-8 keV band (IXPE) and the 0.1-100 GeV band.
The separation between the model predictions in the two energy bands, particularly evident in the HSP case, highlights the importance of polarimetry in X-rays and $\gamma$-rays in constraining the dominant emission processes. 
The rest of the comparisons between different bands can be found on the \href{https://doi.org/10.5281/zenodo.20327371}{Zenodo} platform.

According to our results, simultaneous observations of HSP in the X-ray and $\gamma$-ray bands would be crucial to distinguish between the different models, as is highlighted by Fig. \ref{fig:ixpe_vs_fermi}: the three models give very distinct PD predictions in these two bands.
Under the hadronic model, the expected PD in the two bands is comparable for all three SED classes, whereas the GeV PD is significantly lower in the leptonic model, generally $<5\%$.
Any $\gamma$-ray detection of PD higher than 5\% in the LSP sources would allow us to immediately rule out the leptonic model.

\subsection{Requirements for future missions}
\label{sect_future_missions}

To evaluate requirements for future missions, we adopted a MDP parameterisation with the same functional form as that used for IXPE \citep{weisskopf2022imaging, soffitta2023}, eXTP \citep{digesu2020}, and EXPO (similar to eXTP):

\begin{equation}
\mathrm{MDP}(99\%) =
\frac{K}{\sqrt{\left(\frac{\nu F_{\nu}}{10^{-11}}\right) \cdot
\left(\frac{T}{10}\right)}}
,\end{equation}

where $K$ is a constant, $\nu F_{\nu}$ is the specific flux in $erg\ cm^{-2} s^{-1}$, and $T$ is the exposure time expressed in days.
Note that this MDP functional form assumes a negligible background, which may not be the case for some instruments due to technology limitations, such as COSI, for which the background contribution must be taken into account.

We fitted the normalisation $K$ of the MDP function for each instrument imposing that, for an exposure of 14 days, 25\% of the sources in each SED class fall above the MDP line (i.e. the brightest 25\% of sources are detectable by the instrument).
We report in Table \ref{tab:requirements_flux} the flux sensitivity that future detectors need to achieve in order to detect 25\% of the sources in each SED class.
This was computed by setting $\mathrm{MDP}=5\%$ in our fitted MDP function.
The 5\% threshold for PD was picked based on the leptonic model results: in most cases, the PD prediction for leptonic models is below 5\%, and therefore this represents an optimal threshold to discern between different models (i.e. if we observe all sources with PD>5\%, it is likely that the leptonic model can be ruled out).
Such an improvement in detector sensitivity could prove crucial to deepening our understanding of jet dynamics.

\begin{table}[t]
\centering
\caption{Flux requirements for future missions in the 0.2-5 MeV and 0.1-100 GeV energy bands.}
\label{tab:requirements_flux}

\textbf{MeV band}

\vspace{0.3em}
\begin{tabular}{lcc}
\hline
\hline
SED class & Flux (14 d) & Flux (28 d) \\
          & [erg\,cm$^{-2}$\,s$^{-1}$] & [erg\,cm$^{-2}$\,s$^{-1}$] \\
\hline
\multirow{3}{*}{LSP}
 & $4.712 \cdot 10^{-12}$ &  $2.356 \cdot 10^{-12}$ \\
 & $1.220 \cdot 10^{-10}$ &  $6.100 \cdot 10^{-10}$ \\
 & $7.140 \cdot 10^{-11}$ &  $3.570 \cdot 10^{-11}$ \\
\hline
\multirow{3}{*}{ISP}
 & $6.461 \cdot 10^{-12}$ & $3.230 \cdot 10^{-12}$ \\
 & $2.273 \cdot 10^{-11}$ & $1.137 \cdot 10^{-11}$ \\
 & $1.875 \cdot 10^{-11}$ & $9.374 \cdot 10^{-12}$ \\
\hline
\multirow{3}{*}{HSP}
 & $1.882 \cdot 10^{-10}$ & $9.412 \cdot 10^{-11}$ \\
 & $1.116 \cdot 10^{-10}$ & $5.579 \cdot 10^{-11}$ \\
 & $1.825 \cdot 10^{-10}$ & $9.126 \cdot 10^{-11}$ \\
\hline
\end{tabular}

\vspace{1em}

\textbf{GeV band}

\vspace{0.3em}

\begin{tabular}{lcc}
\hline
\hline
SED class & Flux (14 d) & Flux (28 d) \\
          & [erg\,cm$^{-2}$\,s$^{-1}$] & [erg\,cm$^{-2}$\,s$^{-1}$] \\
\hline
\multirow{3}{*}{LSP}
 & - & - \\
 & $2.181 \cdot 10^{-10}$ & $1.091 \cdot 10^{-10}$ \\
 & $1.860 \cdot 10^{-10}$ & $9.300 \cdot 10^{-11}$ \\
\hline
\multirow{3}{*}{ISP}
 & $2.354 \cdot 10^{-12}$ & $1.177 \cdot 10^{-12}$ \\
 & $4.948 \cdot 10^{-11}$ & $2.474 \cdot 10^{-11}$ \\
 & $4.882 \cdot 10^{-11}$ & $2.441 \cdot 10^{-11}$ \\
\hline
\multirow{3}{*}{HSP}
 & $3.703 \cdot 10^{-13}$ & $1.852 \cdot 10^{-13}$ \\
 & $7.248 \cdot 10^{-11}$ & $3.624 \cdot 10^{-11}$ \\
 & $1.065 \cdot 10^{-11}$ & $5.326 \cdot 10^{-12}$ \\
\hline
\end{tabular}

\tablefoot{\centering{Flux sensitivity required to detect 25\% of the sources in the three SED classes with relative exposure times in the 0.2-5 MeV and 0.1-100 GeV energy bands.
The three rows for each SED class represent, in order, leptonic, hadronic, and hybrid models.
For the LSP leptonic model in the GeV band, PD predictions were too low to obtain detections above 5\%.}}

\end{table}

\section{Conclusions}
\label{sect_conclusions}

In this work, we built a framework that can be used to give predictions on instrument performance for current and future high-energy polarimetry missions, which we apply to a statistically complete sample of blazars.
Although we considered only upcoming and proposed missions, our framework can be readily applied to any future high-energy polarisation experiment.

In the present analysis, we consider energy bands in both the X-ray range (0.275 keV for StokeSAT, 2-8 keV for IXPE, 0.5-10 keV for eXTP, and 6-35 keV for EXPO) and the $\gamma$-ray range (0.2-5 MeV for COSI and 1-100 GeV for a Fermi-like instrument).
Our polarisation predictions consider three models that are commonly invoked to explain the high-energy emission from blazars; namely, a leptonic, a hadronic, and a hybrid model.

By analysing the variability of the polarisation angle, we estimated the optimal exposure times required to obtain polarisation detections while minimising depolarisation effects due to PA variability.
We find that the stability of the PA varies depending on the subclass of blazar; 14 days appears to satisfy all subclasses, and in some cases it can be extended up to 28 days.
However, we note that this is an average estimate that should be treated with caution, since blazar variability is often unpredictable and can be significantly different on a blazar-by-blazar basis.
Nevertheless, our analysis provides a useful benchmark for current and future studies.

We compared our estimates to the MDP of each instrument in order to assess the detectability of the PD in different blazars, and estimate a detection duty cycle, i.e. the probability of detecting a source in a blind survey.
Our results show that StokeSAT would have the highest probability of detecting sources of all classes (HSP, LSP, and ISP) in a blind survey, with very high duty cycles, followed by eXTP and EXPO (particularly sensitive to HSP).

Additionally, we used our population model to establish the necessary sensitivity for future mega-electronvolt and giga-electronvolt missions to measure polarisation from blazars.
Given  that the field is completely unexplored, we set the criteria such as 25\% of the population could be detected within an MDP of 5\%, necessary to differentiate between models.
We find that in the case of leptonic emission, instruments need to reach sensitivities of approximately $10^{-12} \ erg/cm^2s$, or down to $\sim10^{-13} \ erg/cm^2s$ in the giga-electronvolt band for HSP sources.
For hadronic models, optimal sensitivities are of the order of $\sim10^{-11} \ erg/cm^2s$, while hybrid scenarios require sensitivities of $\sim10^{-12} \ erg/cm^2s$.
Overall, mega-electronvolt and giga-electronvolt instruments capable of reaching sensitivities of $\sim10^{-11} \ erg/cm^2s$ would have a good chance of measuring the PD of these sources and, through simultaneous observations in both bands, would enable discrimination between the proposed emission models.

Detecting a large population of sources in the X-ray and $\gamma$-ray energy range is essential to discriminate between competing emission models.
Our results therefore demonstrate the need for next-generation instruments that can be decisive in advancing our understanding of multi-messenger emission from relativistic jets.

\section*{Data availability}

Supplementary tables with the description of our sample and observations and the duty cycles computed for the instruments, as well as supplementary plots for the comparison between different bands and the duty cycles of the instruments over our sample, have been uploaded on the Zenodo platform (\url{https://doi.org/10.5281/zenodo.20327371}).
The supplementary tables are also available in electronic form at the CDS (\url{http://cdsweb.u-strasbg.fr/cgi-bin/qcat?J/A+A/}).

\begin{acknowledgements}

    We thank Casey DeRoo for kindly providing the StokeSAT sensitivity curves used in this work.

    We thank Olivier Hervet for useful discussions about the fitting tool Bjet-MCMC used for this work.
      
      We thank Garrett Latiolais e Mohammad Ali Boroumand for the work on COSI simulations that enabled the calculations made in this paper.

      This research has made use of data from the RoboPol programme, a collaboration between Caltech, the University of Crete, IA-FORTH, IUCAA, the MPIfR, and the Nicolaus Copernicus University, which was conducted at Skinakas Observatory in Crete, Greece.

      Part of this work is based on archival data, software or online services provided by the Space Science Data Center - ASI.
      
      Some of the data used in this work were obtained from the MMDC.

      S. C. and I. L. were funded by the European Union ERC-2022-STG - BOOTES - 101076343. Views and opinions expressed are however those of the author(s) only and do not necessarily reflect those of the European Union or the European Research Council Executive Agency. Neither the European Union nor the granting authority can be held responsible for them.

      J.O.-S. acknowledges founding from the Istituto Nazionale di Fisica Nucleare Cap. U.1.01.01.01.009.

      This research was partially funded by the Deutsche Forschungsgemeinschaft (DFG, German Research Foundation) as part of the DFG Research Unit FOR5195 – project number 443220636.

\end{acknowledgements}

\bibliographystyle{aa} 
\bibliography{bibliography.bib}

\begin{thebibliography}{59}
\expandafter\ifx\csname natexlab\endcsname\relax\def\natexlab#1{#1}\fi

\bibitem[{{Agudo} {et~al.}(2025){Agudo}, {Liodakis}, {Otero-Santos}, {Middei}, {Marscher}, {Jorstad}, {Zhang}, {Li}, {Di Gesu}, {Romani}, {Kim}, {Fenu}, {Marshall}, {Pacciani}, {Escudero Pedrosa}, {Aceituno}, {Ag{\'\i}s-Gonz{\'a}lez}, {Bonnoli}, {Casanova}, {Morcuende}, {Piirola}, {Sota}, {Kouch}, {Lindfors}, {McCall}, {Jermak}, {Steele}, {Borman}, {Grishina}, {Hagen-Thorn}, {Kopatskaya}, {Larionova}, {Morozova}, {Savchenko}, {Shishkina}, {Troitskiy}, {Troitskaya}, {Vasilyev}, {Zhovtan}, {Myserlis}, {Gurwell}, {Keating}, {Rao}, {Kang}, {Lee}, {Kim}, {Cheong}, {Jeong}, {Angelakis}, {Kraus}, {Blinov}, {Maharana}, {Bachev}, {Jormanainen}, {Nilsson}, {Fallah Ramazani}, {Casadio}, {Fuentes}, {Traianou}, {Thum}, {G{\'o}mez}, {Antonelli}, {Bachetti}, {Baldini}, {Baumgartner}, {Bellazzini}, {Bianchi}, {Bongiorno}, {Bonino}, {Brez}, {Bucciantini}, {Capitanio}, {Castellano}, {Cavazzuti}, {Chen}, {Ciprini}, {Costa}, {De Rosa}, {Del Monte}, {Di Lalla}, {Di Marco}, {Donnarumma}, {Doroshenko}, {Dov{\v{c}}iak}, {Ehlert},
  {Enoto}, {Evangelista}, {Fabiani}, {Ferrazzoli}, {Garc{\'\i}a}, {Gunji}, {Hayashida}, {Heyl}, {Iwakiri}, {Kaaret}, {Karas}, {Kislat}, {Kitaguchi}, {Kolodziejczak}, {Krawczynski}, {La Monaca}, {Latronico}, {Maldera}, {Manfreda}, {Marin}, {Marinucci}, {Massaro}, {Matt}, {Mitsuishi}, {Mizuno}, {Muleri}, {Negro}, {Ng}, {O'Dell}, {Omodei}, {Oppedisano}, {Papitto}, {Pavlov}, {Peirson}, {Perri}, {Pesce-Rollins}, {Petrucci}, {Pilia}, {Possenti}, {Poutanen}, {Puccetti}, {Ramsey}, {Rankin}, {Ratheesh}, {Roberts}, {Sgr{\`o}}, {Slane}, {Soffitta}, {Spandre}, {Swartz}, {Tamagawa}, {Tavecchio}, {Taverna}, {Tawara}, {Tennant}, {Thomas}, {Tombesi}, {Trois}, {Tsygankov}, {Turolla}, {Vink}, {Weisskopf}, {Wu}, {Xie}, \& {Zane}}]{agudo2025}
{Agudo}, I., {Liodakis}, I., {Otero-Santos}, J., {et~al.} 2025, \apjl, 985, L15

\bibitem[{{Ajello} {et~al.}(2020){Ajello}, {Angioni}, {Axelsson}, {Ballet}, {Barbiellini}, {Bastieri}, {Becerra Gonzalez}, {Bellazzini}, {Bissaldi}, {Bloom}, {Bonino}, {Bottacini}, {Bruel}, {Buson}, {Cafardo}, {Cameron}, {Cavazzuti}, {Chen}, {Cheung}, {Ciprini}, {Costantin}, {Cutini}, {D'Ammando}, {de la Torre Luque}, {de Menezes}, {de Palma}, {Desai}, {Di Lalla}, {Di Venere}, {Dom{\'\i}nguez}, {Dirirsa}, {Ferrara}, {Finke}, {Franckowiak}, {Fukazawa}, {Funk}, {Fusco}, {Gargano}, {Garrappa}, {Gasparrini}, {Giglietto}, {Giordano}, {Giroletti}, {Green}, {Grenier}, {Guiriec}, {Harita}, {Hays}, {Horan}, {Itoh}, {J{\'o}hannesson}, {Kovac'evic'}, {Krauss}, {Kreter}, {Kuss}, {Larsson}, {Leto}, {Li}, {Liodakis}, {Longo}, {Loparco}, {Lott}, {Lovellette}, {Lubrano}, {Madejski}, {Maldera}, {Manfreda}, {Mart{\'\i}-Devesa}, {Massaro}, {Mazziotta}, {Mereu}, {Meyer}, {Migliori}, {Mirabal}, {Mizuno}, {Monzani}, {Morselli}, {Moskalenko}, {Negro}, {Nemmen}, {Nuss}, {Ojha}, {Ojha}, {Omodei}, {Orienti}, {Orlando}, {Ormes},
  {Paliya}, {Pei}, {Pe{\~n}a-Herazo}, {Persic}, {Pesce-Rollins}, {Petrov}, {Piron}, {Poon}, {Principe}, {Rain{\`o}}, {Rando}, {Rani}, {Razzano}, {Razzaque}, {Reimer}, {Reimer}, {Schinzel}, {Serini}, {Sgr{\`o}}, {Siskind}, {Spandre}, {Spinelli}, {Suson}, {Tachibana}, {Thompson}, {Torres}, {Torresi}, {Troja}, {Valverde}, {van Zyl}, \& {Yassine}}]{ajello2020fourth}
{Ajello}, M., {Angioni}, R., {Axelsson}, M., {et~al.} 2020, \apj, 892, 105

\bibitem[{{Ajello} {et~al.}(2022){Ajello}, {Baldini}, {Ballet}, {Bastieri}, {Becerra Gonzalez}, {Bellazzini}, {Berretta}, {Bissaldi}, {Bonino}, {Brill}, {Bruel}, {Buson}, {Caputo}, {Caraveo}, {Cheung}, {Chiaro}, {Cibrario}, {Ciprini}, {Crnogorcevic}, {Cutini}, {D'Ammando}, {De Gaetano}, {Di Lalla}, {Di Venere}, {Dom{\'\i}nguez}, {Ramazani}, {Ferrara}, {Fiori}, {Fukazawa}, {Funk}, {Fusco}, {Gammaldi}, {Gargano}, {Garrappa}, {Gasparrini}, {Giglietto}, {Giordano}, {Giroletti}, {Green}, {Grenier}, {Guiriec}, {Horan}, {Hou}, {Kayanoki}, {Kuss}, {Larsson}, {Latronico}, {Lewis}, {Li}, {Liodakis}, {Longo}, {Loparco}, {Lott}, {Lovellette}, {Lubrano}, {Madejski}, {Maldera}, {Manfreda}, {Mart{\'\i}-Devesa}, {Mazziotta}, {Mereu}, {Michelson}, {Mirabal}, {Mitthumsiri}, {Mizuno}, {Monzani}, {Morselli}, {Moskalenko}, {Negro}, {Ojha}, {Orienti}, {Orlando}, {Ormes}, {Pei}, {Pe{\~n}a-Herazo}, {Persic}, {Pesce-Rollins}, {Petrosian}, {Pillera}, {Poon}, {Porter}, {Principe}, {Rain{\`o}}, {Rando}, {Rani}, {Razzano}, {Razzaque},
  {Reimer}, {Reimer}, {Scotton}, {Serini}, {Sgr{\`o}}, {Siskind}, {Spandre}, {Spinelli}, {Suson}, {Tajima}, {Torres}, {Valverde}, {Yassin}, \& {Zaharijas}}]{4lac2022}
{Ajello}, M., {Baldini}, L., {Ballet}, J., {et~al.} 2022, \apjs, 263, 24

\bibitem[{{Angel} \& {Stockman}(1980)}]{angel1980}
{Angel}, J.~R.~P. \& {Stockman}, H.~S. 1980, \araa, 18, 321

\bibitem[{Ballet {et~al.}(2023)Ballet, Bruel, Burnett, Lott, Collaboration, {et~al.}}]{ballet2023}
Ballet, J., Bruel, P., Burnett, T., {et~al.} 2023, arXiv preprint arXiv:2307.12546

\bibitem[{{Blandford} {et~al.}(2019){Blandford}, {Meier}, \& {Readhead}}]{blandford2019}
{Blandford}, R., {Meier}, D., \& {Readhead}, A. 2019, \araa, 57, 467

\bibitem[{{Blinov} \& {Pavlidou}(2019)}]{blinov2019}
{Blinov}, D. \& {Pavlidou}, V. 2019, Galaxies, 7, 46

\bibitem[{{Blinov} {et~al.}(2018){Blinov}, {Pavlidou}, {Papadakis}, {Kiehlmann}, {Liodakis}, {Panopoulou}, {Angelakis}, {Balokovi{\'c}}, {Hovatta}, {King}, {Kus}, {Kylafis}, {Mahabal}, {Maharana}, {Myserlis}, {Paleologou}, {Papamastorakis}, {Pazderski}, {Pearson}, {Ramaprakash}, {Readhead}, {Reig}, {Tassis}, \& {Zensus}}]{blinov2018}
{Blinov}, D., {Pavlidou}, V., {Papadakis}, I., {et~al.} 2018, \mnras, 474, 1296

\bibitem[{{Blinov} {et~al.}(2016){Blinov}, {Pavlidou}, {Papadakis}, {Hovatta}, {Pearson}, {Liodakis}, {Panopoulou}, {Angelakis}, {Balokovi{\'c}}, {Das}, {Khodade}, {Kiehlmann}, {King}, {Kus}, {Kylafis}, {Mahabal}, {Marecki}, {Modi}, {Myserlis}, {Paleologou}, {Papamastorakis}, {Pazderska}, {Pazderski}, {Rajarshi}, {Ramaprakash}, {Readhead}, {Reig}, {Tassis}, \& {Zensus}}]{blinov2016}
{Blinov}, D., {Pavlidou}, V., {Papadakis}, I.~E., {et~al.} 2016, \mnras, 457, 2252

\bibitem[{{Boettcher}(2012)}]{boettcher2012modeling}
{Boettcher}, M. 2012, arXiv e-prints, arXiv:1205.0539

\bibitem[{{Bonometto} \& {Saggion}(1973)}]{bonometto1973}
{Bonometto}, S. \& {Saggion}, A. 1973, \aap, 23, 9

\bibitem[{{B{\"o}ttcher} {et~al.}(2013){B{\"o}ttcher}, {Reimer}, {Sweeney}, \& {Prakash}}]{boettch2013}
{B{\"o}ttcher}, M., {Reimer}, A., {Sweeney}, K., \& {Prakash}, A. 2013, \apj, 768, 54

\bibitem[{{Capecchiacci} {et~al.}(2025){Capecchiacci}, {Liodakis}, {Middei}, {Kim}, {Di Gesu}, {Agudo}, {Ag{\'\i}s-Gonz{\'a}lez}, {Arbet-Engels}, {Blinov}, {Chen}, {Ehlert}, {Gau}, {Heckmann}, {Hu}, {Jorstad}, {Kaaret}, {Kouch}, {Krawczynski}, {Lindfors}, {Marin}, {Marscher}, {Myserlis}, {O'Dell}, {Pacciani}, {Paneque}, {Perri}, {Puccetti}, {Saade}, {Tavecchio}, {Tennant}, {Traianou}, {Weisskopf}, {Wu}, {Aceituno}, {Bonnoli}, {Casanova}, {Emery}, {Escudero}, {Morcuende}, {Otero-Santos}, {Sota}, {Piirola}, {Borman}, {Kopatskaya}, {Larionova}, {Morozova}, {Shishkina}, {Savchenko}, {Vasilyev}, {Grishina}, {Troitskiy}, {Zhovtan}, {McCall}, {Jermak}, {Steele}, {Bachev}, {Strigachev}, {Imazawa}, {Sasada}, {Fukazawa}, {Kawabata}, {Uemura}, {Mizuno}, {Nakaoka}, {Tochihara}, {Akai}, {Akitaya}, {Berdyugin}, {Kagitani}, {Kravtsov}, {Poutanen}, {Sakanoi}, {{\'A}lvarez-Ortega}, {Casadio}, {Kang}, {Lee}, {Kim}, {Cheong}, {Jeong}, {Song}, {Li}, {Nam}, {Gurwell}, {Keating}, {Rao}, {Angelakis}, {Kraus}, {Benke}, {Debbrecht},
  {Eich}, {Eppel}, {Gokus}, {H{\"a}mmerich}, {He{\ss}d{\"o}rfer}, {Kadler}, {Kirchner}, {Paraschos}, {R{\"o}sch}, \& {Schulga}}]{cape2025}
{Capecchiacci}, S., {Liodakis}, I., {Middei}, R., {et~al.} 2025, \aap, 703, A19

\bibitem[{{Caputo} {et~al.}(2022){Caputo}, {Ajello}, {Kierans}, {Perkins}, {Racusin}, {Baldini}, {Baring}, {Bissaldi}, {Burns}, {Cannady}, {Charles}, {da Silva}, {Fang}, {Fleischhack}, {Fryer}, {Fukazawa}, {Grove}, {Hartmann}, {Howell}, {Jadhav}, {Karwin}, {Kocevski}, {Kurahashi}, {Latronico}, {Lewis}, {Leys}, {Lien}, {Marcotulli}, {Martinez-Castellanos}, {Mazziotta}, {McEnery}, {Metcalfe}, {Murase}, {Negro}, {Parker}, {Phlips}, {Prescod-Weinstein}, {Razzaque}, {Shawhan}, {Sheng}, {Shutt}, {Shy}, {Sleator}, {Steinhebel}, {Striebig}, {Suda}, {Tak}, {Tajima}, {Valverde}, {Venters}, {Wadiasingh}, {Woolf}, {Wulf}, {Zhang}, \& {Zoglauer}}]{amegox2022}
{Caputo}, R., {Ajello}, M., {Kierans}, C.~A., {et~al.} 2022, Journal of Astronomical Telescopes, Instruments, and Systems, 8, 044003

\bibitem[{{Chen} {et~al.}(2024){Chen}, {Liodakis}, {Middei}, {Kim}, {Di Gesu}, {Di Marco}, {Ehlert}, {Errando}, {Negro}, {Jorstad}, {Marscher}, {Wu}, {Agudo}, {Poutanen}, {Mizuno}, {Kouch}, {Lindfors}, {Borman}, {Grishina}, {Kopatskaya}, {Larionova}, {Morozova}, {Savchenko}, {Troitsky}, {Troitskaya}, {Vasilyev}, {Zhovtan}, {Aceituno}, {Bonnoli}, {Casanova}, {Escudero}, {Ag{\'\i}s-Gonz{\'a}lez}, {Husillos}, {Otero Santos}, {Sota}, {Piirola}, {Myserlis}, {Angelakis}, {Kraus}, {Gurwell}, {Keating}, {Rao}, {Kang}, {Lee}, {Kim}, {Cheong}, {Jeong}, {Song}, {Berdyugin}, {Kagitani}, {Kravtsov}, {Nitindala}, {Sakanoi}, {Imazawa}, {Sasada}, {Fukazawa}, {Kawabata}, {Uemura}, {Nakaoka}, {Akitaya}, {Casadio}, {Sievers}, {Antonelli}, {Bachetti}, {Baldini}, {Baumgartner}, {Bellazzini}, {Bianchi}, {Bongiorno}, {Bonino}, {Brez}, {Bucciantini}, {Capitanio}, {Castellano}, {Cavazzuti}, {Ciprini}, {Costa}, {De Rosa}, {Del Monte}, {Di Lalla}, {Donnarumma}, {Doroshenko}, {Dov{\v{c}}iak}, {Enoto}, {Evangelista}, {Fabiani},
  {Ferrazzoli}, {Garcia}, {Gunji}, {Hayashida}, {Heyl}, {Iwakiri}, {Kaaret}, {Karas}, {Kislat}, {Kitaguchi}, {Kolodziejczak}, {Krawczynski}, {La Monaca}, {Latronico}, {Maldera}, {Manfreda}, {Marin}, {Marinucci}, {Marshall}, {Massaro}, {Matt}, {Mitsuishi}, {Muleri}, {Ng}, {O'Dell}, {Omodei}, {Oppedisano}, {Papitto}, {Pavlov}, {Peirson}, {Perri}, {Pesce-Rollins}, {Petrucci}, {Pilia}, {Possenti}, {Puccetti}, {Ramsey}, {Rankin}, {Ratheesh}, {Roberts}, {Romani}, {Sgr{\'o}}, {Slane}, {Soffitta}, {Spandre}, {Swartz}, {Tamagawa}, {Tavecchio}, {Taverna}, {Tawara}, {Tennant}, {Thomas}, {Tombesi}, {Trois}, {Tsygankov}, {Turolla}, {Vink}, {Weisskopf}, {Xie}, \& {Zane}}]{chen2024x}
{Chen}, C.-T.~J., {Liodakis}, I., {Middei}, R., {et~al.} 2024, \apj, 974, 50

\bibitem[{{de Angelis} {et~al.}(2018){de Angelis}, {Tatischeff}, {Grenier}, {McEnery}, {Mallamaci}, {Tavani}, {Oberlack}, {Hanlon}, {Walter}, {Argan}, {von Ballmoos}, {Bulgarelli}, {Bykov}, {Hernanz}, {Kanbach}, {Kuvvetli}, {Pearce}, {Zdziarski}, {Conrad}, {Ghisellini}, {Harding}, {Isern}, {Leising}, {Longo}, {Madejski}, {Martinez}, {Mazziotta}, {Paredes}, {Pohl}, {Rando}, {Razzano}, {Aboudan}, {Ackermann}, {Addazi}, {Ajello}, {Albertus}, {{\'A}lvarez}, {Ambrosi}, {Ant{\'o}n}, {Antonelli}, {Babic}, {Baibussinov}, {Balbo}, {Baldini}, {Balman}, {Bambi}, {Barres de Almeida}, {Barrio}, {Bartels}, {Bastieri}, {Bednarek}, {Bernard}, {Bernardini}, {Bernasconi}, {Bertucci}, {Biland}, {Bissaldi}, {Boettcher}, {Bonvicini}, {Bosch-Ramon}, {Bottacini}, {Bozhilov}, {Bretz}, {Branchesi}, {Brdar}, {Bringmann}, {Brogna}, {Budtz J{\o}rgensen}, {Busetto}, {Buson}, {Busso}, {Caccianiga}, {Camera}, {Campana}, {Caraveo}, {Cardillo}, {Carlson}, {Celestin}, {Cerme{\~n}o}, {Chen}, {Cheung}, {Churazov}, {Ciprini}, {Coc},
  {Colafrancesco}, {Coleiro}, {Collmar}, {Coppi}, {Curado da Silva}, {Cutini}, {D'Ammando}, {de Lotto}, {de Martino}, {De Rosa}, {Del Santo}, {Delgado}, {Diehl}, {Dietrich}, {Dolgov}, {Dom{\'\i}nguez}, {Dominis Prester}, {Donnarumma}, {Dorner}, {Doro}, {Dutra}, {Elsaesser}, {Fabrizio}, {Fern{\'a}ndez-Barral}, {Fioretti}, {Foffano}, {Formato}, {Fornengo}, {Foschini}, {Franceschini}, {Franckowiak}, {Funk}, {Fuschino}, {Gaggero}, {Galanti}, {Gargano}, {Gasparrini}, {Gehrz}, {Giammaria}, {Giglietto}, {Giommi}, {Giordano}, {Giroletti}, {Ghirlanda}, {Godinovic}, {Gouiff{\'e}s}, {Grove}, {Hamadache}, {Hartmann}, {Hayashida}, {Hryczuk}, {Jean}, {Johnson}, {Jos{\'e}}, {Kaufmann}, {Khelifi}, {Kiener}, {Kn{\"o}dlseder}, {Kole}, {Kopp}, {Kozhuharov}, {Labanti}, {Lalkovski}, {Laurent}, {Limousin}, {Linares}, {Lindfors}, {Lindner}, {Liu}, {Lombardi}, {Loparco}, {L{\'o}pez-Coto}, {L{\'o}pez Moya}, {Lott}, {Lubrano}, {Malyshev}, {Mankuzhiyil}, {Mannheim}, {March{\~a}}, {Marcian{\`o}}, {Marcote}, {Mariotti}, {Marisaldi},
  {McBreen}, {Mereghetti}, {Merle}, {Mignani}, {Minervini}, {Moiseev}, {Morselli}, {Moura}, {Nakazawa}, {Nava}, {Nieto}, {Orienti}, {Orio}, {Orlando}, {Orleanski}, {Paiano}, {Paoletti}, {Papitto}, {Pasquato}, {Patricelli}, {P{\'e}rez-Garc{\'\i}a}, {Persic}, {Piano}, {Pichel}, {Pimenta}, {Pittori}, {Porter}, {Poutanen}, {Prandini}, {Prantzos}, {Produit}, {Profumo}, \& {Queiroz}}]{eastrogam2018}
{de Angelis}, A., {Tatischeff}, V., {Grenier}, I.~A., {et~al.} 2018, Journal of High Energy Astrophysics, 19, 1

\bibitem[{{de Jaeger} {et~al.}(2023){de Jaeger}, {Shappee}, {Kochanek}, {Hinkle}, {Garrappa}, {Liodakis}, {Franckowiak}, {Stanek}, {Beacom}, \& {Prieto}}]{jaeger2023}
{de Jaeger}, T., {Shappee}, B.~J., {Kochanek}, C.~S., {et~al.} 2023, \mnras, 519, 6349

\bibitem[{{Di Gesu} {et~al.}(2022){Di Gesu}, {Donnarumma}, {Tavecchio}, {Agudo}, {Barnounin}, {Cibrario}, {Di Lalla}, {Di Marco}, {Escudero}, {Errando}, {Jorstad}, {Kim}, {Kouch}, {Liodakis}, {Lindfors}, {Madejski}, {Marshall}, {Marscher}, {Middei}, {Muleri}, {Myserlis}, {Negro}, {Omodei}, {Pacciani}, {Paggi}, {Perri}, {Puccetti}, {Antonelli}, {Bachetti}, {Baldini}, {Baumgartner}, {Bellazzini}, {Bianchi}, {Bongiorno}, {Bonino}, {Brez}, {Bucciantini}, {Capitanio}, {Castellano}, {Cavazzuti}, {Ciprini}, {Costa}, {De Rosa}, {Del Monte}, {Doroshenko}, {Dov{\v{c}}iak}, {Ehlert}, {Enoto}, {Evangelista}, {Fabiani}, {Ferrazzoli}, {Garcia}, {Gunji}, {Hayashida}, {Heyl}, {Iwakiri}, {Karas}, {Kitaguchi}, {Kolodziejczak}, {Krawczynski}, {La Monaca}, {Latronico}, {Maldera}, {Manfreda}, {Marin}, {Marinucci}, {Massaro}, {Matt}, {Mitsuishi}, {Mizuno}, {Ng}, {O'Dell}, {Oppedisano}, {Papitto}, {Pavlov}, {Peirson}, {Pesce-Rollins}, {Petrucci}, {Pilia}, {Possenti}, {Poutanen}, {Ramsey}, {Rankin}, {Ratheesh}, {Romani}, {Sgr{\`o}},
  {Slane}, {Soffitta}, {Spandre}, {Tamagawa}, {Taverna}, {Tawara}, {Tennant}, {Thomas}, {Tombesi}, {Trois}, {Tsygankov}, {Turolla}, {Vink}, {Weisskopf}, {Wu}, {Xie}, \& {Zane}}]{di2022x}
{Di Gesu}, L., {Donnarumma}, I., {Tavecchio}, F., {et~al.} 2022, \apjl, 938, L7

\bibitem[{{Di Gesu} {et~al.}(2020){Di Gesu}, {Ferrazzoli}, {Donnarumma}, {Soffitta}, {Costa}, {Muleri}, {Pesce-Rollins}, \& {Marin}}]{digesu2020}
{Di Gesu}, L., {Ferrazzoli}, R., {Donnarumma}, I., {et~al.} 2020, \aap, 643, A52

\bibitem[{{Di Gesu} {et~al.}(2023){Di Gesu}, {Marshall}, {Ehlert}, {Kim}, {Donnarumma}, {Tavecchio}, {Liodakis}, {Kiehlmann}, {Agudo}, {Jorstad}, {Muleri}, {Marscher}, {Puccetti}, {Middei}, {Perri}, {Pacciani}, {Negro}, {Romani}, {Di Marco}, {Blinov}, {Bourbah}, {Kontopodis}, {Mandarakas}, {Romanopoulos}, {Skalidis}, {Vervelaki}, {Casadio}, {Escudero}, {Myserlis}, {Gurwell}, {Rao}, {Keating}, {Kouch}, {Lindfors}, {Aceituno}, {Bernardos}, {Bonnoli}, {Casanova}, {Garc{\'\i}a-Comas}, {Ag{\'\i}s-Gonz{\'a}lez}, {Husillos}, {Marchini}, {Sota}, {Imazawa}, {Sasada}, {Fukazawa}, {Kawabata}, {Uemura}, {Mizuno}, {Nakaoka}, {Akitaya}, {Savchenko}, {Vasilyev}, {G{\'o}mez}, {Antonelli}, {Barnouin}, {Bonino}, {Cavazzuti}, {Costamante}, {Chen}, {Cibrario}, {De Rosa}, {Di Pierro}, {Errando}, {Kaaret}, {Karas}, {Krawczynski}, {Lisalda}, {Madejski}, {Malacaria}, {Marin}, {Marinucci}, {Massaro}, {Matt}, {Mitsuishi}, {O'Dell}, {Paggi}, {Peirson}, {Petrucci}, {Ramsey}, {Tennant}, {Wu}, {Bachetti}, {Baldini}, {Baumgartner},
  {Bellazzini}, {Bianchi}, {Bongiorno}, {Brez}, {Bucciantini}, {Capitanio}, {Castellano}, {Ciprini}, {Costa}, {Del Monte}, {Di Lalla}, {Doroshenko}, {Dov{\v{c}}iak}, {Enoto}, {Evangelista}, {Fabiani}, {Ferrazzoli}, {Garcia}, {Gunji}, {Hayashida}, {Heyl}, {Iwakiri}, {Kislat}, {Kitaguchi}, {Kolodziejczak}, {La Monaca}, {Latronico}, {Maldera}, {Manfreda}, {Ng}, {Omodei}, {Oppedisano}, {Papitto}, {Pavlov}, {Pesce-Rollins}, {Pilia}, {Possenti}, {Poutanen}, {Rankin}, {Ratheesh}, {Roberts}, {Sgr{\`o}}, {Slane}, {Soffitta}, {Spandre}, {Swartz}, {Tamagawa}, {Taverna}, {Tawara}, {Thomas}, {Tombesi}, {Trois}, {Tsygankov}, {Turolla}, {Vink}, {Weisskopf}, {Xie}, \& {Zane}}]{digesu2023}
{Di Gesu}, L., {Marshall}, H.~L., {Ehlert}, S.~R., {et~al.} 2023, Nature Astronomy, 7, 1245

\bibitem[{{Ehlert} {et~al.}(2022){Ehlert}, {Ferrazzoli}, {Marinucci}, {Marshall}, {Middei}, {Pacciani}, {Perri}, {Petrucci}, {Puccetti}, {Barnouin}, {Bianchi}, {Liodakis}, {Madejski}, {Marin}, {Marscher}, {Matt}, {Poutanen}, {Wu}, {Agudo}, {Antonelli}, {Bachetti}, {Baldini}, {Baumgartner}, {Bellazzini}, {Bongiorno}, {Bonino}, {Brez}, {Bucciantini}, {Capitanio}, {Castellano}, {Cavazzuti}, {Ciprini}, {Costa}, {De Rosa}, {Del Monte}, {Di Gesu}, {Di Lalla}, {Di Marco}, {Donnarumma}, {Doroshenko}, {Dov{\v{c}}iak}, {Enoto}, {Evangelista}, {Fabiani}, {Garcia}, {Gunji}, {Hayashida}, {Heyl}, {Iwakiri}, {Jorstad}, {Karas}, {Kitaguchi}, {Kolodziejczak}, {Krawczynski}, {La Monaca}, {Latronico}, {Maldera}, {Manfreda}, {Massaro}, {Mitsuishi}, {Mizuno}, {Muleri}, {Negro}, {Ng}, {O'Dell}, {Omodei}, {Oppedisano}, {Papitto}, {Pavlov}, {Peirson}, {Pesce-Rollins}, {Pilia}, {Possenti}, {Ramsey}, {Rankin}, {Ratheesh}, {Romani}, {Sgr{\`o}}, {Slane}, {Soffitta}, {Spandre}, {Tamagawa}, {Tavecchio}, {Taverna}, {Tawara}, {Tennant},
  {Thomas}, {Tombesi}, {Trois}, {Tsygankov}, {Turolla}, {Vink}, {Weisskopf}, {Xie}, {Zane}, {IXPE Collaboration}, {Rodi}, {Jourdain}, \& {Roques}}]{ehlert2022}
{Ehlert}, S.~R., {Ferrazzoli}, R., {Marinucci}, A., {et~al.} 2022, \apj, 935, 116

\bibitem[{{Errando} {et~al.}(2024){Errando}, {Liodakis}, {Marscher}, {Marshall}, {Middei}, {Negro}, {Peirson}, {Perri}, {Puccetti}, {Rabinowitz}, {Agudo}, {Jorstad}, {Savchenko}, {Blinov}, {Bourbah}, {Kiehlmann}, {Kontopodis}, {Mandarakas}, {Romanopoulos}, {Skalidis}, {Vervelaki}, {Aceituno}, {Bernardos}, {Bonnoli}, {Casanova}, {Ag{\'\i}s-Gonz{\'a}lez}, {Husillos}, {Marchini}, {Sota}, {Kouch}, {Lindfors}, {Casadio}, {Escudero}, {Myserlis}, {Imazawa}, {Sasada}, {Fukazawa}, {Kawabata}, {Uemura}, {Mizuno}, {Nakaoka}, {Akitaya}, {Gurwell}, {Keating}, {Rao}, {Ingram}, {Massaro}, {Antonelli}, {Bonino}, {Cavazzuti}, {Chen}, {Cibrario}, {Ciprini}, {De Rosa}, {Di Gesu}, {Di Pierro}, {Donnarumma}, {Ehlert}, {Fenu}, {Gau}, {Karas}, {Kim}, {Krawczynski}, {Laurenti}, {Lisalda}, {L{\'o}pez-Coto}, {Madejski}, {Marin}, {Marinucci}, {Mitsuishi}, {Muleri}, {Pacciani}, {Paggi}, {Petrucci}, {Rodriguez Cavero}, {Romani}, {Tavecchio}, {Tugliani}, {Wu}, {Bachetti}, {Baldini}, {Baumgartner}, {Bellazzini}, {Bianchi}, {Bongiorno},
  {Brez}, {Bucciantini}, {Capitanio}, {Castellano}, {Costa}, {Del Monte}, {Di Lalla}, {Di Marco}, {Doroshenko}, {Dov{\v{c}}iak}, {Enoto}, {Evangelista}, {Fabiani}, {Ferrazzoli}, {Garcia}, {Gunji}, {Hayashida}, {Heyl}, {Iwakiri}, {Kaaret}, {Kislat}, {Kitaguchi}, {Kolodziejczak}, {La Monaca}, {Latronico}, {Maldera}, {Manfreda}, {Matt}, {Ng}, {O'Dell}, {Omodei}, {Oppedisano}, {Papitto}, {Pavlov}, {Pesce-Rollins}, {Pilia}, {Possenti}, {Poutanen}, {Ramsey}, {Rankin}, {Ratheesh}, {Roberts}, {Sgr{\`o}}, {Slane}, {Soffitta}, {Spandre}, {Swartz}, {Tamagawa}, {Taverna}, {Tawara}, {Tennant}, {Thomas}, {Tombesi}, {Trois}, {Tsygankov}, {Turolla}, {Vink}, {Weisskopf}, {Xie}, \& {Zane}}]{errando2024detection}
{Errando}, M., {Liodakis}, I., {Marscher}, A.~P., {et~al.} 2024, \apj, 963, 5

\bibitem[{{Hervet} {et~al.}(2024){Hervet}, {Johnson}, \& {Youngquist}}]{bjetmcmc}
{Hervet}, O., {Johnson}, C.~A., \& {Youngquist}, A. 2024, \apj, 962, 140

\bibitem[{{Hovatta} \& {Lindfors}(2019)}]{hovatta2019}
{Hovatta}, T. \& {Lindfors}, E. 2019, \nar, 87, 101541

\bibitem[{{IceCube Collaboration} {et~al.}(2018){IceCube Collaboration}, {Aartsen}, {Ackermann}, {Adams}, {Aguilar}, {Ahlers}, {Ahrens}, {Al Samarai}, {Altmann}, {Andeen}, {Anderson}, {Ansseau}, {Anton}, {Arg{\"u}elles}, {Auffenberg}, {Axani}, {Bagherpour}, {Bai}, {Barron}, {Barwick}, {Baum}, {Bay}, {Beatty}, {Becker Tjus}, {Becker}, {BenZvi}, {Berley}, {Bernardini}, {Besson}, {Binder}, {Bindig}, {Blaufuss}, {Blot}, {Bohm}, {B{\"o}rner}, {Bos}, {B{\"o}ser}, {Botner}, {Bourbeau}, {Bourbeau}, {Bradascio}, {Braun}, {Brenzke}, {Bretz}, {Bron}, {Brostean-Kaiser}, {Burgman}, {Busse}, {Carver}, {Cheung}, {Chirkin}, {Christov}, {Clark}, {Classen}, {Coenders}, {Collin}, {Conrad}, {Coppin}, {Correa}, {Cowen}, {Cross}, {Dave}, {Day}, {de Andr{\'e}}, {De Clercq}, {DeLaunay}, {Dembinski}, {De Ridder}, {Desiati}, {de Vries}, {de Wasseige}, {de With}, {DeYoung}, {D{\'\i}az-V{\'e}lez}, {di Lorenzo}, {Dujmovic}, {Dumm}, {Dunkman}, {Dvorak}, {Eberhardt}, {Ehrhardt}, {Eichmann}, {Eller}, {Evenson}, {Fahey}, {Fazely}, {Felde},
  {Filimonov}, {Finley}, {Flis}, {Franckowiak}, {Friedman}, {Fritz}, {Gaisser}, {Gallagher}, {Gerhardt}, {Ghorbani}, {Glauch}, {Gl{\"u}senkamp}, {Goldschmidt}, {Gonzalez}, {Grant}, {Griffith}, {Haack}, {Hallgren}, {Halzen}, {Hanson}, {Hebecker}, {Heereman}, {Helbing}, {Hellauer}, {Hickford}, {Hignight}, {Hill}, {Hoffman}, {Hoffmann}, {Hoinka}, {Hokanson-Fasig}, {Hoshina}, {Huang}, {Huber}, {Hultqvist}, {H{\"u}nnefeld}, {Hussain}, {In}, {Iovine}, {Ishihara}, {Jacobi}, {Japaridze}, {Jeong}, {Jero}, {Jones}, {Kalaczynski}, {Kang}, {Kappes}, {Kappesser}, {Karg}, {Karle}, {Katz}, {Kauer}, {Keivani}, {Kelley}, {Kheirandish}, {Kim}, {Kim}, {Kintscher}, {Kiryluk}, {Kittler}, {Klein}, {Koirala}, {Kolanoski}, {K{\"o}pke}, {Kopper}, {Kopper}, {Koschinsky}, {Koskinen}, {Kowalski}, {Krings}, {Kroll}, {Kr{\"u}ckl}, {Kunwar}, {Kurahashi}, {Kuwabara}, {Kyriacou}, {Labare}, {Lanfranchi}, {Larson}, {Lauber}, {Leonard}, {Lesiak-Bzdak}, {Leuermann}, {Liu}, {Lozano Mariscal}, {Lu}, {L{\"u}nemann}, {Luszczak}, {Madsen}, {Maggi},
  {Mahn}, {Mancina}, {Maruyama}, {Mase}, {Maunu}, {Meagher}, {Medici}, {Meier}, {Menne}, {Merino}, {Meures}, {Miarecki}, {Micallef}, {Moment{\'e}}, {Montaruli}, {Moore}, {Morse}, {Moulai}, {Nahnhauer}, {Nakarmi}, {Naumann}, \& {Neer}}]{icecube2018}
{IceCube Collaboration}, {Aartsen}, M.~G., {Ackermann}, M., {et~al.} 2018, Science, 361, eaat1378

\bibitem[{{Kim} {et~al.}(2024){Kim}, {Di Gesu}, {Liodakis}, {Marscher}, {Jorstad}, {Middei}, {Marshall}, {Pacciani}, {Agudo}, {Tavecchio}, {Cibrario}, {Tugliani}, {Bonino}, {Negro}, {Puccetti}, {Tombesi}, {Costa}, {Donnarumma}, {Soffitta}, {Mizuno}, {Fukazawa}, {Kawabata}, {Nakaoka}, {Uemura}, {Imazawa}, {Sasada}, {Akitaya}, {Jos{\`e} Aceituno}, {Bonnoli}, {Casanova}, {Myserlis}, {Sievers}, {Angelakis}, {Kraus}, {Yeon Cheong}, {Jeong}, {Kang}, {Kim}, {Lee}, {Ag{\`\i}s-Gonz{\`a}lez}, {Sota}, {Escudero}, {Gurwell}, {Keating}, {Rao}, {Kouch}, {Lindfors}, {Bourbah}, {Kiehlmann}, {Kontopodis}, {Mandarakas}, {Romanopoulos}, {Skalidis}, {Vervelaki}, {Savchenko}, {Antonelli}, {Bachetti}, {Baldini}, {Baumgartner}, {Bellazzini}, {Bianchi}, {Bongiorno}, {Brez}, {Bucciantini}, {Capitanio}, {Castellano}, {Cavazzuti}, {Chen}, {Ciprini}, {De Rosa}, {Del Monte}, {Di Lalla}, {Di Marco}, {Doroshenko}, {Dov{\v{c}}iak}, {Ehlert}, {Enoto}, {Evangelista}, {Fabiani}, {Ferrazzoli}, {Garcia}, {Gunji}, {Hayashida}, {Heyl}, {Iwakiri},
  {Kaaret}, {Karas}, {Kislat}, {Kitaguchi}, {Kolodziejczak}, {Krawczynski}, {La Monaca}, {Latronico}, {Maldera}, {Manfreda}, {Marin}, {Marinucci}, {Massaro}, {Matt}, {Mitsuishi}, {Muleri}, {Ng}, {O'Dell}, {Omodei}, {Oppedisano}, {Papitto}, {Pavlov}, {Peirson}, {Perri}, {Pesce-Rollins}, {Petrucci}, {Pilia}, {Possenti}, {Poutanen}, {Ramsey}, {Rankin}, {Ratheesh}, {Roberts}, {Romani}, {Sgr{\'o}}, {Slane}, {Spandre}, {Swartz}, {Tamagawa}, {Taverna}, {Tawara}, {Tennant}, {Thomas}, {Trois}, {Tsygankov}, {Turolla}, {Vink}, {Weisskopf}, {Wu}, {Xie}, \& {Zane}}]{kim2024magnetic}
{Kim}, D.~E., {Di Gesu}, L., {Liodakis}, I., {et~al.} 2024, \aap, 681, A12

\bibitem[{{Kouch} {et~al.}(2026){Kouch}, {Hovatta}, {Lindfors}, {Liodakis}, {Koljonen}, \& {Paggi}}]{kouch2025}
{Kouch}, P.~M., {Hovatta}, T., {Lindfors}, E., {et~al.} 2026, \aap, 708, A383

\bibitem[{{Kouch} {et~al.}(2024){Kouch}, {Liodakis}, {Middei}, {Kim}, {Tavecchio}, {Marscher}, {Marshall}, {Ehlert}, {Di Gesu}, {Jorstad}, {Agudo}, {Madejski}, {Romani}, {Errando}, {Lindfors}, {Nilsson}, {Toppari}, {Potter}, {Imazawa}, {Sasada}, {Fukazawa}, {Kawabata}, {Uemura}, {Mizuno}, {Nakaoka}, {Akitaya}, {McCall}, {Jermak}, {Steele}, {Myserlis}, {Gurwell}, {Keating}, {Rao}, {Kang}, {Lee}, {Kim}, {Cheong}, {Jeong}, {Angelakis}, {Kraus}, {Aceituno}, {Bonnoli}, {Casanova}, {Escudero}, {Ag{\'\i}s-Gonz{\'a}lez}, {Husillos}, {Morcuende}, {Otero-Santos}, {Sota}, {Bachev}, {Antonelli}, {Bachetti}, {Baldini}, {Baumgartner}, {Bellazzini}, {Bianchi}, {Bongiorno}, {Bonino}, {Brez}, {Bucciantini}, {Capitanio}, {Castellano}, {Cavazzuti}, {Chen}, {Ciprini}, {Costa}, {De Rosa}, {Del Monte}, {Di Lalla}, {Di Marco}, {Donnarumma}, {Doroshenko}, {Dov{\v{c}}iak}, {Enoto}, {Evangelista}, {Fabiani}, {Ferrazzoli}, {Garcia}, {Gunji}, {Hayashida}, {Heyl}, {Iwakiri}, {Kaaret}, {Karas}, {Kislat}, {Kitaguchi}, {Kolodziejczak},
  {Krawczynski}, {La Monaca}, {Latronico}, {Maldera}, {Manfreda}, {Marin}, {Marinucci}, {Massaro}, {Matt}, {Mitsuishi}, {Muleri}, {Negro}, {Ng}, {O'Dell}, {Omodei}, {Oppedisano}, {Papitto}, {Pavlov}, {Peirson}, {Perri}, {Pesce-Rollins}, {Petrucci}, {Pilia}, {Possenti}, {Poutanen}, {Puccetti}, {Ramsey}, {Rankin}, {Ratheesh}, {Roberts}, {Sgr{\`o}}, {Slane}, {Soffitta}, {Spandre}, {Swartz}, {Tamagawa}, {Taverna}, {Tawara}, {Tennant}, {Thomas}, {Tombesi}, {Trois}, {Tsygankov}, {Turolla}, {Vink}, {Weisskopf}, {Wu}, {Xie}, \& {Zane}}]{kouch2024ixpe}
{Kouch}, P.~M., {Liodakis}, I., {Middei}, R., {et~al.} 2024, \aap, 689, A119

\bibitem[{{Krawczynski}(2012)}]{krawczynski2012}
{Krawczynski}, H. 2012, \apj, 744, 30

\bibitem[{Latiolais {et~al.}(2026)Latiolais, Otero-Santos, Negro, Marcotulli, Boroumand, Gallego, Karwin, Martinez-Castellanos, Kocevski, Ajello, Capecchiacci, Liodakis, Bhavanam, Boggs, Hartmann, Kierans, Lewis, Sciaccaluga, Tomsick, Zhang, \& Zoglauer}]{Latiolais2026}
Latiolais, G.~A., Otero-Santos, J., Negro, M., {et~al.} 2026, The Astrophysical Journal, 1004, 80

\bibitem[{{Liodakis} {et~al.}(2025{\natexlab{a}}){Liodakis}, {Chakraborty}, {Marin}, {Ehlert}, {Barnouin}, {Kouch}, {Nilsson}, {Lindfors}, {Pursimo}, {Paraschos}, {Middei}, {Trindade Falc{\~a}o}, {Jorstad}, {Agudo}, {Kovalev}, {Casey}, {Di Gesu}, {Kaaret}, {Kim}, {Kislat}, {Ratheesh}, {Saade}, {Tombesi}, {Marscher}, {Jos{\'e} Aceituno}, {Bonnoli}, {Casanova}, {Emery}, {Escudero Pedrosa}, {Morcuende}, {Otero-Santos}, {Sota}, {Piirola}, {Bachev}, {Strigachev}, {Borman}, {Grishina}, {Hagen-Thorn}, {Kopatskaya}, {Larionova}, {Morozova}, {Savchenko}, {Shishkina}, {Troitskiy}, {Troitskaya}, {Vasilyev}, {Zhovtan}, {Myserlis}, {Gurwell}, {Keating}, {Rao}, {Kang}, {Lee}, {Kim}, {Yeon Cheong}, {Jeong}, {Song}, {Li}, {Nam}, {{\'A}lvarez-Ortega}, {Casadio}, {Angelakis}, {Kraus}, {Jormanainen}, {Fallah Ramazani}, {Chen}, {Costa}, {Churazov}, {Ferrazzoli}, {Galanti}, {Khabibullin}, {O'Dell}, {Pacciani}, {Roncadelli}, {Roberts}, {Soffitta}, {Swartz}, {Tavecchio}, {Weisskopf}, \& {Zhuravleva}}]{liodakis2025lsp}
{Liodakis}, I., {Chakraborty}, S., {Marin}, F., {et~al.} 2025{\natexlab{a}}, \apjl, 994, L9

\bibitem[{{Liodakis} {et~al.}(2022){Liodakis}, {Marscher}, {Agudo}, {Berdyugin}, {Bernardos}, {Bonnoli}, {Borman}, {Casadio}, {Casanova}, {Cavazzuti}, {Rodriguez Cavero}, {Di Gesu}, {Di Lalla}, {Donnarumma}, {Ehlert}, {Errando}, {Escudero}, {Garc{\'\i}a-Comas}, {Ag{\'\i}s-Gonz{\'a}lez}, {Husillos}, {Jormanainen}, {Jorstad}, {Kagitani}, {Kopatskaya}, {Kravtsov}, {Krawczynski}, {Lindfors}, {Larionova}, {Madejski}, {Marin}, {Marchini}, {Marshall}, {Morozova}, {Massaro}, {Masiero}, {Mawet}, {Middei}, {Millar-Blanchaer}, {Myserlis}, {Negro}, {Nilsson}, {O'Dell}, {Omodei}, {Pacciani}, {Paggi}, {Panopoulou}, {Peirson}, {Perri}, {Petrucci}, {Poutanen}, {Puccetti}, {Romani}, {Sakanoi}, {Savchenko}, {Sota}, {Tavecchio}, {Tinyanont}, {Vasilyev}, {Weaver}, {Zhovtan}, {Antonelli}, {Bachetti}, {Baldini}, {Baumgartner}, {Bellazzini}, {Bianchi}, {Bongiorno}, {Bonino}, {Brez}, {Bucciantini}, {Capitanio}, {Castellano}, {Ciprini}, {Costa}, {De Rosa}, {Del Monte}, {Di Marco}, {Doroshenko}, {Dov{\v{c}}iak}, {Enoto},
  {Evangelista}, {Fabiani}, {Ferrazzoli}, {Garcia}, {Gunji}, {Hayashida}, {Heyl}, {Iwakiri}, {Karas}, {Kitaguchi}, {Kolodziejczak}, {La Monaca}, {Latronico}, {Maldera}, {Manfreda}, {Marinucci}, {Matt}, {Mitsuishi}, {Mizuno}, {Muleri}, {Ng}, {Oppedisano}, {Papitto}, {Pavlov}, {Pesce-Rollins}, {Pilia}, {Possenti}, {Ramsey}, {Rankin}, {Ratheesh}, {Sgr{\'o}}, {Slane}, {Soffitta}, {Spandre}, {Tamagawa}, {Taverna}, {Tawara}, {Tennant}, {Thomas}, {Tombesi}, {Trois}, {Tsygankov}, {Turolla}, {Vink}, {Weisskopf}, {Wu}, {Xie}, \& {Zane}}]{liodakis2022polarized}
{Liodakis}, I., {Marscher}, A.~P., {Agudo}, I., {et~al.} 2022, \nat, 611, 677

\bibitem[{{Liodakis} {et~al.}(2018){Liodakis}, {Romani}, {Filippenko}, {Kiehlmann}, {Max-Moerbeck}, {Readhead}, \& {Zheng}}]{liodakis2018}
{Liodakis}, I., {Romani}, R.~W., {Filippenko}, A.~V., {et~al.} 2018, \mnras, 480, 5517

\bibitem[{{Liodakis} {et~al.}(2019){Liodakis}, {Romani}, {Filippenko}, {Kocevski}, \& {Zheng}}]{liodakis2019}
{Liodakis}, I., {Romani}, R.~W., {Filippenko}, A.~V., {Kocevski}, D., \& {Zheng}, W. 2019, \apj, 880, 32

\bibitem[{{Liodakis} {et~al.}(2025{\natexlab{b}}){Liodakis}, {Zhang}, {Boula}, {Middei}, {Otero-Santos}, {Blinov}, {Agudo}, {B{\"o}ttcher}, {Chen}, {Ehlert}, {Jorstad}, {Kaaret}, {Krawczynski}, {Peirson}, {Romani}, {Tavecchio}, {Weisskopf}, {Kouch}, {Lindfors}, {Nilsson}, {McCall}, {Jermak}, {Steele}, {Myserlis}, {Gurwell}, {Keating}, {Rao}, {Kang}, {Lee}, {Kim}, {Yeon Cheong}, {Jeong}, {Angelakis}, {Kraus}, {Jos{\'e} Aceituno}, {Bonnoli}, {Casanova}, {Escudero}, {Ag{\'\i}s-Gonz{\'a}lez}, {Morcuende}, {Sota}, {Bachev}, {Grishina}, {Kopatskaya}, {Larionova}, {Morozova}, {Savchenko}, {Shishkina}, {Troitskiy}, {Troitskaya}, \& {Vasilyev}}]{liodakis2025}
{Liodakis}, I., {Zhang}, H., {Boula}, S., {et~al.} 2025{\natexlab{b}}, \aap, 698, L19

\bibitem[{{Maksym} {et~al.}(2025){Maksym}, {Liodakis}, {Saade}, {Kim}, {Middei}, {Di Gesu}, {Kiehlmann}, {Matzeu}, {Agudo}, {Marscher}, {Ehlert}, {Jorstad}, {Kaaret}, {Marshall}, {Pacciani}, {Perri}, {Puccetti}, {Kouch}, {Lindfors}, {Aceituno}, {Bonnoli}, {Casanova}, {Escudero}, {G{\'o}mez}, {Ag{\'\i}s-Gonz{\'a}lez}, {Husillos}, {Morcuende}, {Otero-Santos}, {Sota}, {Piirola}, {Imazawa}, {Sasada}, {Fukazawa}, {Kawabata}, {Uemura}, {Mizuno}, {Nakaoka}, {Akitaya}, {McCall}, {Jermak}, {Steele}, {Borman}, {Grishina}, {Hagen-Thorn}, {Kopatskaya}, {Larionova}, {Morozova}, {Savchenko}, {Shishkina}, {Troitskiy}, {Troitskaya}, {Vasilyev}, {Zhovtan}, {Myserlis}, {Gurwell}, {Keating}, {Rao}, {Pauley}, {Angelakis}, {Kraus}, {Berdyugin}, {Kagitani}, {Kravtsov}, {Poutanen}, {Sakanoi}, {Kang}, {Lee}, {Kim}, {Cheong}, {Jeong}, {Song}, {Blinov}, {Shablovinskaya}, {Antonelli}, {Bachetti}, {Baldini}, {Baumgartner}, {Bellazzini}, {Bianchi}, {Bongiorno}, {Bonino}, {Brez}, {Bucciantini}, {Capitanio}, {Castellano}, {Cavazzuti},
  {Chen}, {Ciprini}, {Costa}, {De Rosa}, {Del Monte}, {Di Lalla}, {Di Marco}, {Donnarumma}, {Doroshenko}, {Dov{\v{c}}iak}, {Enoto}, {Evangelista}, {Fabiani}, {Ferrazzoli}, {Garcia}, {Gunji}, {Hayashida}, {Heyl}, {Iwakiri}, {Karas}, {Kislat}, {Kitaguchi}, {Kolodziejczak}, {Krawczynski}, {La Monaca}, {Latronico}, {Maldera}, {Manfreda}, {Marin}, {Marinucci}, {Massaro}, {Matt}, {Mitsuishi}, {Muleri}, {Negro}, {Ng}, {O'Dell}, {Omodei}, {Oppedisano}, {Papitto}, {Pavlov}, {Peirson}, {Pesce-Rollins}, {Petrucci}, {Pilia}, {Possenti}, {Ramsey}, {Rankin}, {Ratheesh}, {Roberts}, {Romani}, {Sgr{\'o}}, {Slane}, {Soffitta}, {Spandre}, {Swartz}, {Tamagawa}, {Tavecchio}, {Taverna}, {Tawara}, {Tennant}, {Thomas}, {Tombesi}, {Trois}, {Tsygankov}, {Turolla}, {Vink}, {Weisskopf}, {Wu}, {Xie}, \& {Zane}}]{maksym2024two}
{Maksym}, W.~P., {Liodakis}, I., {Saade}, M.~L., {et~al.} 2025, \apj, 986, 230

\bibitem[{{Mannheim} \& {Biermann}(1992)}]{mannheim1992}
{Mannheim}, K. \& {Biermann}, P.~L. 1992, \aap, 253, L21

\bibitem[{{Marshall} {et~al.}(2022){Marshall}, {Heine}, {Garner}, {Masterson}, {Miller}, {Guenther}, \& {Bongiorno}}]{gosox2022}
{Marshall}, H., {Heine}, S., {Garner}, A., {et~al.} 2022, in 44th COSPAR Scientific Assembly. Held 16-24 July, Vol.~44, 1871

\bibitem[{{Marshall} {et~al.}(2024){Marshall}, {Liodakis}, {Marscher}, {Di Lalla}, {Jorstad}, {Kim}, {Middei}, {Negro}, {Omodei}, {Peirson}, {Perri}, {Puccetti}, {Laurenti}, {Agudo}, {Bonnoli}, {Berdyugin}, {Cavazzuti}, {Rodriguez Cavero}, {Donnarumma}, {Di Gesu}, {Jormanainen}, {Krawczynski}, {Lindfors}, {Madjeski}, {Marin}, {Massaro}, {Pacciani}, {Poutanen}, {Tavecchio}, {Kouch}, {Aceituno}, {Bernardos}, {Casanova}, {Garc{\'\i}a-Comas}, {Ag{\'\i}s-Gonz{\'a}lez}, {Husillos}, {Marchini}, {Sota}, {Blinov}, {Bourbah}, {Kielhmann}, {Kontopodis}, {Mandarakas}, {Romanopoulos}, {Skalidis}, {Vervelaki}, {Borman}, {Kopatskaya}, {Larionova}, {Morozova}, {Savchenko}, {Vasilyev}, {Zhovtan}, {Casadio}, {Escudero}, {Kramer}, {Myserlis}, {Trainou}, {Imazawa}, {Sasada}, {Fukazawa}, {Kawabata}, {Uemura}, {Mizuno}, {Nakaoka}, {Akitaya}, {Masiero}, {Mawet}, {Panopoulou}, {Tinyanont}, {Kagitani}, {Kravtsov}, {Sakanoi}, {Dattolo}, {Gurwell}, {Keating}, {Rao}, {Cheong}, {Jeong}, {Kang}, {Kim}, {Lee}, {Angelakis}, {Kraus},
  {Hales}, {Kameno}, {Kneissl}, {Messias}, {Nagai}, {Antonelli}, {Bachetti}, {Baldini}, {Baumgartner}, {Bellazzini}, {Bianchi}, {Bongiorno}, {Bonino}, {Brez}, {Bucciantini}, {Capitanio}, {Castellano}, {Chen}, {Ciprini}, {Costa}, {De Rosa}, {Del Monte}, {Di Marco}, {Doroshenko}, {Dov{\v{c}}iak}, {Ehlert}, {Enoto}, {Evangelista}, {Fabiani}, {Ferrazzoli}, {Garcia}, {Gunji}, {Hayashida}, {Heyl}, {Iwakiri}, {Kaaret}, {Karas}, {Kislat}, {Kitaguchi}, {Kolodziejczak}, {La Monaca}, {Latronico}, {Maldera}, {Manfreda}, {Marinucci}, {Matt}, {Mitsuishi}, {Muleri}, {Ng}, {O'Dell}, {Oppedisano}, {Papitto}, {Pavlov}, {Pesce-Rollins}, {Petrucci}, {Pilia}, {Possenti}, {Ramsey}, {Rankin}, {Ratheesh}, {Roberts}, {Romani}, {Sgr{\`o}}, {Slane}, {Soffitta}, {Spandre}, {Swartz}, {Tamagawa}, {Taverna}, {Tawara}, {Tennant}, {Thomas}, {Tombesi}, {Trois}, {Tsygankov}, {Turolla}, {Vink}, {Weisskopf}, {Wu}, {Xie}, \& {Zane}}]{marshall2024observations}
{Marshall}, H.~L., {Liodakis}, I., {Marscher}, A.~P., {et~al.} 2024, \apj, 972, 74

\bibitem[{{Middei} {et~al.}(2023{\natexlab{a}}){Middei}, {Liodakis}, {Perri}, {Puccetti}, {Cavazzuti}, {Di Gesu}, {Ehlert}, {Madejski}, {Marscher}, {Marshall}, {Muleri}, {Negro}, {Jorstad}, {Ag{\'\i}s-Gonz{\'a}lez}, {Agudo}, {Bonnoli}, {Bernardos}, {Casanova}, {Garc{\'\i}a-Comas}, {Husillos}, {Marchini}, {Sota}, {Kouch}, {Lindfors}, {Borman}, {Kopatskaya}, {Larionova}, {Morozova}, {Savchenko}, {Vasilyev}, {Zhovtan}, {Casadio}, {Escudero}, {Myserlis}, {Hales}, {Kameno}, {Kneissl}, {Messias}, {Nagai}, {Blinov}, {Bourbah}, {Kiehlmann}, {Kontopodis}, {Mandarakas}, {Romanopoulos}, {Skalidis}, {Vervelaki}, {Masiero}, {Mawet}, {Millar-Blanchaer}, {Panopoulou}, {Tinyanont}, {Berdyugin}, {Kagitani}, {Kravtsov}, {Sakanoi}, {Imazawa}, {Sasada}, {Fukazawa}, {Kawabata}, {Uemura}, {Mizuno}, {Nakaoka}, {Akitaya}, {Gurwell}, {Rao}, {Di Lalla}, {Cibrario}, {Donnarumma}, {Kim}, {Omodei}, {Pacciani}, {Poutanen}, {Tavecchio}, {Antonelli}, {Bachetti}, {Baldini}, {Baumgartner}, {Bellazzini}, {Bianchi}, {Bongiorno}, {Bonino},
  {Brez}, {Bucciantini}, {Capitanio}, {Castellano}, {Ciprini}, {Costa}, {De Rosa}, {Del Monte}, {Di Marco}, {Doroshenko}, {Dov{\v{c}}iak}, {Enoto}, {Evangelista}, {Fabiani}, {Ferrazzoli}, {Garcia}, {Gunji}, {Hayashida}, {Heyl}, {Iwakiri}, {Karas}, {Kitaguchi}, {Kolodziejczak}, {Krawczynski}, {La Monaca}, {Latronico}, {Maldera}, {Manfreda}, {Marin}, {Marinucci}, {Massaro}, {Matt}, {Mitsuishi}, {Ng}, {O'Dell}, {Oppedisano}, {Papitto}, {Pavlov}, {Peirson}, {Pesce-Rollins}, {Petrucci}, {Pilia}, {Possenti}, {Ramsey}, {Rankin}, {Ratheesh}, {Romani}, {Sgr{\'o}}, {Slane}, {Soffitta}, {Spandre}, {Tamagawa}, {Taverna}, {Tawara}, {Tennant}, {Thomas}, {Tombesi}, {Trois}, {Tsygankov}, {Turolla}, {Vink}, {Weisskopf}, {Wu}, {Xie}, \& {Zane}}]{middei2022x}
{Middei}, R., {Liodakis}, I., {Perri}, M., {et~al.} 2023{\natexlab{a}}, \apjl, 942, L10

\bibitem[{{Middei} {et~al.}(2023{\natexlab{b}}){Middei}, {Perri}, {Puccetti}, {Liodakis}, {Di Gesu}, {Marscher}, {Rodriguez Cavero}, {Tavecchio}, {Donnarumma}, {Laurenti}, {Jorstad}, {Agudo}, {Marshall}, {Pacciani}, {Kim}, {Aceituno}, {Bonnoli}, {Casanova}, {Ag{\'\i}s-Gonz{\'a}lez}, {Sota}, {Casadio}, {Escudero}, {Myserlis}, {Sievers}, {Kouch}, {Lindfors}, {Gurwell}, {Keating}, {Rao}, {Kang}, {Lee}, {Kim}, {Cheong}, {Jeong}, {Angelakis}, {Kraus}, {Antonelli}, {Bachetti}, {Baldini}, {Baumgartner}, {Bellazzini}, {Bianchi}, {Bongiorno}, {Bonino}, {Brez}, {Bucciantini}, {Capitanio}, {Castellano}, {Cavazzuti}, {Chen}, {Ciprini}, {Costa}, {De Rosa}, {Del Monte}, {Di Lalla}, {Di Marco}, {Doroshenko}, {Dov{\v{c}}iak}, {Ehlert}, {Enoto}, {Evangelista}, {Fabiani}, {Ferrazzoli}, {Garc{\'\i}a}, {Gunji}, {Hayashida}, {Heyl}, {Iwakiri}, {Kaaret}, {Karas}, {Kislat}, {Kitaguchi}, {Kolodziejczak}, {Krawczynski}, {La Monaca}, {Latronico}, {Maldera}, {Manfreda}, {Marin}, {Marinucci}, {Massaro}, {Matt}, {Mitsuishi}, {Mizuno},
  {Muleri}, {Negro}, {Ng}, {O'Dell}, {Omodei}, {Oppedisano}, {Papitto}, {Pavlov}, {Peirson}, {Pesce-Rollins}, {Petrucci}, {Pilia}, {Possenti}, {Poutanen}, {Ramsey}, {Rankin}, {Ratheesh}, {Roberts}, {Romani}, {Sgr{\`o}}, {Slane}, {Soffitta}, {Spandre}, {Swartz}, {Tamagawa}, {Taverna}, {Tawara}, {Tennant}, {Thomas}, {Tombesi}, {Trois}, {Tsygankov}, {Turolla}, {Vink}, {Weisskopf}, {Wu}, {Xie}, \& {Zane}}]{middei2023}
{Middei}, R., {Perri}, M., {Puccetti}, S., {et~al.} 2023{\natexlab{b}}, \apjl, 953, L28

\bibitem[{{M{\"u}cke} \& {Protheroe}(2001)}]{mucke2001}
{M{\"u}cke}, A. \& {Protheroe}, R.~J. 2001, Astroparticle Physics, 15, 121

\bibitem[{{Nolan} {et~al.}(2012){Nolan}, {Abdo}, {Ackermann}, {Ajello}, {Allafort}, {Antolini}, {Atwood}, {Axelsson}, {Baldini}, {Ballet}, {Barbiellini}, {Bastieri}, {Bechtol}, {Belfiore}, {Bellazzini}, {Berenji}, {Bignami}, {Blandford}, {Bloom}, {Bonamente}, {Bonnell}, {Borgland}, {Bottacini}, {Bouvier}, {Brandt}, {Bregeon}, {Brigida}, {Bruel}, {Buehler}, {Burnett}, {Buson}, {Caliandro}, {Cameron}, {Campana}, {Ca{\~n}adas}, {Cannon}, {Caraveo}, {Casandjian}, {Cavazzuti}, {Ceccanti}, {Cecchi}, {{\c{C}}elik}, {Charles}, {Chekhtman}, {Cheung}, {Chiang}, {Chipaux}, {Ciprini}, {Claus}, {Cohen-Tanugi}, {Cominsky}, {Conrad}, {Corbet}, {Cutini}, {D'Ammando}, {Davis}, {de Angelis}, {DeCesar}, {DeKlotz}, {De Luca}, {den Hartog}, {de Palma}, {Dermer}, {Digel}, {Silva}, {Drell}, {Drlica-Wagner}, {Dubois}, {Dumora}, {Enoto}, {Escande}, {Fabiani}, {Falletti}, {Favuzzi}, {Fegan}, {Ferrara}, {Focke}, {Fortin}, {Frailis}, {Fukazawa}, {Funk}, {Fusco}, {Gargano}, {Gasparrini}, {Gehrels}, {Germani}, {Giebels}, {Giglietto},
  {Giommi}, {Giordano}, {Giroletti}, {Glanzman}, {Godfrey}, {Grenier}, {Grondin}, {Grove}, {Guillemot}, {Guiriec}, {Gustafsson}, {Hadasch}, {Hanabata}, {Harding}, {Hayashida}, {Hays}, {Hill}, {Horan}, {Hou}, {Hughes}, {Iafrate}, {Itoh}, {J{\'o}hannesson}, {Johnson}, {Johnson}, {Johnson}, {Johnson}, {Kamae}, {Katagiri}, {Kataoka}, {Katsuta}, {Kawai}, {Kerr}, {Kn{\"o}dlseder}, {Kocevski}, {Kuss}, {Lande}, {Landriu}, {Latronico}, {Lemoine-Goumard}, {Lionetto}, {Llena Garde}, {Longo}, {Loparco}, {Lott}, {Lovellette}, {Lubrano}, {Madejski}, {Marelli}, {Massaro}, {Mazziotta}, {McConville}, {McEnery}, {Mehault}, {Michelson}, {Minuti}, {Mitthumsiri}, {Mizuno}, {Moiseev}, {Mongelli}, {Monte}, {Monzani}, {Morselli}, {Moskalenko}, {Murgia}, {Nakamori}, {Naumann-Godo}, {Norris}, {Nuss}, {Nymark}, {Ohno}, {Ohsugi}, {Okumura}, {Omodei}, {Orlando}, {Ormes}, {Ozaki}, {Paneque}, {Panetta}, {Parent}, {Perkins}, {Pesce-Rollins}, {Pierbattista}, {Pinchera}, {Piron}, {Pivato}, {Porter}, {Racusin}, {Rain{\`o}}, {Rando}, {Razzano},
  {Razzaque}, {Reimer}, {Reimer}, {Reposeur}, {Ritz}, {Rochester}, {Romani}, {Roth}, {Rousseau}, {Ryde}, {Sadrozinski}, {Salvetti}, {Sanchez}, {Saz Parkinson}, {Sbarra}, {Scargle}, {Schalk}, {Sgr{\`o}}, {Shaw}, {Shrader}, \& {Siskind}}]{nolan2012}
{Nolan}, P.~L., {Abdo}, A.~A., {Ackermann}, M., {et~al.} 2012, \apjs, 199, 31

\bibitem[{{Pacciani} {et~al.}(2025){Pacciani}, {Kim}, {Middei}, {Marshall}, {Marscher}, {Liodakis}, {Agudo}, {Jorstad}, {Poutanen}, {Errando}, {Di Gesu}, {Negro}, {Tavecchio}, {Wu}, {Chen}, {Muleri}, {Antonelli}, {Donnarumma}, {Ehlert}, {Massaro}, {O'Dell}, {Perri}, {Puccetti}, {Aceituno}, {Bonnoli}, {Casanova}, {Escudero}, {Ag{\'\i}s-Gonz{\'a}lez}, {Husillos}, {Morcuende}, {Otero-Santos}, {Sota}, {Kouch}, {Lindfors}, {Borman}, {G{\'o}mez}, {Kopatskaya}, {Larionova}, {Morozova}, {Savchenko}, {Vasilyev}, {Zhovtan}, {Blinov}, {Gourni}, {Kiehlmann}, {Kourtidis}, {Mandarakas}, {Palaiologou}, {Triantafyllou}, {Vervelaki}, {Myserlis}, {Gurwell}, {Keating}, {Rao}, {Angelakis}, {Kraus}, {Bachetti}, {Baldini}, {Baumgartner}, {Bellazzini}, {Bianchi}, {Bongiorno}, {Bonino}, {Brez}, {Bucciantini}, {Capitanio}, {Castellano}, {Cavazzuti}, {Ciprini}, {Costa}, {De Rosa}, {Del Monte}, {Di Lalla}, {Di Marco}, {Doroshenko}, {Dov{\v{c}}iak}, {Enoto}, {Evangelista}, {Fabiani}, {Ferrazzoli}, {Garcia}, {Gunji}, {Hayashida}, {Heyl},
  {Iwakiri}, {Kaaret}, {Karas}, {Kislat}, {Kitaguchi}, {Kolodziejczak}, {Krawczynski}, {La Monaca}, {Latronico}, {Maldera}, {Manfreda}, {Marin}, {Marinucci}, {Matt}, {Mitsuishi}, {Mizuno}, {Ng}, {Omodei}, {Oppedisano}, {Papitto}, {Pavlov}, {Peirson}, {Pesce-Rollins}, {Petrucci}, {Pilia}, {Possenti}, {Ramsey}, {Rankin}, {Ratheesh}, {Roberts}, {Romani}, {Sgr{\'o}}, {Slane}, {Soffitta}, {Spandre}, {Swartz}, {Tamagawa}, {Taverna}, {Tawara}, {Tennant}, {Thomas}, {Tombesi}, {Trois}, {Tsygankov}, {Turolla}, {Vink}, {Weisskopf}, {Xie}, \& {Zane}}]{pacciani2025}
{Pacciani}, L., {Kim}, D.~E., {Middei}, R., {et~al.} 2025, \apj, 983, 78

\bibitem[{{Pavlidou} {et~al.}(2014){Pavlidou}, {Angelakis}, {Myserlis}, {Blinov}, {King}, {Papadakis}, {Tassis}, {Hovatta}, {Pazderska}, {Paleologou}, {Balokovi{\'c}}, {Feiler}, {Fuhrmann}, {Khodade}, {Kus}, {Kylafis}, {Modi}, {Panopoulou}, {Papamastorakis}, {Pazderski}, {Pearson}, {Rajarshi}, {Ramaprakash}, {Readhead}, {Reig}, \& {Zensus}}]{pavlidou2014}
{Pavlidou}, V., {Angelakis}, E., {Myserlis}, I., {et~al.} 2014, \mnras, 442, 1693

\bibitem[{{Peirson} {et~al.}(2023){Peirson}, {Negro}, {Liodakis}, {Middei}, {Kim}, {Marscher}, {Marshall}, {Pacciani}, {Romani}, {Wu}, {Di Marco}, {Di Lalla}, {Omodei}, {Jorstad}, {Agudo}, {Kouch}, {Lindfors}, {Aceituno}, {Bernardos}, {Bonnoli}, {Casanova}, {Garc{\'\i}a-Comas}, {Ag{\'\i}s-Gonz{\'a}lez}, {Husillos}, {Marchini}, {Sota}, {Casadio}, {Escudero}, {Myserlis}, {Sievers}, {Gurwell}, {Rao}, {Imazawa}, {Sasada}, {Fukazawa}, {Kawabata}, {Uemura}, {Mizuno}, {Nakaoka}, {Akitaya}, {Cheong}, {Jeong}, {Kang}, {Kim}, {Lee}, {Angelakis}, {Kraus}, {Cibrario}, {Donnarumma}, {Poutanen}, {Tavecchio}, {Antonelli}, {Bachetti}, {Baldini}, {Baumgartner}, {Bellazzini}, {Bianchi}, {Bongiorno}, {Bonino}, {Brez}, {Bucciantini}, {Capitanio}, {Castellano}, {Cavazzuti}, {Chen}, {Ciprini}, {Costa}, {De Rosa}, {Del Monte}, {Di Gesu}, {Doroshenko}, {Dov{\v{c}}iak}, {Ehlert}, {Enoto}, {Evangelista}, {Fabiani}, {Ferrazzoli}, {Garcia}, {Gunji}, {Hayashida}, {Heyl}, {Iwakiri}, {Kaaret}, {Karas}, {Kitaguchi}, {Kolodziejczak},
  {Krawczynski}, {La Monaca}, {Latronico}, {Madejski}, {Maldera}, {Manfreda}, {Marin}, {Marinucci}, {Massaro}, {Matt}, {Mitsuishi}, {Muleri}, {Ng}, {O'Dell}, {Oppedisano}, {Papitto}, {Pavlov}, {Perri}, {Pesce-Rollins}, {Petrucci}, {Pilia}, {Possenti}, {Puccetti}, {Ramsey}, {Rankin}, {Ratheesh}, {Roberts}, {Sgr{\'o}}, {Slane}, {Soffitta}, {Spandre}, {Swartz}, {Tamagawa}, {Taverna}, {Tawara}, {Tennant}, {Thomas}, {Tombesi}, {Trois}, {Tsygankov}, {Turolla}, {Vink}, {Weisskopf}, {Xie}, \& {Zane}}]{peirson2023bllac}
{Peirson}, A.~L., {Negro}, M., {Liodakis}, I., {et~al.} 2023, \apjl, 948, L25

\bibitem[{{Raiteri}(2025)}]{raiteri2025}
{Raiteri}, C.~M. 2025, \aapr, 33, 8

\bibitem[{{Ramaprakash} {et~al.}(2019){Ramaprakash}, {Rajarshi}, {Das}, {Khodade}, {Modi}, {Panopoulou}, {Maharana}, {Blinov}, {Angelakis}, {Casadio}, {Fuhrmann}, {Hovatta}, {Kiehlmann}, {King}, {Kylafis}, {Kougentakis}, {Kus}, {Mahabal}, {Marecki}, {Myserlis}, {Paterakis}, {Paleologou}, {Liodakis}, {Papadakis}, {Papamastorakis}, {Pavlidou}, {Pazderski}, {Pearson}, {Readhead}, {Reig}, {S{\l}owikowska}, {Tassis}, \& {Zensus}}]{robopol}
{Ramaprakash}, A.~N., {Rajarshi}, C.~V., {Das}, H.~K., {et~al.} 2019, \mnras, 485, 2355

\bibitem[{{Rani} {et~al.}(2013){Rani}, {Krichbaum}, {Fuhrmann}, {B{\"o}ttcher}, {Lott}, {Aller}, {Aller}, {Angelakis}, {Bach}, {Bastieri}, {Falcone}, {Fukazawa}, {Gabanyi}, {Gupta}, {Gurwell}, {Itoh}, {Kawabata}, {Krips}, {L{\"a}hteenm{\"a}ki}, {Liu}, {Marchili}, {Max-Moerbeck}, {Nestoras}, {Nieppola}, {Quintana-Lacaci}, {Readhead}, {Richards}, {Sasada}, {Sievers}, {Sokolovsky}, {Stroh}, {Tammi}, {Tornikoski}, {Uemura}, {Ungerechts}, {Urano}, \& {Zensus}}]{rani2013}
{Rani}, B., {Krichbaum}, T.~P., {Fuhrmann}, L., {et~al.} 2013, \aap, 552, A11

\bibitem[{{Soffitta} {et~al.}(2023){Soffitta}, {Baldini}, {Baumgartner}, {Bellazzini}, {Bongiorno}, {Bucciantini}, {Costa}, {Dov{\v{c}}iak}, {Ehlert}, {Kaaret}, {Kolodziejczak}, {Latronico}, {Marin}, {Marscher}, {Marshall}, {Matt}, {Muleri}, {O'Dell}, {Poutanen}, {Ramsey}, {Romani}, {Slane}, {Tennant}, {Turolla}, {Weisskopf}, {Agudo}, {Antonelli}, {Bachetti}, {Bianchi}, {Bonino}, {Brez}, {Capitanio}, {Castellano}, {Cavazzuti}, {Chen}, {Ciprini}, {De Rosa}, {Del Monte}, {Di Gesu}, {Di Lalla}, {Di Marco}, {Donnarumma}, {Doroshenko}, {Enoto}, {Evangelista}, {Fabiani}, {Ferrazzoli}, {Garcia}, {Gunji}, {Hayashida}, {Heyl}, {Iwakiri}, {Jorstad}, {Karas}, {Kislat}, {Kitaguchi}, {Krawczynski}, {La Monaca}, {Liodakis}, {Maldera}, {Manfreda}, {Marinucci}, {Massaro}, {Mitsuishi}, {Mizuno}, {Negro}, {Ng}, {Omodei}, {Oppedisano}, {Papitto}, {Pavlov}, {Peirson}, {Perri}, {Pesce-Rollins}, {Petrucci}, {Pilia}, {Possenti}, {Puccetti}, {Rankin}, {Ratheesh}, {Roberts}, {Sgr{\`o}}, {Spandre}, {Swartz}, {Tamagawa}, {Tavecchio},
  {Taverna}, {Tawara}, {Thomas}, {Tombesi}, {Trois}, {Tsygankov}, {Vink}, {Wu}, {Xie}, \& {Zane}}]{soffitta2023}
{Soffitta}, P., {Baldini}, L., {Baumgartner}, W., {et~al.} 2023, in Society of Photo-Optical Instrumentation Engineers (SPIE) Conference Series, Vol. 12678, UV, X-Ray, and Gamma-Ray Space Instrumentation for Astronomy XXIII, ed. O.~H. {Siegmund} \& K.~{Hoadley}, 1267803

\bibitem[{{Tavecchio} {et~al.}(2025){Tavecchio}, {Bolis}, {Sobacchi}, {Boula}, \& {Sciaccaluga}}]{Tavecchio2025}
{Tavecchio}, F., {Bolis}, F., {Sobacchi}, E., {Boula}, S., \& {Sciaccaluga}, A. 2025, \aap, 700, A185

\bibitem[{{Tomsick} {et~al.}(2021){Tomsick}, {Boggs}, {Zoglauer}, {Lazar}, {Beechert}, {Gulick}, {Roberts}, {Siegert}, {Wulf}, {Sleator}, {Grove}, {Phlips}, {Brandt}, {Smale}, {Kierans}, {Hartmann}, {Leising}, {Ajello}, {Jean}, {von Ballmoos}, {Malzac}, {Burns}, {Fryer}, {Chang}, {Tavecchio}, \& {Takahashi}}]{tomsick2021}
{Tomsick}, J., {Boggs}, S., {Zoglauer}, A., {et~al.} 2021, in American Astronomical Society Meeting Abstracts, Vol. 237, American Astronomical Society Meeting Abstracts \#237, 315.01

\bibitem[{{Tomsick} {et~al.}(2014){Tomsick}, {Jean}, {Chang}, {Boggs}, {Zoglauer}, {Von Ballmoos}, {Amman}, {Chiu}, {Chang}, {Chou}, {Kierans}, {Lin}, {Lowell}, {Shang}, {Tseng}, \& {Yang}}]{tomsick2014}
{Tomsick}, J., {Jean}, P., {Chang}, H.-K., {et~al.} 2014, in 40th COSPAR Scientific Assembly, Vol.~40, PSB.1--14--14

\bibitem[{{Urry} \& {Padovani}(1995)}]{urry1995}
{Urry}, C.~M. \& {Padovani}, P. 1995, \pasp, 107, 803

\bibitem[{{Weisskopf} {et~al.}(2022){Weisskopf}, {Soffitta}, {Baldini}, {Ramsey}, {O'Dell}, {Romani}, {Matt}, {Deininger}, {Baumgartner}, {Bellazzini}, {Costa}, {Kolodziejczak}, {Latronico}, {Marshall}, {Muleri}, {Bongiorno}, {Tennant}, {Bucciantini}, {Dovciak}, {Marin}, {Marscher}, {Poutanen}, {Slane}, {Turolla}, {Kalinowski}, {Di Marco}, {Fabiani}, {Minuti}, {La Monaca}, {Pinchera}, {Rankin}, {Sgro'}, {Trois}, {Xie}, {Alexander}, {Allen}, {Amici}, {Andersen}, {Antonelli}, {Antoniak}, {Attin{\`a}}, {Barbanera}, {Bachetti}, {Baggett}, {Bladt}, {Brez}, {Bonino}, {Boree}, {Borotto}, {Breeding}, {Brienza}, {Bygott}, {Caporale}, {Cardelli}, {Carpentiero}, {Castellano}, {Castronuovo}, {Cavalli}, {Cavazzuti}, {Ceccanti}, {Centrone}, {Citraro}, {D'Amico}, {D'Alba}, {Di Gesu}, {Del Monte}, {Dietz}, {Di Lalla}, {Persio}, {Dolan}, {Donnarumma}, {Evangelista}, {Ferrant}, {Ferrazzoli}, {Ferrie}, {Footdale}, {Forsyth}, {Foster}, {Garelick}, {Gunji}, {Gurnee}, {Head}, {Hibbard}, {Johnson}, {Kelly}, {Kilaru}, {Lefevre},
  {Roy}, {Loffredo}, {Lorenzi}, {Lucchesi}, {Maddox}, {Magazzu}, {Maldera}, {Manfreda}, {Mangraviti}, {Marengo}, {Marrocchesi}, {Massaro}, {Mauger}, {McCracken}, {McEachen}, {Mize}, {Mereu}, {Mitchell}, {Mitsuishi}, {Morbidini}, {Mosti}, {Nasimi}, {Negri}, {Negro}, {Nguyen}, {Nitschke}, {Nuti}, {Onizuka}, {Oppedisano}, {Orsini}, {Osborne}, {Pacheco}, {Paggi}, {Painter}, {Pavelitz}, {Pentz}, {Piazzolla}, {Perri}, {Pesce-Rollins}, {Peterson}, {Pilia}, {Profeti}, {Puccetti}, {Ranganathan}, {Ratheesh}, {Reedy}, {Root}, {Rubini}, {Ruswick}, {Sanchez}, {Sarra}, {Santoli}, {Scalise}, {Sciortino}, {Schroeder}, {Seek}, {Sosdian}, {Spandre}, {Speegle}, {Tamagawa}, {Tardiola}, {Tobia}, {Thomas}, {Valerie}, {Vimercati}, {Walden}, {Weddendorf}, {Wedmore}, {Welch}, {Zanetti}, \& {Zanetti}}]{weisskopf2022imaging}
{Weisskopf}, M.~C., {Soffitta}, P., {Baldini}, L., {et~al.} 2022, Journal of Astronomical Telescopes, Instruments, and Systems, 8, 026002

\bibitem[{{Zhang} \& {B{\"o}ttcher}(2013)}]{zhangboettch2013}
{Zhang}, H. \& {B{\"o}ttcher}, M. 2013, \apj, 774, 18

\bibitem[{{Zhang} {et~al.}(2024){Zhang}, {B{\"o}ttcher}, \& {Liodakis}}]{zhang2024}
{Zhang}, H., {B{\"o}ttcher}, M., \& {Liodakis}, I. 2024, \apj, 967, 93

\bibitem[{{Zhang} {et~al.}(2019{\natexlab{a}}){Zhang}, {Fang}, {Li}, {Giannios}, {B{\"o}ttcher}, \& {Buson}}]{zhang2019}
{Zhang}, H., {Fang}, K., {Li}, H., {et~al.} 2019{\natexlab{a}}, \apj, 876, 109

\bibitem[{{Zhang} {et~al.}(2019{\natexlab{b}}){Zhang}, {Santangelo}, {Feroci}, {Xu}, {Lu}, {Chen}, {Feng}, {Zhang}, {Brandt}, {Hernanz}, {Baldini}, {Bozzo}, {Campana}, {De Rosa}, {Dong}, {Evangelista}, {Karas}, {Meidinger}, {Meuris}, {Nandra}, {Pan}, {Pareschi}, {Orleanski}, {Huang}, {Schanne}, {Sironi}, {Spiga}, {Svoboda}, {Tagliaferri}, {Tenzer}, {Vacchi}, {Zane}, {Walton}, {Wang}, {Winter}, {Wu}, {in't Zand}, {Ahangarianabhari}, {Ambrosi}, {Ambrosino}, {Barbera}, {Basso}, {Bayer}, {Bellazzini}, {Bellutti}, {Bertucci}, {Bertuccio}, {Borghi}, {Cao}, {Cadoux}, {Campana}, {Ceraudo}, {Chen}, {Chen}, {Chevenez}, {Civitani}, {Cui}, {Cui}, {Dauser}, {Del Monte}, {Di Cosimo}, {Diebold}, {Doroshenko}, {Dovciak}, {Du}, {Ducci}, {Fan}, {Favre}, {Fuschino}, {G{\'a}lvez}, {Gao}, {Ge}, {Gevin}, {Grassi}, {Gu}, {Gu}, {Han}, {Hong}, {Hu}, {Ji}, {Jia}, {Jiang}, {Kennedy}, {Kreykenbohm}, {Kuvvetli}, {Labanti}, {Latronico}, {Li}, {Li}, {Li}, {Li}, {Li}, {Limousin}, {Liu}, {Liu}, {Lu}, {Luo}, {Macera}, {Malcovati},
  {Martindale}, {Michalska}, {Meng}, {Minuti}, {Morbidini}, {Muleri}, {Paltani}, {Perinati}, {Picciotto}, {Piemonte}, {Qu}, {Rachevski}, {Rashevskaya}, {Rodriguez}, {Schanz}, {Shen}, {Sheng}, {Song}, {Song}, {Sgro}, {Sun}, {Tan}, {Uttley}, {Wang}, {Wang}, {Wang}, {Wang}, {Wang}, {Wang}, {Watts}, {Wen}, {Wilms}, {Xiong}, {Yang}, {Yang}, {Yang}, {Yu}, {Zhang}, {Zampa}, {Zampa}, {Zdziarski}, {Zhang}, {Zhang}, {Zhang}, {Zhang}, {Zhang}, {Zhang}, {Zhang}, {Zhang}, {Zhao}, {Zheng}, {Zhou}, {Zorzi}, \& {Zwart}}]{extp2019}
{Zhang}, S., {Santangelo}, A., {Feroci}, M., {et~al.} 2019{\natexlab{b}}, Science China Physics, Mechanics, and Astronomy, 62, 29502

\end{thebibliography}

\newpage

\begin{appendix}
\onecolumn

\section{Sample and observations}
\label{appendix_sample}

We report a detailed description of sources and observations.
The complete tables can be found on the \href{https://doi.org/10.5281/zenodo.20327371}{Zenodo} platform.
For all tables in this section, the source name in the Robopol catalogue has the prefix "RBPL" in front of the name reported in column "RBPL name", while the source name in the 4FGL catalogue has the prefix "4FGL" in front of the name reported in column "4FGL name".
The column "Optical PD" refers to the median optical PD measured by Robopol for that source, which was used to scale our PD models by matching the values at the central frequency of the Robopol band. The error is calculated as the standard error of the median.
The column "Fermi flux" indicates the flux reported in the 4FGL catalogue.
The column "Observations" indicates the dates of the first and the last observation by Robopol. Note that the observations do not cover the entire period indicated in the column, but rather intervals within that time span.

\begin{table*}[h]
\caption{LSP sources.}
\centering
\begin{tabular}{c c c c c c c}
\hline\hline
\noalign{\vskip 0.3mm}
Rbpl name & 4FGL name & Observations [MJD] & Optical PD [\%] & Error & Fermi flux [$erg \ cm^2$ $s^{-1}$] & Error \\
\noalign{\vskip 0.2mm}
\hline
\noalign{\vskip 0.4mm}
J0136+4751 & J0137.0+4751 & 56550.09 - 57637.00 & 7.90 & 0.86 & $2.89 \cdot 10^{-11}$ & $8.513 \cdot 10^{-13}$ \\
J0217+0837 & J0217.2+0837 & 56525.05 - 57637.03 & 5.30 & 0.50 & $1.94 \cdot 10^{-11}$ & $8.452 \cdot 10^{-13}$ \\
J0259+0747 & J0259.4+0746 & 56538.11 - 57349.85 & 24.55 & 2.33 & $1.86 \cdot 10^{-11}$ & $8.986 \cdot 10^{-13}$ \\
J0405-1308 & J0405.6-1308 & 56594.06 - 57345.96 & 1.45 & 0.21 & $1.109 \cdot 10^{-11}$ & $4.619 \cdot 10^{-13}$ \\
J0423-0120 & J0423.3-0120 & 56593.04 - 57345.99 & 9.30 & 1.65 & $3.504 \cdot 10^{-11}$ & $8.83 \cdot 10^{-13}$ \\
\multicolumn{7}{c}{\dots} \\
\hline
\end{tabular}
\tablefoot{\centering{The full table is available at the CDS.}}
\label{table:lsp_sample}
\end{table*}

\begin{table*}[h]
\caption{ISP sources.}
\centering
\begin{tabular}{c c c c c c c}
\hline\hline
\noalign{\vskip 0.3mm}
Rbpl name & 4FGL name & Observations [MJD] & Optical PD [\%] & Error & Fermi flux [$erg \ cm^2$ $s^{-1}$] & Error \\
\noalign{\vskip 0.2mm}
\hline
\noalign{\vskip 0.4mm}
J0114+1325 & J0114.8+1326 & 56524.12 - 57637.09 & 6.00 & 0.56 & $1.299 \cdot 10^{-11}$ & $5.741 \cdot 10^{-13}$ \\
J0211+1051 & J0211.2+1051 & 56525.06 - 57637.05 & 13.60 & 1.23 & $5.864 \cdot 10^{-11}$ & $1.362 \cdot 10^{-12}$ \\
J0848+6606 & J0849.1+6607 & 56464.80 - 57349.99 & 7.85 & 0.62 & $4.507 \cdot 10^{-12}$ & $4.229 \cdot 10^{-13}$ \\
J1037+5711 & J1037.7+5711 & 56465.85 - 57350.06 & 3.50 & 0.49 & $3.696 \cdot 10^{-11}$ & $1.159 \cdot 10^{-12}$ \\
J1058+5628 & J1058.6+5627 & 56484.81 - 57346.17 & 3.95 & 0.66 & $3.014 \cdot 10^{-11}$ & $8.762 \cdot 10^{-13}$ \\
\multicolumn{7}{c}{\dots} \\
\hline
\end{tabular}
\tablefoot{\centering{The full table is available at the CDS.}}
\label{table:isp_sample}
\end{table*}

\begin{table*}[h]
\caption{HSP sources.}
\centering
\begin{tabular}{c c c c c c c}
\hline\hline
\noalign{\vskip 0.3mm}
Rbpl name & 4FGL name & Observations [MJD] & Optical PD [\%] & Error & Fermi flux [$erg \ cm^2$ $s^{-1}$] & Error \\
\noalign{\vskip 0.2mm}
\hline
\noalign{\vskip 0.4mm}
J0045+2127 & J0045.3+2128 & 56523.10 - 57633.01 & 4.30 & 0.41 & $1.464 \cdot 10^{-11}$ & $8.213 \cdot 10^{-13}$ \\
J0303-2407 & J0303.4-2407 & 56599.93 - 57331.92 & 6.85 & 0.61 & $4.845 \cdot 10^{-11}$ & $1.272 \cdot 10^{-12}$ \\
J1555+1111 & J1555.7+1111 & 56443.93 - 57928.00 & 2.90 & 0.11 & $1.851 \cdot 10^{-10}$ & $3.029 \cdot 10^{-12}$ \\
J1653+3945 & J1653.8+3945 & 56445.92 - 57285.75 & 2.00 & 0.14 & $1.177 \cdot 10^{-10}$ & $1.893 \cdot 10^{-12}$ \\
J1725+1152 & J1725.0+1152 & 56438.03 - 57636.74 & 6.70 & 0.50 & $3.897 \cdot 10^{-11}$ & $1.335 \cdot 10^{-12}$ \\
\multicolumn{7}{c}{\dots} \\
\hline
\end{tabular}
\tablefoot{\centering{The full table is available at the CDS.}}
\label{table:hsp_sample}
\end{table*}

\clearpage

\section{Polarisation models}
\label{appendix_polarisation_models}

We report here the polarisation models used for ISP and HSP sources, obtained from J0211+1051 and Mrk~501.

\begin{figure}[h!]
    \centering
    \begin{subfigure}{0.49\linewidth}
        \centering
        \includegraphics[width=\linewidth]{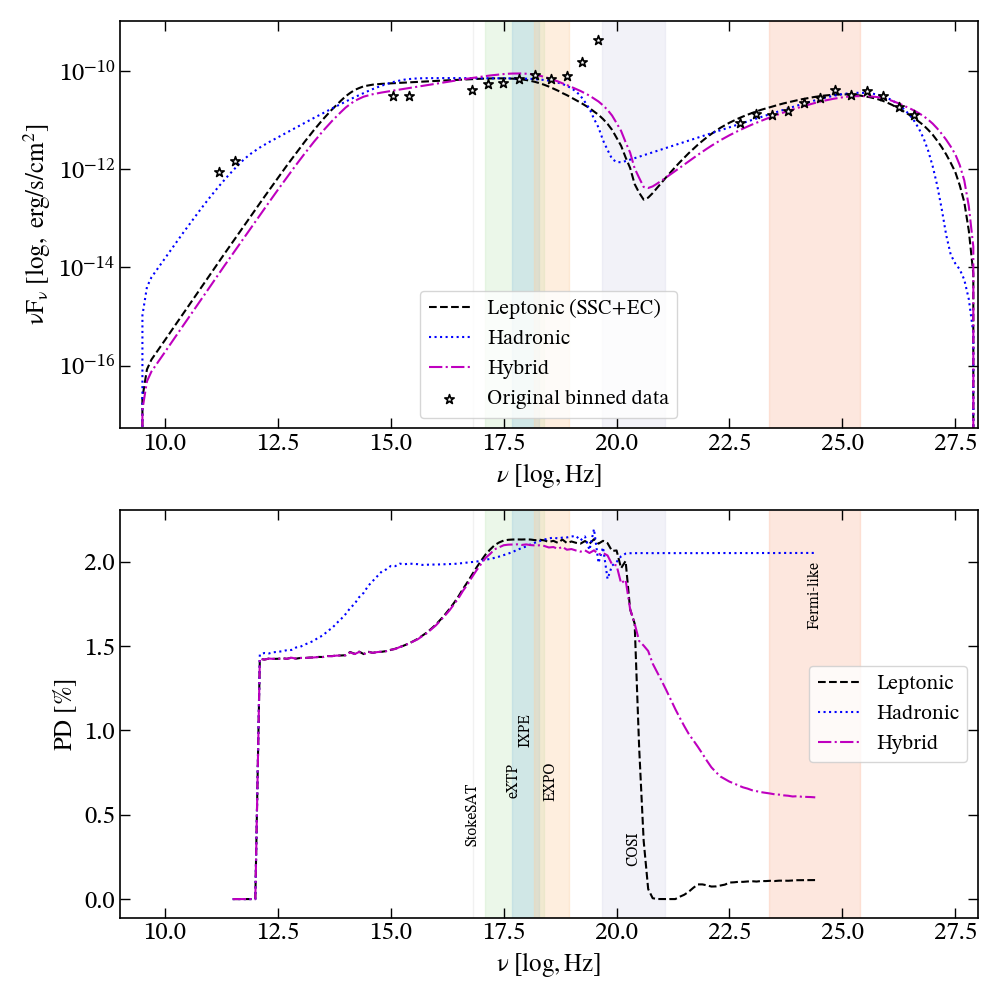}
        \caption{Mrk~501 (HSP)}
        \label{fig:mrk501_models}
    \end{subfigure}
    \hfill
    \begin{subfigure}{0.49\linewidth}
        \centering
        \includegraphics[width=\linewidth]{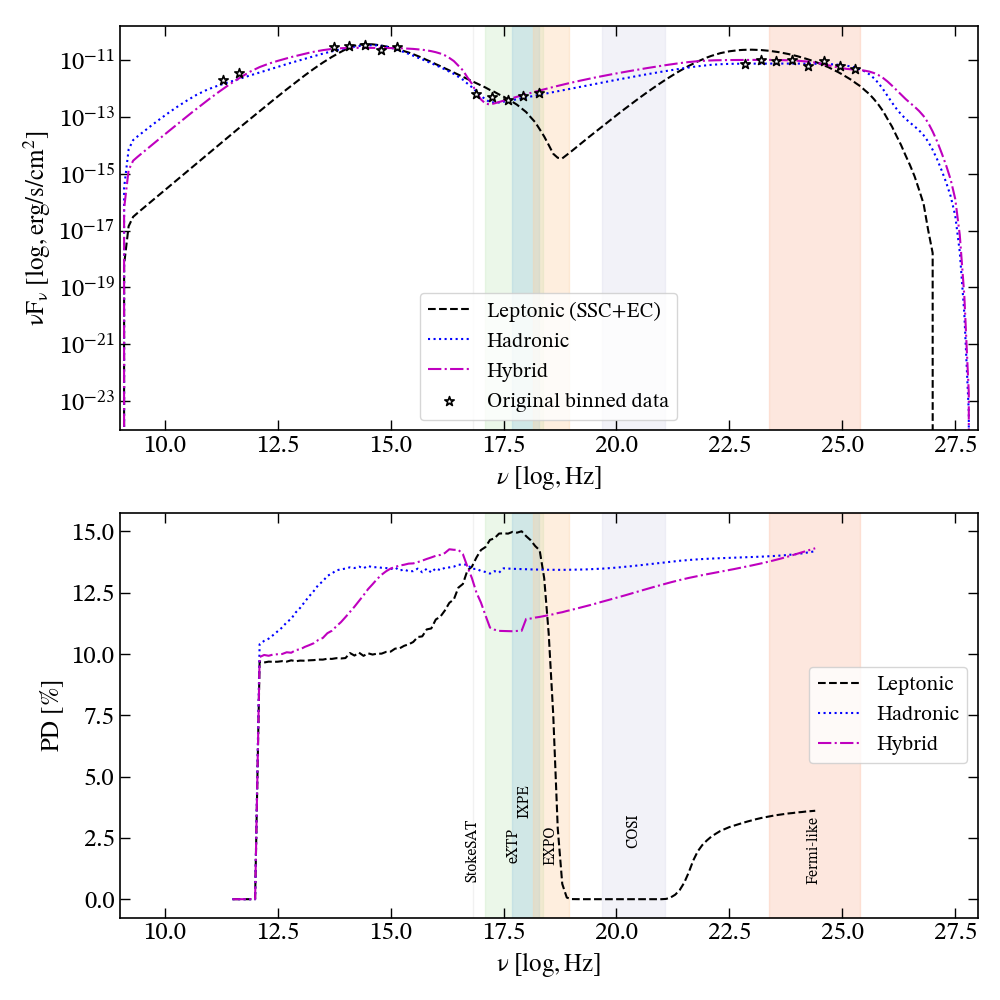}
        \caption{BZB~J0211+1051 (ISP)}
        \label{fig:bzb_models}
    \end{subfigure}

    \caption{
    Reference models used in this work (HSP on the left, ISP on the right).
    LSP can be found in the text (Fig. \ref{fig:bllac_model}).
    The upper panels show the fitted SED obtained by applying the leptonic (black dashed line), hadronic (blue dotted line), and hybrid (magenta dash-dot line) models, with the black stars representing the frequency-binned data.
    The lower panels show the PD predicted by the three models.
    }
    \label{fig:reference_models}
\end{figure}

\section{Polarisation detectability}
\label{appendix_mdp_plots}

We report here the plots representing the MDP curves of each instrument and the predictions given by our models for each class of sources.
For all plots, blue, yellow, and red points represent predictions from the leptonic, hybrid, and hadronic models, respectively.
MDP curves represent two week exposures (solid light grey) and four week exposures (dashed dark grey).

\begin{figure*}[!ht]
    \centering
    \begin{subfigure}{0.32\textwidth}
        \centering
        \includegraphics[width=\textwidth]{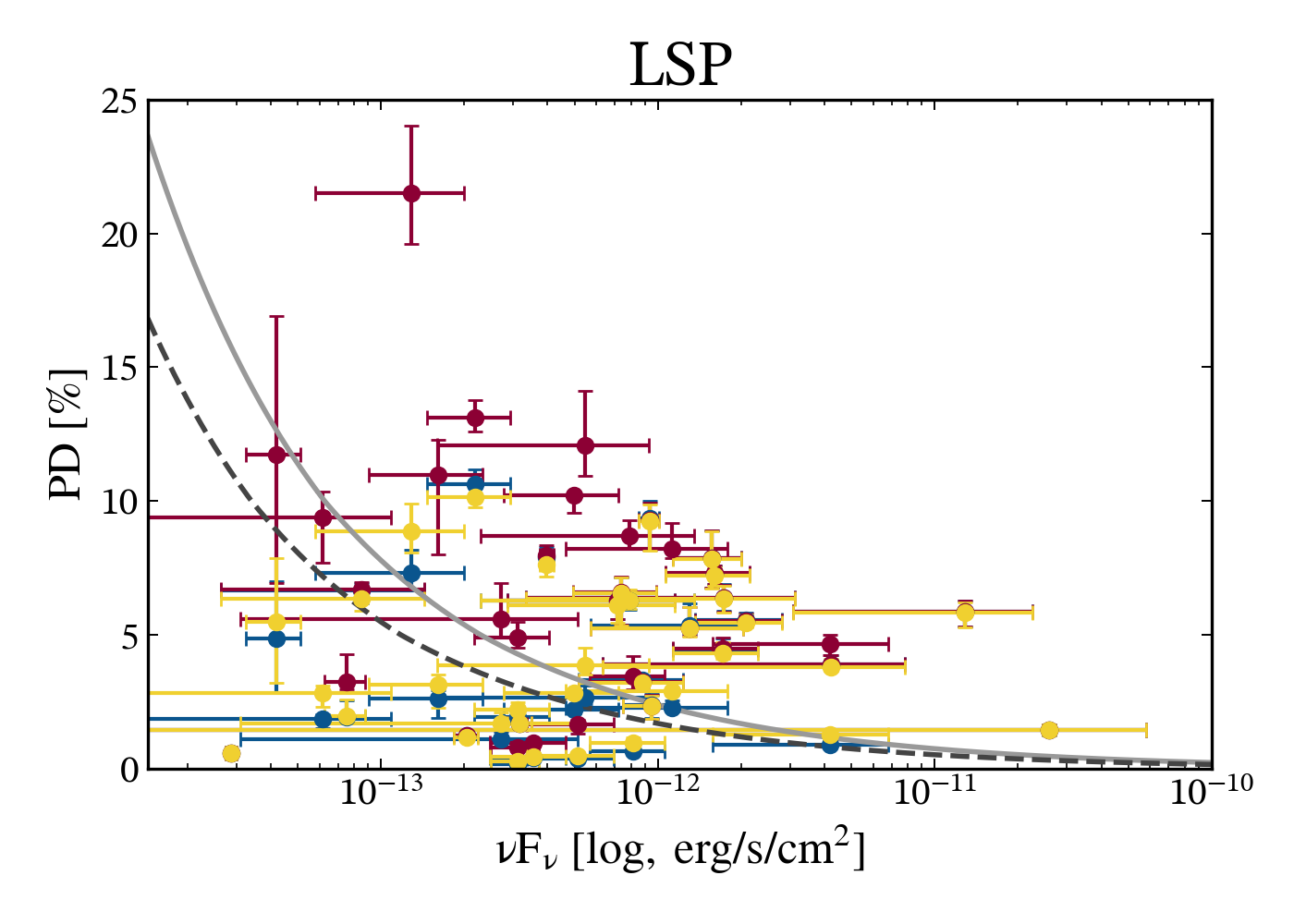}
    \end{subfigure}
    \hfill
    \begin{subfigure}{0.32\textwidth}
        \centering
        \includegraphics[width=\textwidth]{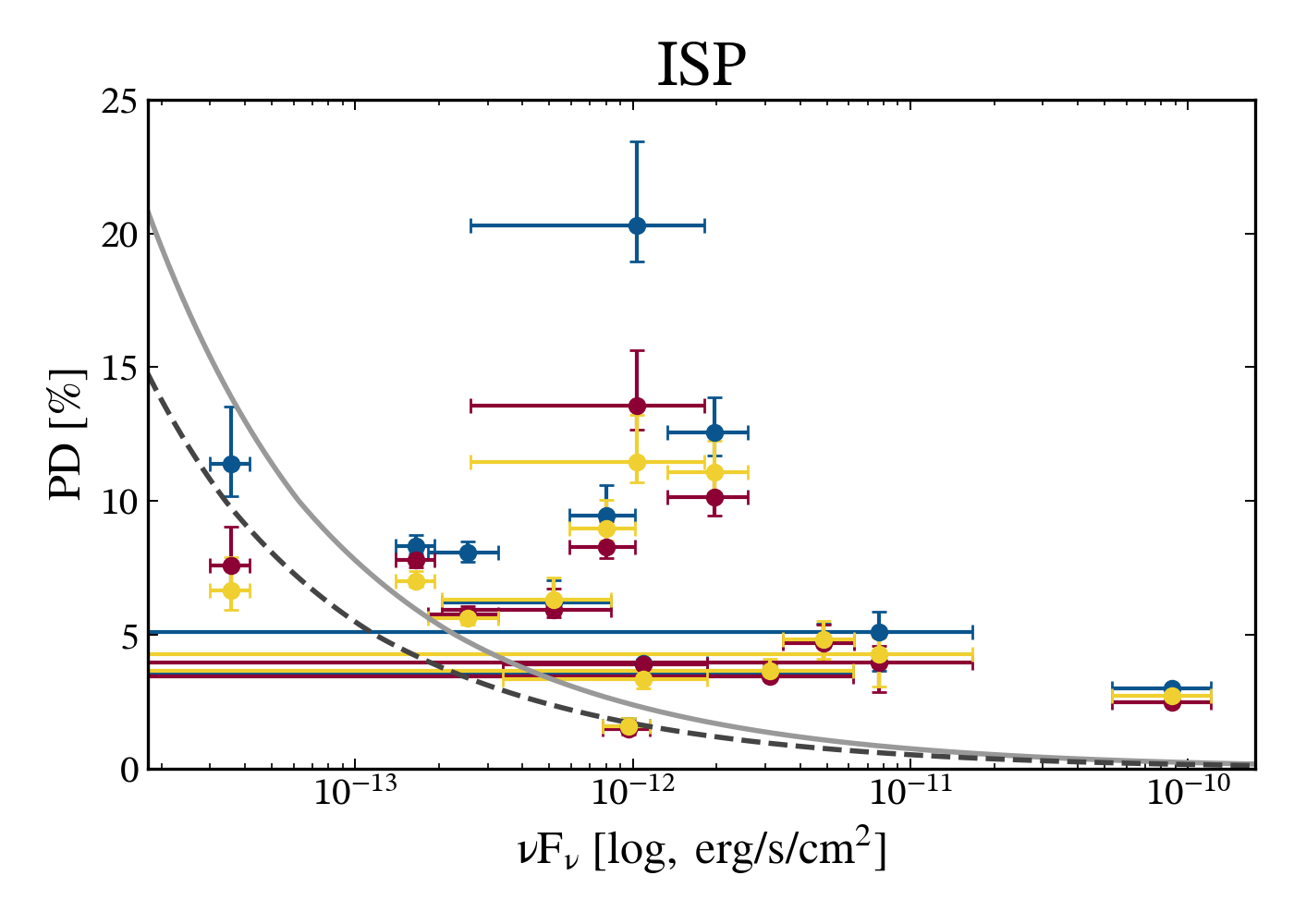}
    \end{subfigure}
    \hfill
    \begin{subfigure}{0.32\textwidth}
        \centering
        \includegraphics[width=\textwidth]{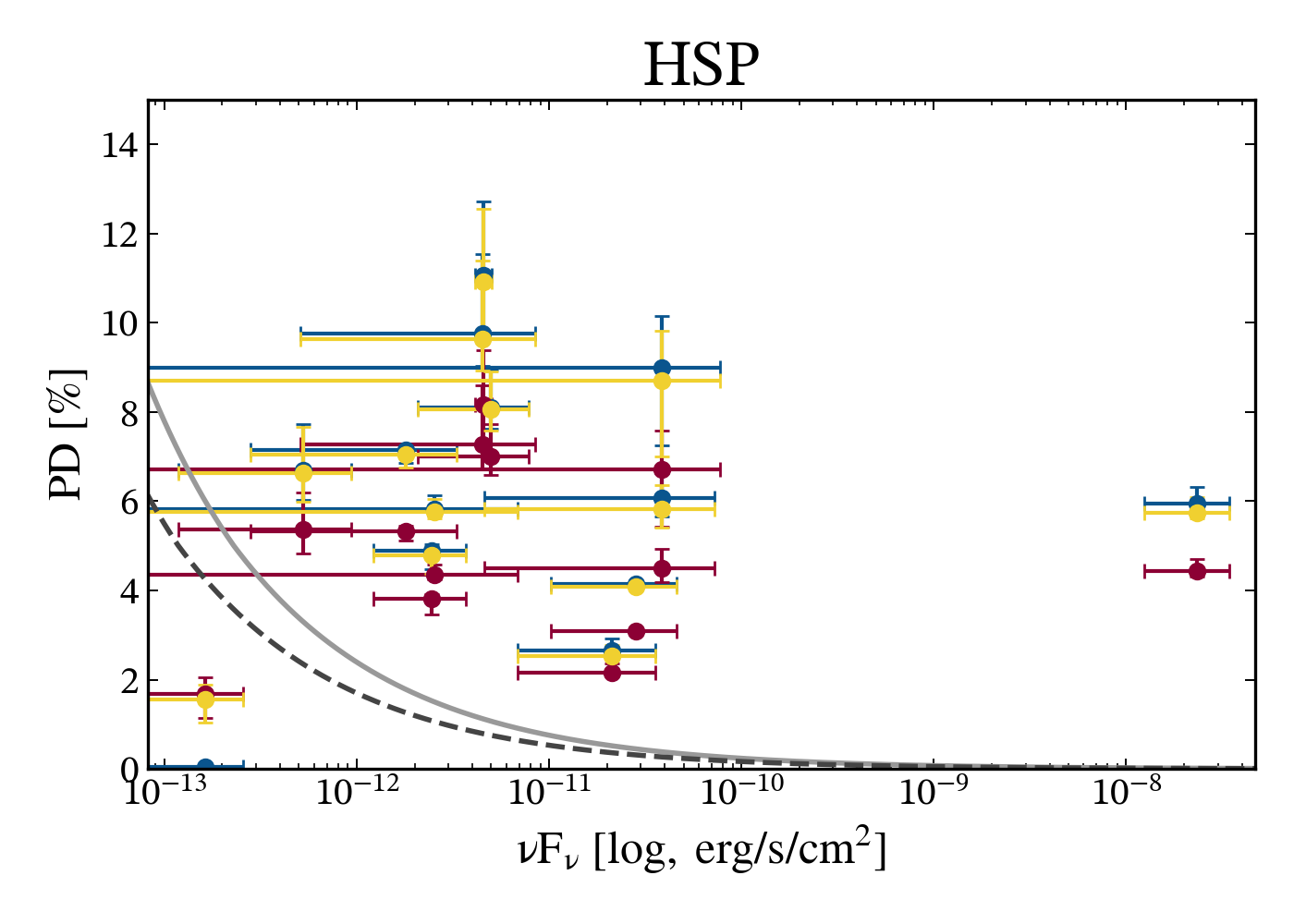}
    \end{subfigure}
    \caption{StokeSAT band (0.270-0.280 keV)}
    \label{fig:mdp_stokesat}
\end{figure*}

\begin{figure*}[!ht]
    \centering
    \begin{subfigure}{0.32\textwidth}
        \centering
        \includegraphics[width=\textwidth]{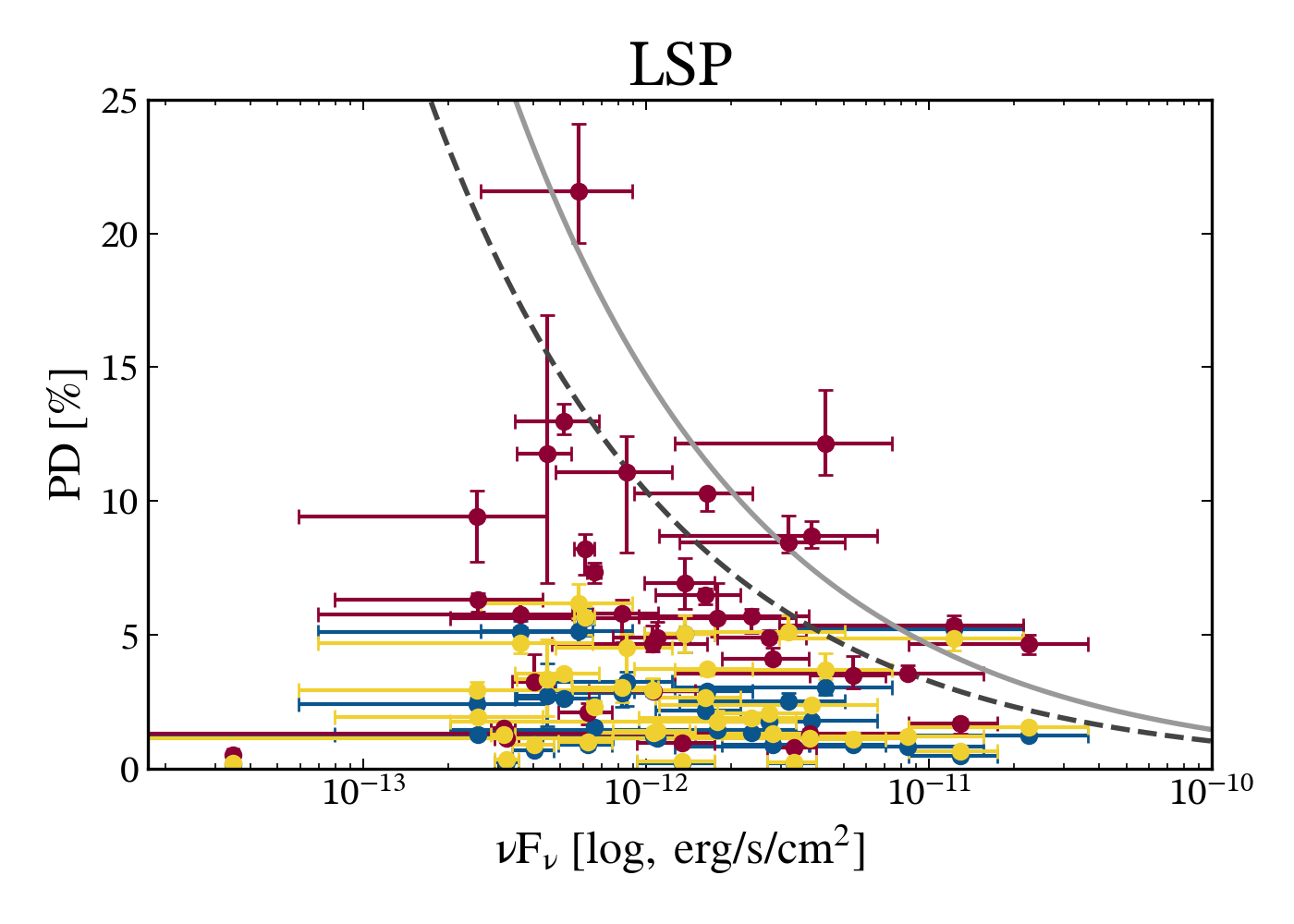}
    \end{subfigure}
    \hfill
    \begin{subfigure}{0.32\textwidth}
        \centering
        \includegraphics[width=\textwidth]{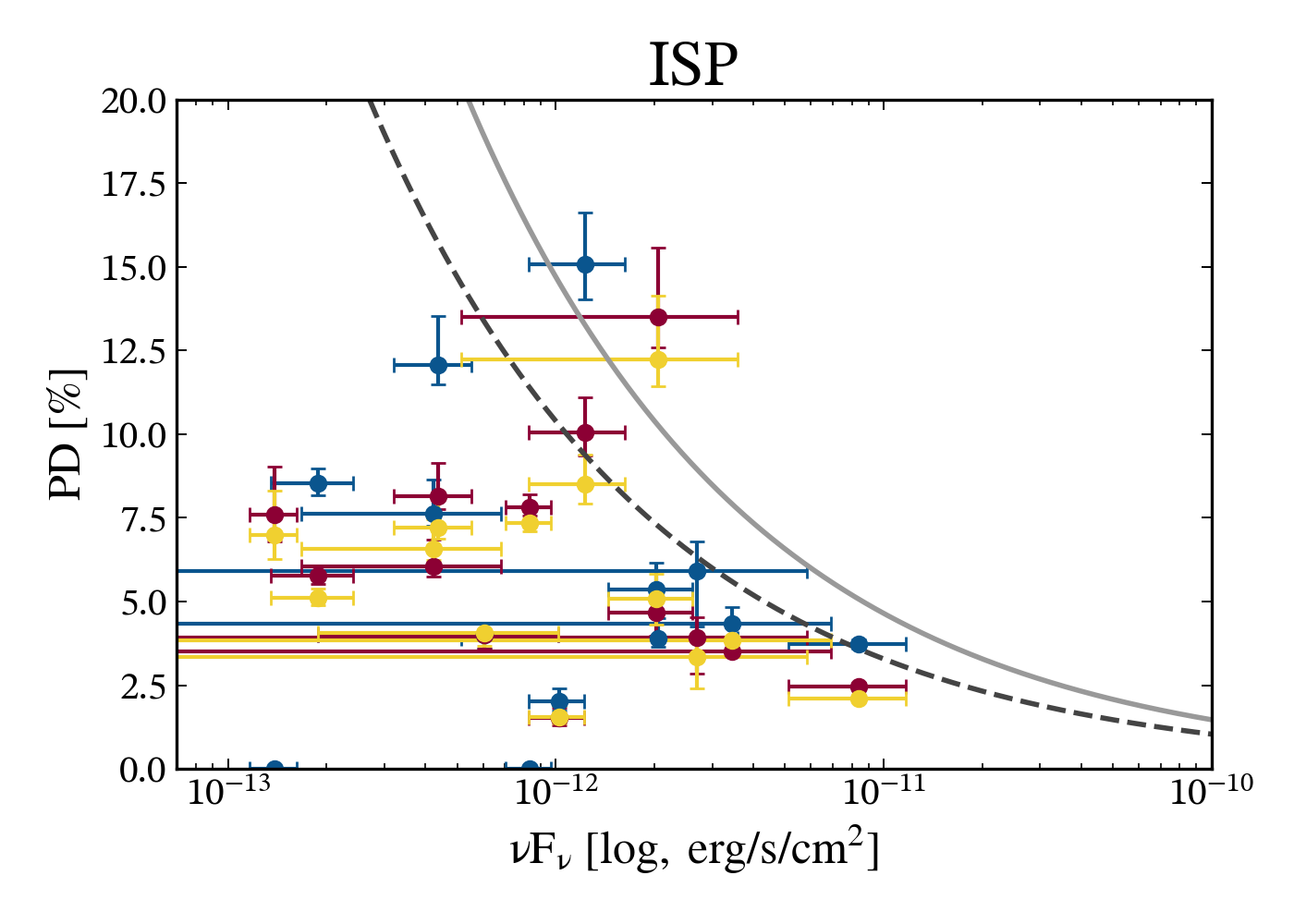}
    \end{subfigure}
    \hfill
    \begin{subfigure}{0.32\textwidth}
        \centering
        \includegraphics[width=\textwidth]{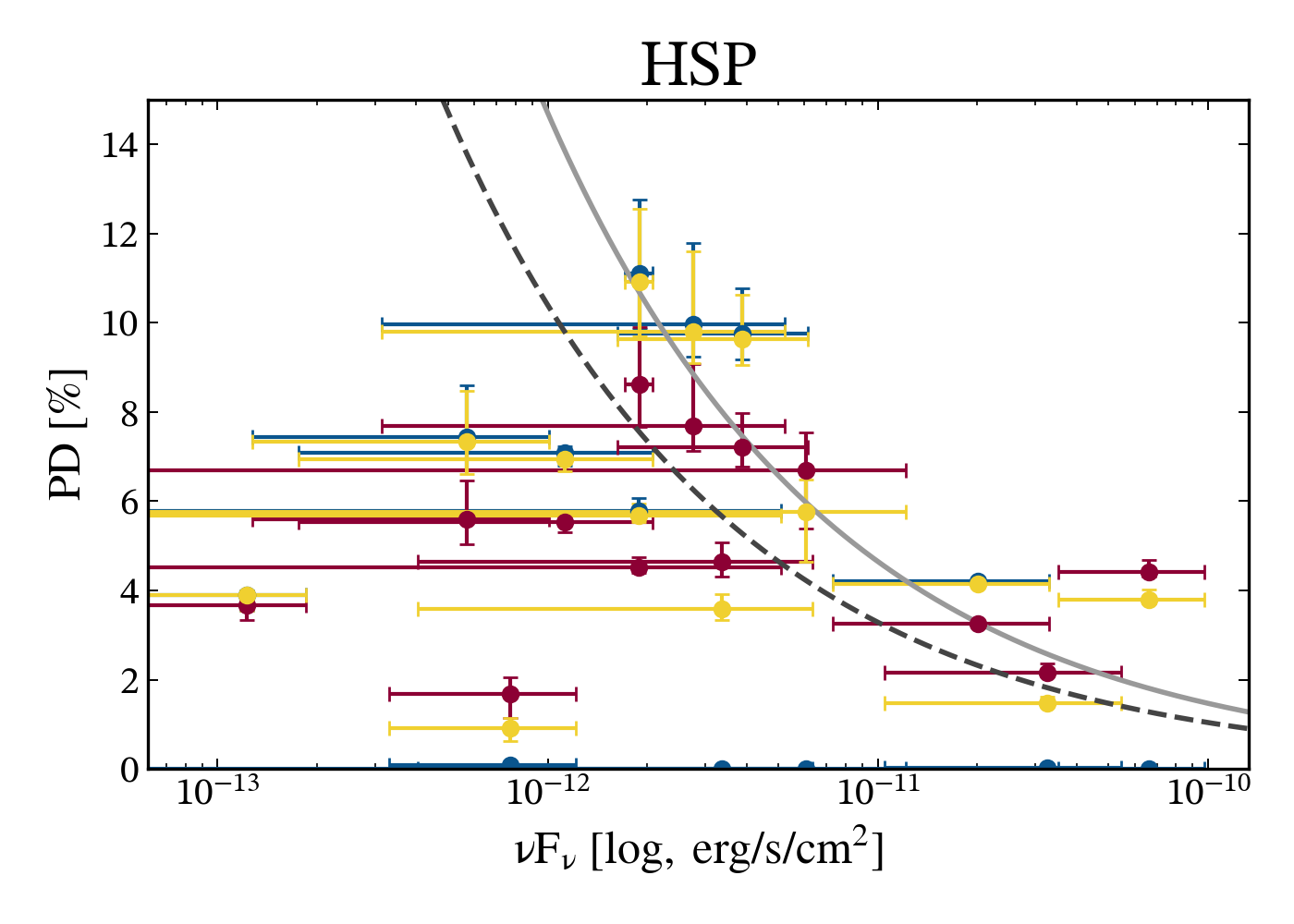}
    \end{subfigure}
    \caption{IXPE band (2-8 keV)}
    \label{fig:mdp_ixpe}
\end{figure*}

\begin{figure*}[!ht]
    \centering
    \begin{subfigure}{0.32\textwidth}
        \centering
        \includegraphics[width=\textwidth]{Images/LSP_all_models_eXTP_MDP.png}
    \end{subfigure}
    \hfill
    \begin{subfigure}{0.32\textwidth}
        \centering
        \includegraphics[width=\textwidth]{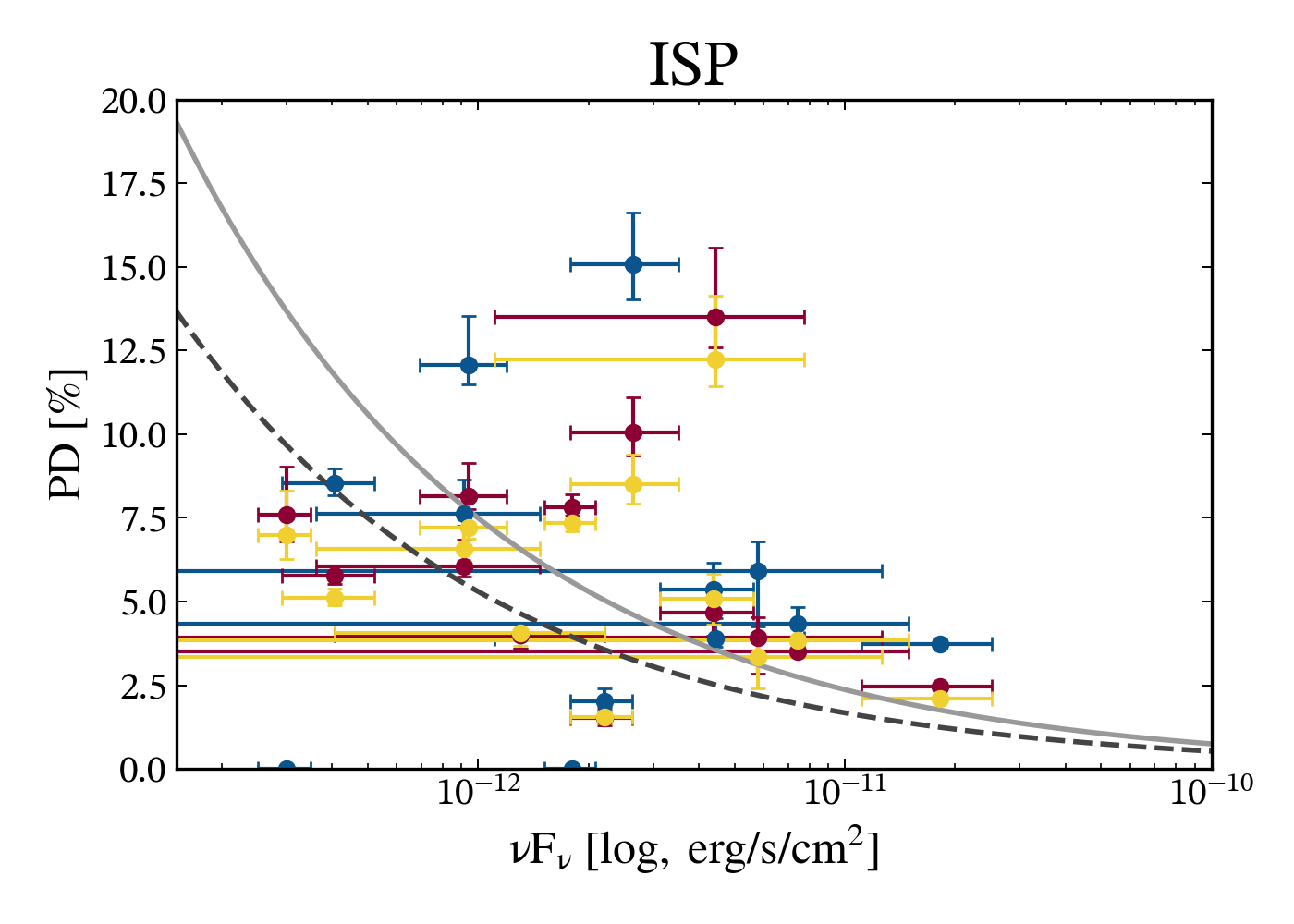}
    \end{subfigure}
    \hfill
    \begin{subfigure}{0.32\textwidth}
        \centering
        \includegraphics[width=\textwidth]{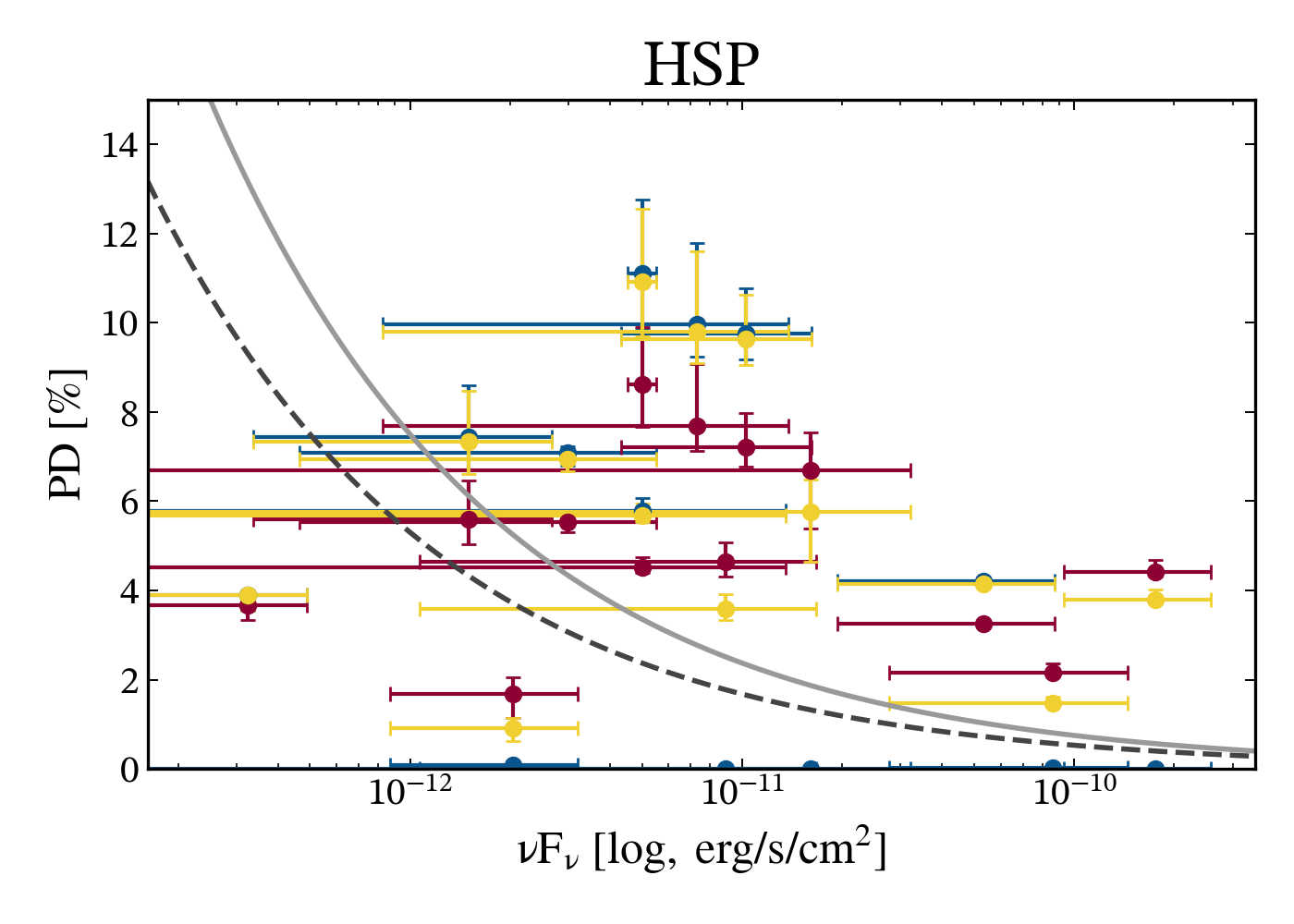}
    \end{subfigure}
    \caption{eXTP band (0.5-10 keV)}
    \label{fig:mdp_extp}
\end{figure*}

\begin{figure*}[!ht]
    \centering
    \begin{subfigure}{0.32\textwidth}
        \centering
        \includegraphics[width=\textwidth]{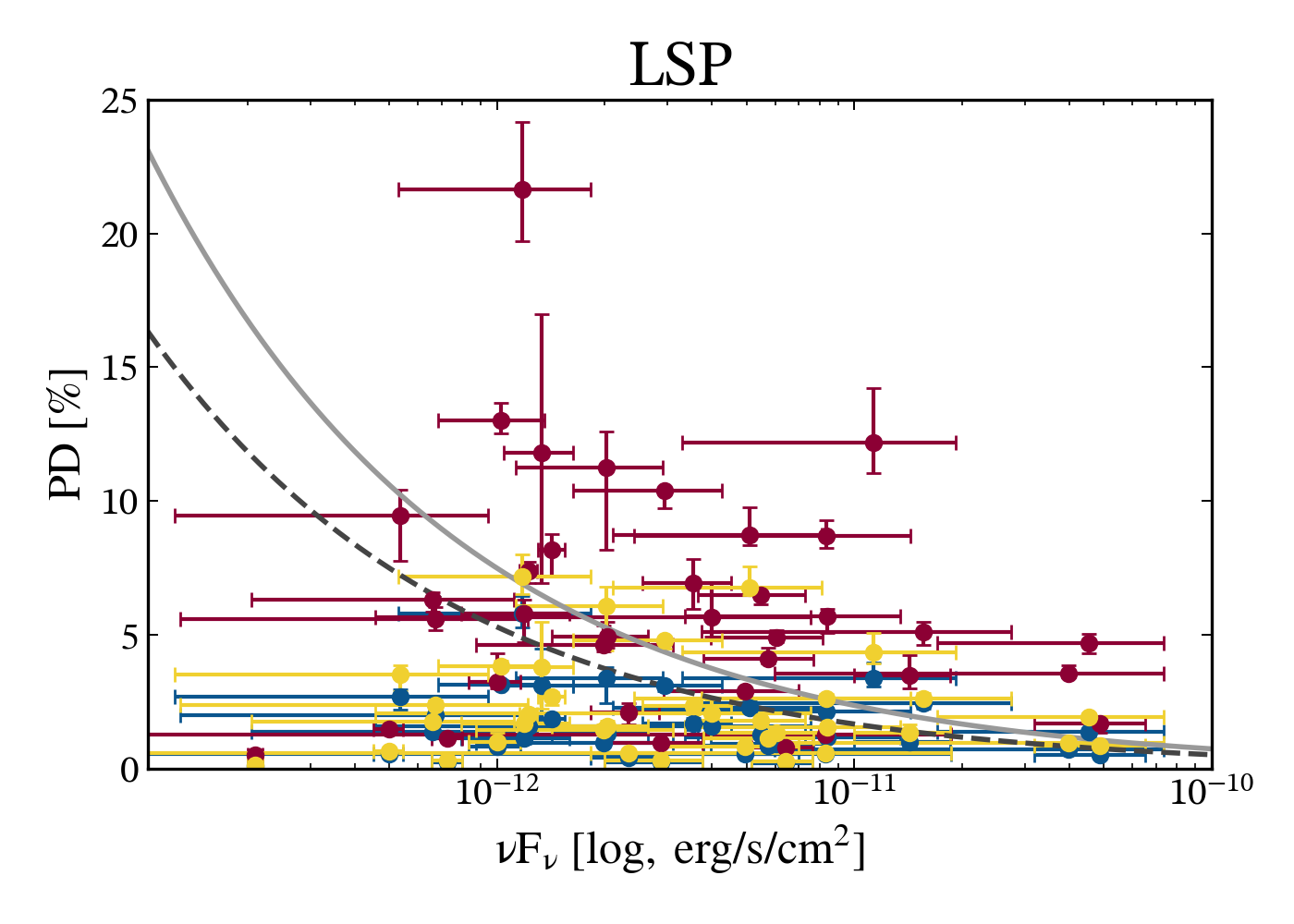}
    \end{subfigure}
    \hfill
    \begin{subfigure}{0.32\textwidth}
        \centering
        \includegraphics[width=\textwidth]{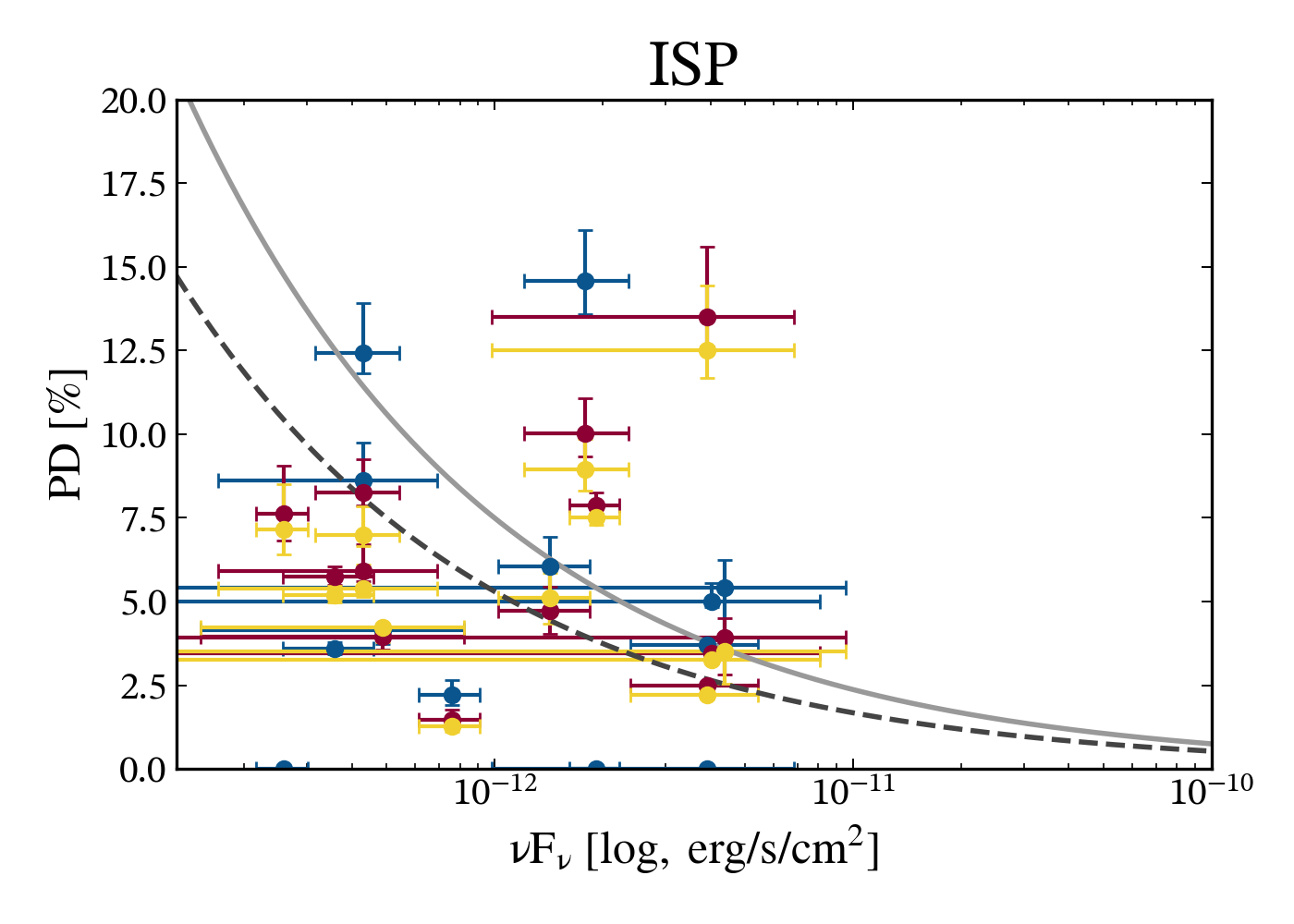}
    \end{subfigure}
    \hfill
    \begin{subfigure}{0.32\textwidth}
        \centering
        \includegraphics[width=\textwidth]{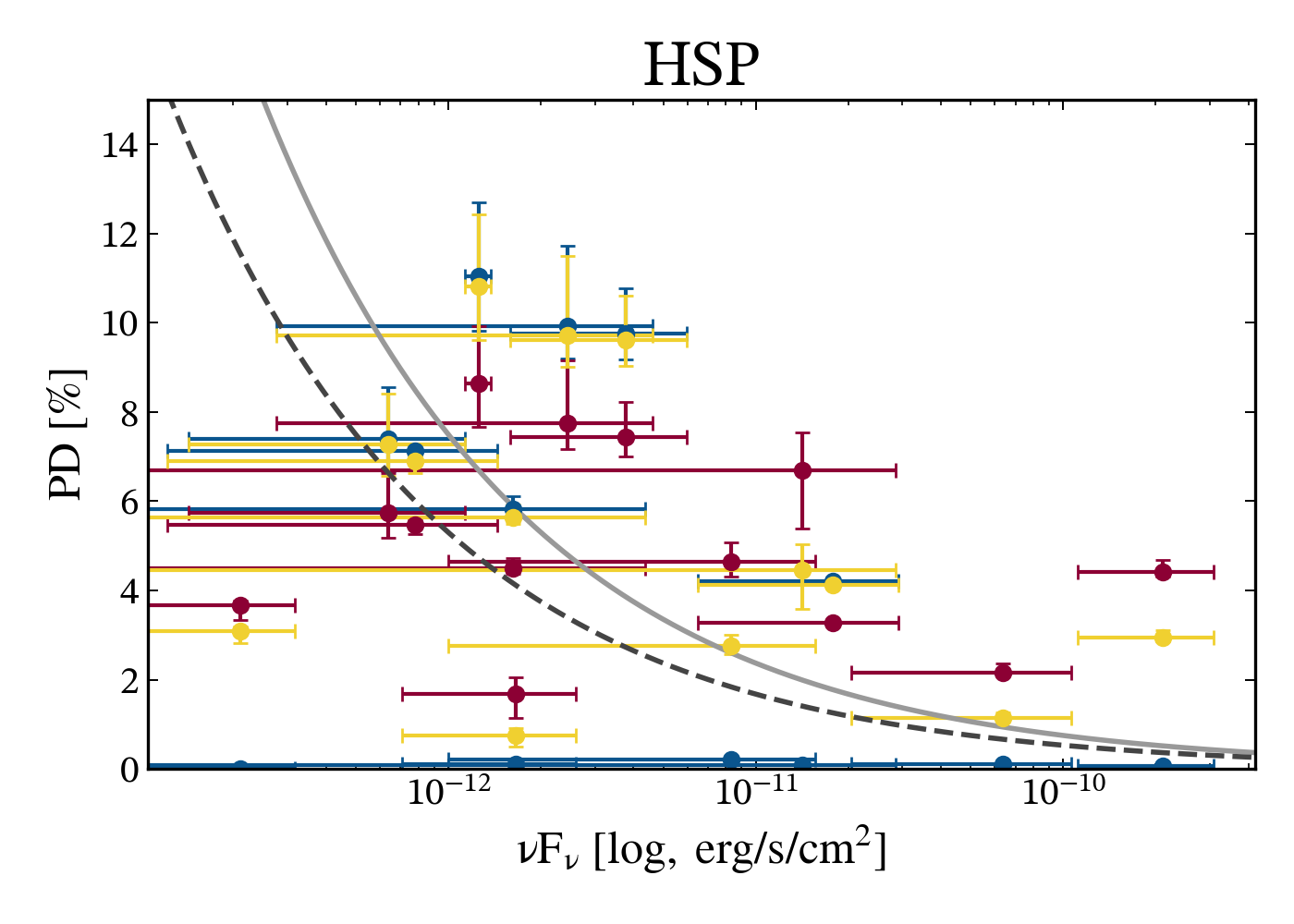}
    \end{subfigure}
    \caption{EXPO band (6-35 keV)}
    \label{fig:mdp_expo}
\end{figure*}

\begin{figure*}[!ht]
    \centering
    \begin{subfigure}{0.32\textwidth}
        \centering
        \includegraphics[width=\textwidth]{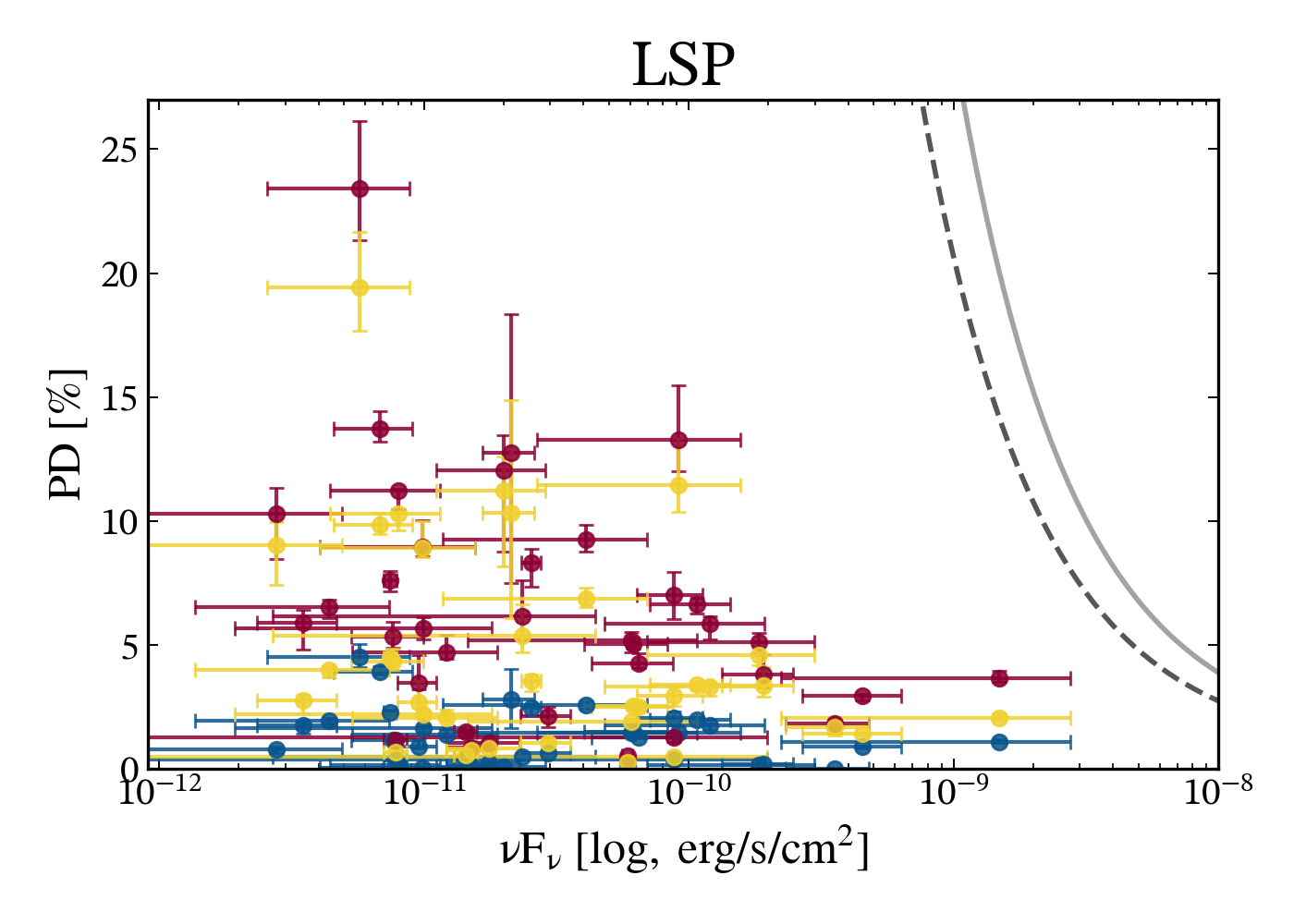}
    \end{subfigure}
    \hfill
    \begin{subfigure}{0.32\textwidth}
        \centering
        \includegraphics[width=\textwidth]{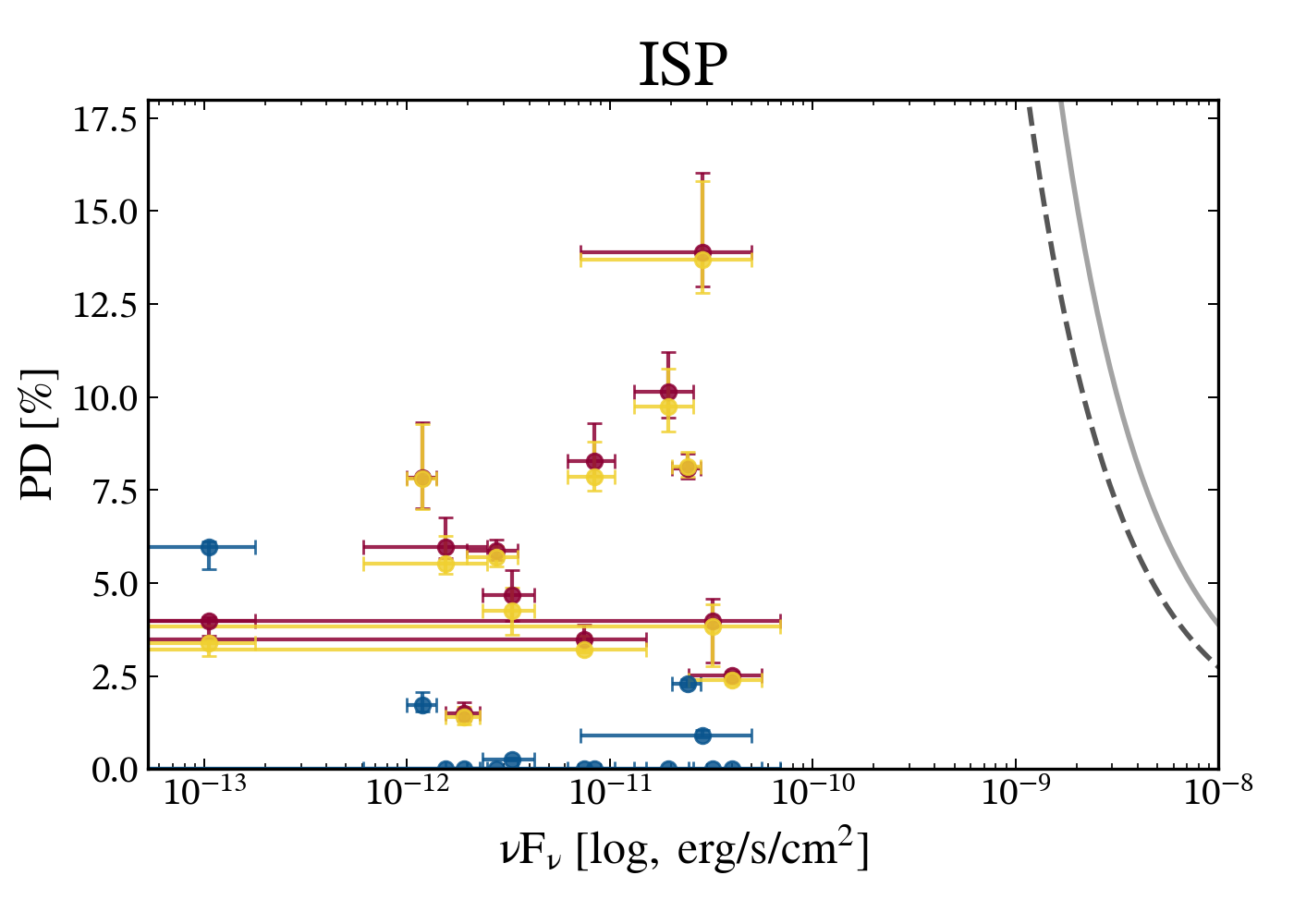}
    \end{subfigure}
    \hfill
    \begin{subfigure}{0.32\textwidth}
        \centering
        \includegraphics[width=\textwidth]{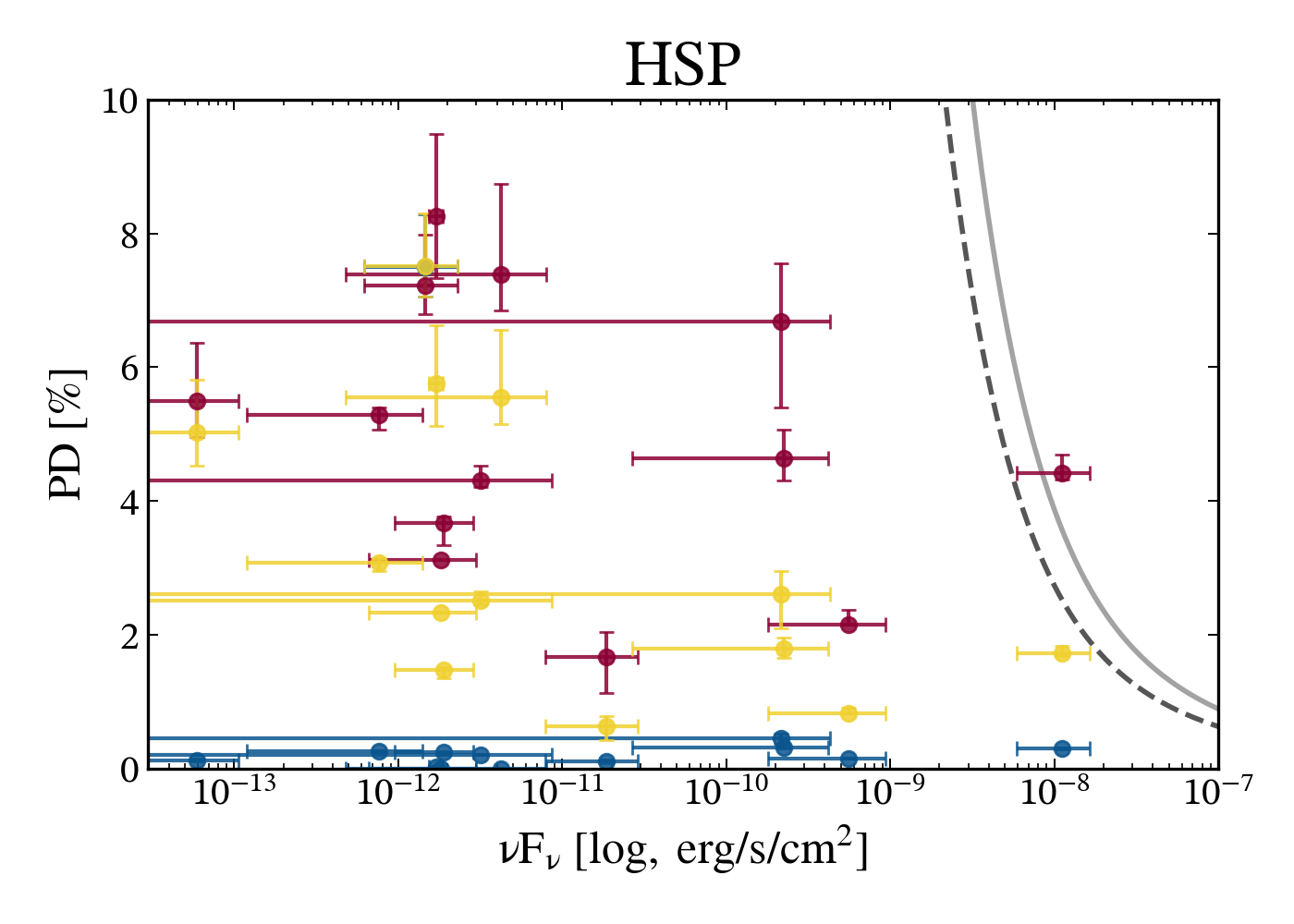}
    \end{subfigure}
    \caption{COSI band (0.2-5 MeV)}
    \label{fig:mdp_cosi}
\end{figure*}

\end{appendix}

\end{document}